\documentclass{bmcart}
\pdfoutput=1

\usepackage[utf8]{inputenc} 

\usepackage{lscape}
\usepackage{graphicx}
\usepackage[all]{xy}

\usepackage{enumerate}
\usepackage{amsmath,amssymb}
\usepackage{mathrsfs}
\usepackage{mathbbol}
\usepackage{graphicx}
\usepackage{scrextend}   
\usepackage{url}
\usepackage{lipsum}
\usepackage{cite}
\usepackage{xcolor}
\usepackage{authblk}
\usepackage{afterpage}

\usepackage[normalem]{ulem}

\usepackage{dsfont}
\usepackage{cancel}

\usepackage{pbox}
\usepackage{bbm}

\newcommand{\vect}[1]{\boldsymbol{#1}}
\newcommand{\blue}[1]{\textcolor[rgb]{0,0,0}{#1}}

\newcommand{\blanco}[1]{\textcolor[rgb]{1,1,1}{#1}}

\startlocaldefs
\endlocaldefs

\begin{document}

\begin{frontmatter}

\begin{fmbox}
\dochead{Research}


\title{Spatio-Chromatic Information available from different Neural Layers via Gaussianization}


\author[
   addressref={aff1},                   
   corref={aff1},                       
   email={jesus.malo@uv.es}   
]{\inits{J}\fnm{Jes\'us} \snm{Malo}}


\address[id=aff1]{
  \orgname{Image Processing Lab, Universitat de Val\`{e}ncia}, 
  \street{Catedr\'{a}tico Escardino},                    %
  \postcode{46980}                                
  \city{Paterna, Valencia},                              
  \cny{Spain}                                    
}


\begin{artnotes}
\end{artnotes}

\end{fmbox}


\begin{abstractbox}

\begin{abstract} 

\emph{How much visual information about the retinal images can be extracted from the different layers of the visual pathway?}
This question depends on the complexity of the visual input, the set of transforms applied to this multivariate input, and
the noise of the sensors in the considered layer.
Separate subsystems (e.g. opponent channels, spatial filters, nonlinearities of the texture sensors) have been suggested to
be organized for optimal information transmission.
However, the efficiency of these different layers has not been measured when they operate together on
colorimetrically calibrated natural images and using multivariate information-theoretic units
over the joint spatio-chromatic array of responses.

In this work we present a statistical tool to address this question in an appropriate (multivariate) way.
Specifically, we propose an empirical estimate of the \blue{transmitted information} by the system based on a recent Gaussianization technique
that reduces the challenging multivariate PDF estimation problem to a set of \blue{simpler} univariate estimations.
Total correlation measured using the proposed estimator is consistent with predictions based on the analytical Jacobian
of a standard spatio-chromatic model of the retina-cortex pathway.
\blue{If the noise at certain representation is proportional to the dynamic range of the response, and one assumes sensors of equivalent noise level, transmitted information shows the following trends: (1)~progressively deeper representations are better in terms of the amount of information about the input, (2)~the transmitted information up to the cortical representation follows the PDF of natural scenes over the chromatic and achromatic dimensions of the stimulus space, (3)~the contribution of spatial transforms to capture visual information is substantially bigger than the contribution of chromatic transforms, and (4)~nonlinearities of the responses contribute substantially to the transmitted information but less than the linear transforms.}


\end{abstract}




\end{abstractbox}
%

\end{frontmatter}



\section{Introduction}


\blue{\emph{Seeing} is all about extracting information about a scene from the neural responses induced by the image of the scene.
Therefore, both visual neuroscientists and image coding engineers are interested in similar information theoretic tools.}

Neuroscience has a long tradition in using information theory both to quantify the performance of neurons~\cite{McKay52} and
to formulate principles that explain observed behavior using the so-called \emph{Efficient Coding Hypothesis}~\cite{Barlow59,Barlow01}.
Information theory is useful at many scales of sensory processing~\cite{Dimitrov11,Friston12,Bialek16}.

On the one hand, a large body of literature studies information transmission starting from a spiking neuron \cite{Strong98},
\blue{including all sorts of additional constraints such as energy or size \cite{Rehn+JCN+2007,Perge09,Laughlin13,Jolivet15,Laughlin15}}. From these
 estimates of transmitted information in single-cells, different summation strategies~\cite{Borghuis08} or independence assumptions~\cite{Koch06}
are considered to give global estimates of the amount of information transmitted by a set of sensors.
This detailed low-level descriptions are certainly the basis for higher-level
behavioral effects, but psychophysics is usually described with more abstract models.
In fact, cascades of \emph{linear+nonlinear} layers of more schematic neurons~\cite{Heeger92,Carandini94,Lennie08,Carandini12}
describe a wide range of psychophysical phenomena, including color~\cite{Abrams07,Fairchild13}, spatial texture~\cite{Watson90,Watson97},
and motion~\cite{Simoncelli98motion}.
On the other hand, these more schematic neural layers where inputs and outputs are arrays of continuous real values (not spike trains) are also studied according to the \emph{Efficient Coding Hypothesis}.
Information-theoretic goals are used to derive physiological behavior but also psychophysical effects. Examples include
the derivation of chromatic opponent channels~\cite{Buchsbaum83},
the saturation of achromatic and chromatic channels~\cite{Laughlin83,MacLeod03b,Laparra12},
the emergence of linear texture sensors (not only achromatic~\cite{Hancock92,Olshausen96}, but also spatio-chromatic~\cite{Ruderman98color,Doi03}, and equipped with chromatic adaptation~\cite{Gutmann14}),
and linear sensors sensitive to motion~\cite{Hyvarinen09}.
Finally, the saturation of the responses of these spatio-temporal sensors has also been derived
from information theoretic arguments in the case of achromatic textures~\cite{Schwartz01,Malo06b} and motion~\cite{Laparra15}.

\vspace{0.25cm}

\blue{In this work we will use these higher-level models~\cite{Carandini12,Abrams07,Fairchild13,Watson90,Watson97,Simoncelli98motion} which, being connected to physiology, are more related to
the psychophysics of color and spatial texture.
Information-theoretic study of psychophysical models may address biologically relevant questions like: \emph{What perceptual behavior is more relevant to encode images?, color constancy or contrast adaptation?}.
\emph{What mechanisms contribute to extract more information about the images?, the chromatic opponent channels or the texture filters?}.
\emph{What is the improvement due to the nonlinearities?, being more flexible is always better?}.
\emph{What kind of images are better represented by the visual system?, smooth achromatic shapes or sharp chromatic patterns?.}}

\blue{Quantitative answers to the above questions require reliable estimators of \emph{transmitted information}
over the cascade of layers. These estimators have to work on high-dimensional vectors, and more importantly,
estimations should not depend on the specific expression of the response so that they could be applied to any model or even to
raw experimental data.}

\blue{In this work we introduce a recent estimator of \emph{mutual information} between multivariate
variables based on Gaussianization~\cite{Laparra11,Johnson19} to address relevant sensory questions such as those stated above.}
\blue{Here we illustrate the kind of answers that can be found through this method by exploring
the transmission of spatio-chromatic information through a
cascade of standard linear+nonlinear layers that address in turn basic
psychophysical phenomena on color and texture~\cite{Abrams07,Fairchild13,Watson90,Watson97}.}
\blue{Illustrative study of a model of known analytic Jacobian~\cite{Martinez18,Martinez19} is convenient because, for some redundancy measures such as \emph{total correlation},
the model-free estimations proposed here can be compared to estimations that use the analytical model.}
\blue{Interestingly, here we discuss that the analytical insight obtained from the model Jacobian for \emph{total correlation} (as done in~\cite{Lyu09,Martinez18,Gomez19})
is not always applicable to predict how the \emph{transmitted information} is going to be.
However, this poses no major problem if the estimator at hand does not depend on the model, which is the case in the proposed method.}
%
%

\blue{The conventional way to check the Efficient Coding Hypothesis goes in the \emph{statistics-to-biology} direction.
In that case, biologically plausible behaviors are shown to emerge from the optimization of specific information-theoretic
measures~\cite{Buchsbaum83,Laughlin83,MacLeod03b,Laparra12,Hancock92,Olshausen96,Ruderman98color,Doi03,Gutmann14,Hyvarinen09,Schwartz01,Malo06b,Laparra15}. On the contrary, following~\cite{Malo10}, here we go in the opposite direction: from \emph{biology-to-statistics}.
Here we take a standard psychophysical model that was not explicitly optimized to encode natural images, and we show it is remarkably good
in terms of \emph{transmitted information}.}

\blue{Therefore, by the nature of the approach, our results should not be misunderstood
as a claim of \emph{infomax} as the ultimate organization goal.
Note that we are not proposing to optimize the system according to that goal.
In fact, a large body of literature includes additional constraints to information
maximization~\cite{Rehn+JCN+2007,Perge09,Laughlin13,Jolivet15,Laughlin15,Olshausen96,MacLeod03b,Laparra12,Laparra15,Gutmann14,Sporns12}.
On the contrary, beyond disputes among specific goals, the method proposed here can be used to include \emph{transmitted information} together with other goals in a combined cost function if the \emph{statistics-to-biology} strategy is preferred.}


%

\section{Materials: illustrative vision model and calibrated images}
\label{Materials}

The use of the Gaussianization tool presented in this work to measure the information transmitted through neural layers
is illustrated in a standard spatio-chromatic retina-cortex model fed with color-calibrated natural stimuli.

In this section we first review the elements of this standard model, and then, we show the distribution of natural images in terms of chromatic contrast, achromatic contrast and luminance used in our experiments.

\subsection{A standard spatio-chromatic psychophysical pathway}
\label{model}

The model considered here for illustrative purposes is a cascade of \emph{linear+nonlinear} layers \cite{Carandini12,Martinez18,Martinez19,Gomez19}.
In this setting, the $i$-th layer takes the array of responses coming from a previous layer $\vect{x}^{(i-1)}$, applies a set of
linear receptive fields that lead to the responses, $\vect{r}^{(i)}$, and these outputs interact to lead to the saturated responses, $\vect{x}^{(i)}$:
\vspace{-0.2cm}
\begin{equation}
  \xymatrixcolsep{2pc}
  \xymatrix{ \cdots \vect{x}^{(i-1)} \,\,\,\,\, \ar@/^1pc/[r]^{\scalebox{1.00}{$\mathcal{L}^{(i)}$}} & \vect{r}^{(i)} \ar@/^1pc/[r]^{\scalebox{1.00}{$\mathcal{N}^{(i)}$}} &
             \vect{x}^{(i)} \cdots
  }
  \label{module}
\end{equation}

Specifically, the model used in the simulations below consist of three of such differentiable and invertible \emph{linear+nonlinear} layers:
\vspace{-0.2cm}
\begin{equation}
  \xymatrixcolsep{2pc}
  \xymatrix{ \vect{x}^{(0)} \ar@/^1pc/[r]^{\scalebox{0.85}{LMS}} & \vect{r}^{(1)} \ar@/^1pc/[r]^{\scalebox{0.85}{Adapt.}} &
             \vect{x}^{(1)} \ar@/^1pc/[r]^{\scalebox{0.85}{ATD}} & \vect{r}^{(2)} \ar@/^1pc/[r]^{\scalebox{0.85}{Satur.}} &
             \vect{x}^{(2)} \ar@/^1pc/[r]^{\scalebox{0.85}{DCT-CSF}} & \vect{r}^{(3)} \ar@/^1pc/[r]^{\scalebox{0.85}{Div.Norm.}} &  \vect{x}^{(3)}
  }
  \label{cascade}
\end{equation}
where we can identify the following standard processing elements:

\paragraph{\textbf{Linear spectral integration.}} The spectral image, $\vect{x}^{(0)}$, is analyzed at each spatial location by linear photoreceptors tuned to \emph{Long}, \emph{Medium} and \emph{Short} wavelengths, in particular we use the standard cone fundamentals LMS in \cite{Stockman00}.
The $\vect{r}^{(1)}$ array contains the LMS retinal images.

\paragraph{\textbf{Chromatic adaptation.}}  We use the simplest chromatic adaptation scheme: the classical Von Kries normalization \cite{Fairchild13}, where the linear LMS signals are divided by an estimate of what is considered to be white in the scene.
The $\vect{x}^{(1)}$ array contains the \emph{Von Kries-balanced} LMS retinal images.

\paragraph{\textbf{Linear opponent color space.}} The LMS signals at each spatial location are linearly recombined into an \emph{opponent} representation with Achromatic (luminance), Tritanopic (red-green) and Deuteranopic (yellow-blue) sensors.
Specifically, $\vect{r}^{(2)}$ contains the ATD images of Jameson \& Hurvich opponent sensors \cite{Jameson59,Capilla98,Stockman11}.

\paragraph{\textbf{Weber-like saturation.}} The linear ATD responses saturate \cite{Krauskopf92,Romero93} to give, brightness (nonlinear A), and nonlinear versions of images T and D. This saturation can be modeled in sophisticated ways with psychophysical \cite{CIELab76,ciecam97,Fairchild13} or statistical
grounds \cite{MacLeod03b,Laparra12}, but in $\vect{x}^{(2)}$ we will use a simple dimension-wise nonlinearity using a $\gamma<1$ exponent with
parabolic correction at the origin \cite{Martinez18} to avoid singularities in the Jacobian.
This exponential with fixed $\gamma$ is the simplest model for the Weber-like luminance-brightness
relation and the observed saturation in chromatic opponent channels \cite{Stockman11}.

Up to $\vect{x}^{(2)}$ the model consists of purely chromatic transforms that operate at each spatial location.
In these initial layers spatial context is not considered except for the scarce use made in chromatic adaptation to estimate the \emph{white}.
The final \emph{linear+nonlinear} layer addresses spatio-chromatic texture.

\paragraph{\textbf{Linear texture filters and Contrast Sensitivity.}}
In the simple model considered here, spatial transforms are applied over each A, T, and D, images in parallel. Neglecting the interactions between ATD images is consistent with the results found in analyzing the spatio-chromatic statistics of natural images: the chromatic variation of the statistical filters
follows Von Kries-corrected ATD directions regardless of the spatial distribution
of the receptive field \cite{Gutmann14}.
Here we use a crude local-DCT model for the local-oriented receptive fields in V1: the local oscillations applied to the $A$ part of the image account for the achromatic texture sensors~\cite{Ringach02}, and applied to $T$ and $D$ arrays account for the double-opponent cells~\cite{Shapley11}.
The gain of these linear filters (receptive fields) is weighted according to their frequency using achromatic and chromatic Contrast Sensitivity Functions (CSFs) \cite{Campbell68,Mullen85}.
The weights for the local-DCT functions are based on the CSFs of the Standard Spatial Observer defined for sinusoids \cite{Watson02}, and a procedure to transfer the weights from one domain to another \cite{Malo97}.
The bandwidth of the achromatic and chromatic CSFs is markedly different.
This bandwidth is related to the frequency response of magno, parvo, and konio cells~\cite{Uriegas94,DeAngelis97,Shapley11}.
The array $\vect{r}^{(3)}$ consists of the spatial transforms of the A patch, the T patch, and the D patch,
frequency weighted and stacked one after the other.

\paragraph{\textbf{Nonlinear interactions between texture sensors.}} Following \cite{Heeger92,Carandini94,Carandini12,Watson97,Lennie08}
the saturation of the sensors tuned to chromatic/achromatic textures has been modeled using a psychophysically-tuned Divisive Normalization \cite{Malo06a,Martinez18,Martinez19}.
Note that alternative cortical nonlinearities such as Wilson-Cowan equations~\cite{Wilson73} have been found to be equivalent to Divisive Normalization~\cite{Malo19}.
The same parameters for the normalization have been used for the achromatic part and the chromatic parts of the response. As in the linear case, no interaction has been considered between the A, T and D parts.

\subsection{\blue{Plausibility of the psychophysical model}}
\label{sect_plausible}

\begin{figure}[t]
   \begin{centering}
    \hspace{-0.5cm}
    \includegraphics[width=1.00\textwidth]{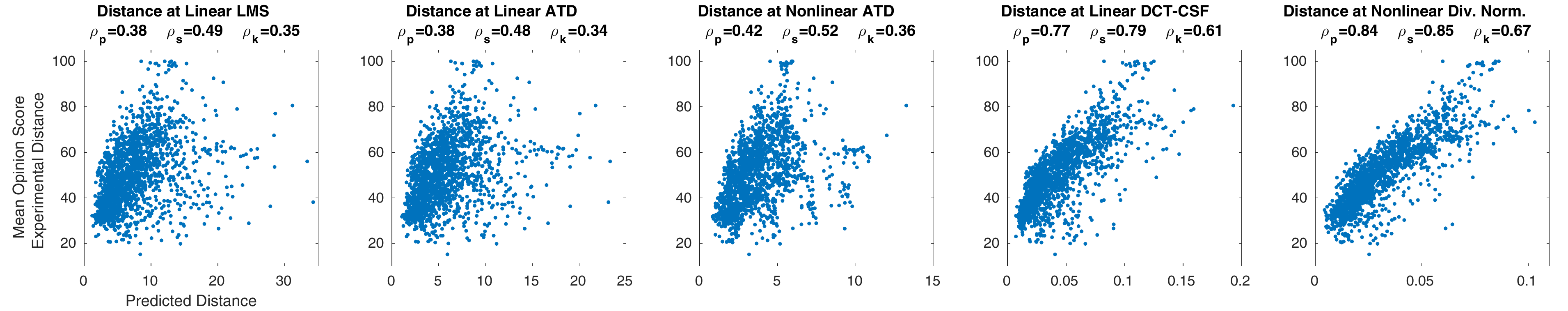}
    \end{centering}
    \vspace{-0.8cm}
    \caption{Correlation with human opinion as additional layers are added to the model. As different illuminations are
    not present in the considered database~\cite{ponomarenko08}, we cannot assess the effect of the Von-Kries normalization on LMS
    (missing intermediate step between linear LMS and linear ATD).}
    \label{biol_plaus1}
\end{figure}

\blue{The specific details of the cascade of standard operations listed above is available here\footnote{\small{Our implementation of this standard model, based on the Matlab libraries Colorlab and Vistalab~\cite{Colorlab02,Vistalab97}, is available at http://isp.uv.es/code/visioncolor/vistamodels.html}}.
In this section we illustrate the plausibility of the specific blocks (with specific parameters)
considered in this work by predicting results in image distortion psychophysics.}


\blue{In image quality databases~\cite{ponomarenko08} human viewers assess the visibility of a range of distortions seen on top of
natural images. A vision model is good if its predictions of visibility correlate with human opinion.
The visibility of a distortion using a psychophysical response model is done by measuring the distance between the response vectors to the original and the distorted images~\cite{Wang04,Laparra10a}.
In this context, we made two simple numerical experiments:
(1) we checked if the consideration of more and more standard blocks
in the model leads to consistent improvements of the correlation with human opinion, and
(2) we checked how substantial modifications to the standard blocks affect its performance.
In particular, (a) we changed the flexibility of the baseline masking model
by decreasing or increasing the semisaturation constant in the denominator of the Divisive Normalization.
This leads to more flexible and more rigid models respectively, and
(b) we generated a totally rigid model by changing all nonlinear parts by
identity transforms.}

\blue{Fig.~\ref{biol_plaus1} confirms that the progressive consideration of the blocks leads to consistent improvements.
Moreover, the resulting model is reasonable:
the final correlation with human behavior in this non-trivial scenario is similar to state-of-the-art image quality metrics~\cite{Watson02,Laparra10,Martinez18,Martinez19,Perceptnet19}.}
\blue{Moreover, Table~\ref{biol_plaus2} shows that the standard baseline model is better than more adaptive and more rigid versions of the model. Of course, the correlation substantially increases with the spatial size of the image patches
and receptive fields, but a model with \emph{exactly the same parameters} on small image patches shows the same trend when considering the progression layers.}
\blue{These results illustrate the meaningfulness of the considered psychophysical blocks and its appropriate behavior
when functioning together.}

\begin{table}[t!]
\caption{Pearson correlation with human viewers using different building blocks (or model layers).}
\label{biol_plaus2}
      \begin{tabular}{lcccccc}
        \hline\\[-0.1cm]
                            & Spatial Extent & $\vect{r}^{(1)}$ & $\vect{r}^{(2)}$ & $\vect{x}^{(2)}$ & $\vect{r}^{(3)}$ & $\vect{x}^{(3)}$ \\ \hline\\[-0.2cm]
        More flexible model     & (0.27 deg) & 0.38 & 0.38 & 0.42 & 0.77 & 0.51 \\
        \textbf{Baseline model} & (0.27 deg) & 0.38 & 0.38 & 0.42 & 0.77 & \textbf{0.84} \\
        More rigid model        & (0.27 deg) & 0.38 & 0.38 & 0.42 & 0.77 & 0.79 \\
        Totally rigid model     & (0.27 deg) & 0.38 & 0.38 & 0.38 & 0.68 & 0.68 \\
        \hline\\[-0.2cm]
        Baseline model          & (0.05 deg) & 0.26 & 0.27 & 0.31 & 0.37 & 0.40 \\
        \hline
      \end{tabular}
\end{table}


Finally, intuition about the information-theoretic performance of the considered model may be obtained by visualizing the geometrical effect of the series of transforms on the manifold of natural images. This intuition is discussed in Appendix~\ref{visual_manifolds}.

\subsection{Calibrated natural images}
\label{images}

\begin{figure}[b!]
   \centering
    \includegraphics[width=0.9\textwidth]{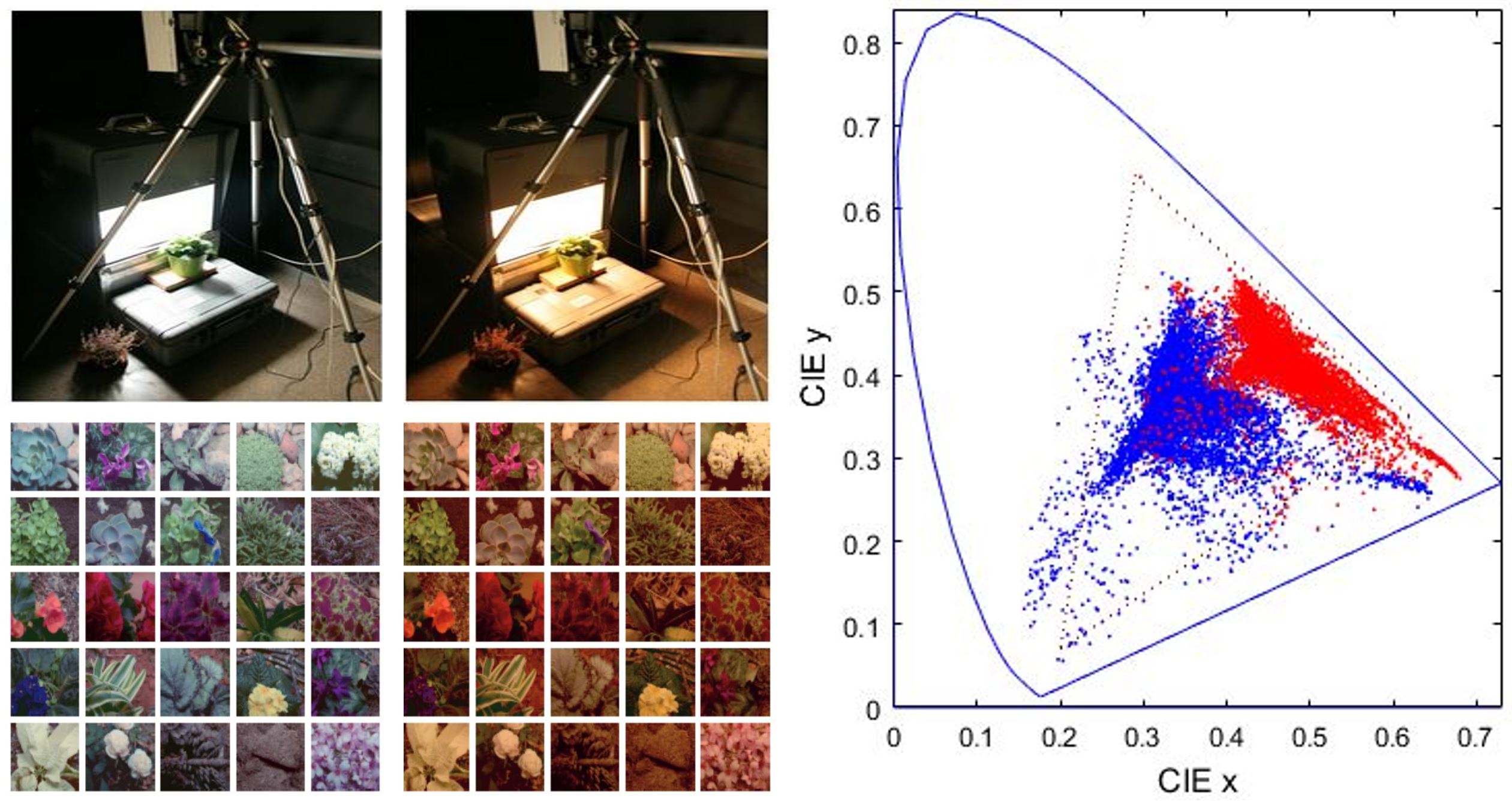}
     \vspace{-0.25cm}
    \caption{Calibrated images used in the experiments: natural scenes under CIE D65 (left image panel) and CIE A (right image panel) illuminations.
    The CIE xy diagram shows the shift in chromaticity (samples under D65 in blue and samples under A in red).}
    \label{data}
\end{figure}

The IPL color image database \cite{Laparra12,Gutmann14,Laparra15} is well suited to study color adaptation because its controlled illumination under CIE A and CIE D65 allows straightforward application of Von Kries adaptation.
With the knowledge of the illumination there is no need for extra approximations such as the gray-world assumption to estimate the white.
Acquisition of the images in the controlled setting and resulting CIE xy data is illustrated in Fig.~\ref{data}.

Alternative calibrated choices could have been the spectral-image datasets~\cite{Foster15,Foster16} or the tristimulus-calibrated dataset~\cite{Parraga09}, which also provide the illumination information from gray spheres located in the scenes.

We extracted $19\cdot10^6$ image patches from the database (expressed in CIE XYZ tristimulus values) and transformed them into the
linear LMS representation.
In fact, in this work we will consider the behavior of the model from $\vect{r}^{(1)}$ on. This amounts to considering that the input to the system is the set of linear LMS images.
In order to keep the dimensionality small for a proper comparison of the empirical and theoretical estimates of information quantities done below, we kept the spatial extent small: only $3\times3$ pixels.
As a result, the input stimuli, $\vect{r}^{(1)}$, and the response arrays live in $27$-dimensional spaces.

In order to check whether the \emph{transmitted information} through the system is adapted to the statistics of the natural input
we consider the distribution of the considered samples over three relevant visual features: the average
luminance, the achromatic contrast, and the chromatic contrast of the pattern in the image.
To define these features LMS images are expressed in the Jameson \& Hurvich ATD space using no chromatic
adaptation. In this representation, average luminance in $cd/m^2$ units is simply the average of the A image. The achromatic contrast is defined as the RMSE contrast of the A image (standard deviation divided by the mean luminance). Similarly, the chromatic contrast is defined as the mean of the RMSE contrasts of the T and D images, where the corresponding standard deviation is divided by the norm of the average color in the patch.

\begin{figure}[t]
   \centering
    \hspace{0cm}
    \includegraphics[width=0.97\textwidth]{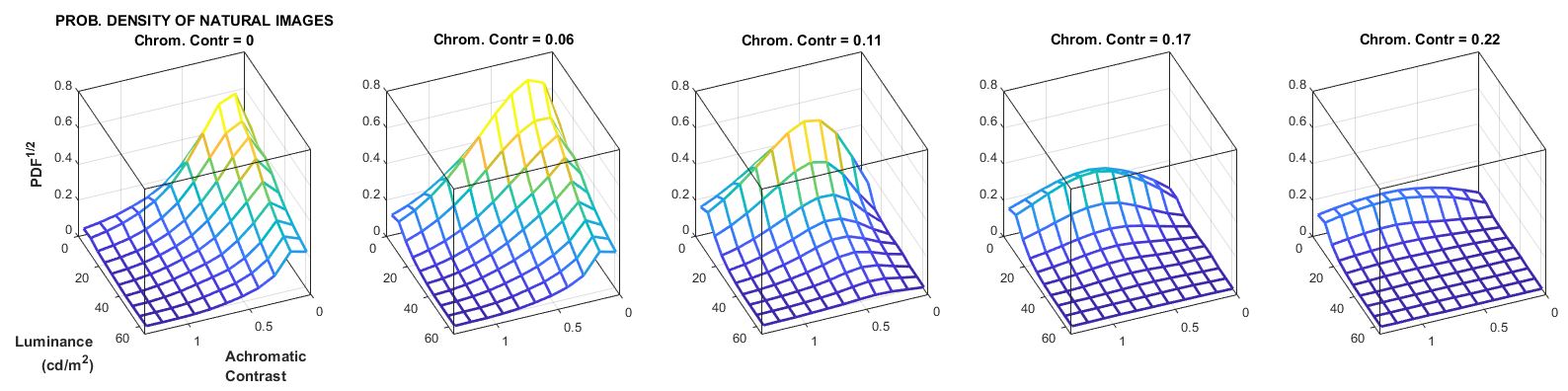}
     \vspace{-0.5cm}
    \caption{Probability of natural scenes as a function of \emph{chromatic contrast}, \emph{achromatic contrast}, and \emph{luminance}:
    dark and smooth patches are more frequent.}
    \label{jointPDF}
\end{figure}

By computing these three features for the $19 \cdot 10^6$ images, we find a markedly
non-uniform density: image patches are mostly dark, and the variance of the deviations
with regard to the mean is small both in luminance and in color (see Fig.~\ref{jointPDF}).
\blue{This distribution shows that in nature some regions of the image space are more frequent that others. The Efficient Coding Hypothesis generically argues that the performance of the visual system should be shaped by this uneven distribution.
Here we propose a quantitative comparison. In the Results section~\ref{results} we plot the information transmitted up to the cortical sensors (at layer $\vect{x}^{(3)}$ in Eq.~\ref{cascade}) over the luminance/contrast dimensions to check if the transmitted information for specific images is related to how frequent these images are}.

\section{Methods: transmitted information from Gaussianization}

In a sensory system where the input undergoes certain deterministic transform, $S$, but the sensors are noisy:
\vspace{-0.4cm}
\begin{equation}
  \xymatrixcolsep{2pc}
  \xymatrix{\vect{r} \, \ar@/^1.25pc/[r]^{\scalebox{1.00}{$S$}} & \,\,\;\;\;\;\;\;\;\;\;\;\;\;\;\;\;\; \vect{x} = s(\vect{r}) + \vect{n}
  }
  \label{setting}
\end{equation}
\vspace{-0.7cm}

\noindent
\blue{the mutual information between input and output, $I(\vect{r},\vect{x})$, is the information about the input vector $\vect{r}$ that is shared by the response vector $\vect{x}$~\cite{Cover06} (chapt.2).
We will refer to $I(\vect{r},\vect{x})$ as the information about the input captured by the response, or as the information transmitted from the input to the response. This mutual information $I(\vect{r},\vect{x})$ is also the information about the input available at the representation $\vect{x}$.}


In the context of the \emph{Efficient Coding Hypothesis} it is interesting to compare
different image representations $s(\vect{r})$
in terms of the amount of information they capture about the input for some given noise, $\vect{n}$.
This would tell us about the relative contribution of the different operations in a model to the get a better representation of the visual signal.
%
%
Ideally, we'd like to describe the trends of this performance measure, $I(\vect{r},\vect{x})$, from the analytical description of the response, $s(\vect{r})$,
but this may not be possible \blue{as discussed later in Section~\ref{discussion}}. Therefore, empirical estimations from stimulus-response pairs are very interesting.

Estimation of $I(\vect{r},\vect{x})$
directly from samples and using the definitions based on the PDFs is not straightforward:
it implies the estimation of multivariate PDFs and this challenging problem would introduce substantial bias in the results.
This is also true for surrogates for performance such as the reduction of redundancy measured by the reduction in \emph{total correlation} $\Delta T(\vect{r},\vect{x})$~\cite{Lyu09}.

Using equivalences of $I$ with other quantities its estimation can be reduced to entropy estimations or total correlation estimations. Definitions based on entropy could use the estimator of Kozachenko-Leonenko \cite{Kozachenko87}, as done in \cite{Ivan13,FosterJOSA18} to compute the information shared by corresponding scenes under different illuminations.

In this work we solve the estimation of the transmitted information up to different layers of the visual pathway through the relation between the \emph{transmitted information}, $I$, and the \emph{total correlation} $T$.
In particular, we use a novel estimator of $T$ which only relies on (easy-to-compute) univariate density estimations: the Rotation-Based Iterative Gaussianization (RBIG) \cite{Johnson19}.


\blue{The RBIG is a cascade of $L$ \emph{nonlinear+linear} layers, and the $l$-th layer is made of marginal Gaussianizations, $\Psi^{(l)}(\vect{x}^{(l)})$, followed by a rotation, $R^{(l)}$.
Each of such layers is applied on the output of the previous layer:}
\begin{equation}
       \blue{\vect{x}^{(l+1)} = R^{(l)} \cdot \Psi^{(l)}(\vect{x}^{(l)})}
       \label{rbig}
\end{equation}
For a big enough number of layers, this invertible architecture is able to transform any input PDF, $p(\vect{x}^{(0)})$,
into a zero-mean unit-covariance multivariate Gaussian even if the chosen rotations are random \cite{Laparra11}.
\blue{Theoretical convergence is obtained when the number of layers tends to infinity.
However, in practical situations early stopping criteria can be proposed taking into account the uncertainty associated with a finite number of samples \cite{Laparra11}.}
Convergence even with random rotations implies that both elements of the transform are straightforward: univariate equalizations and random rotations.
\blue{The differentiable and invertible nature of RBIG make it a member of the Normalizing Flow family~\cite{WorkshopINNF19,WorkshopINNF20}.
Within this general family, differentiable transforms with the ability to remove all the structure of the PDF of the input data
are referred to as \emph{density destructors}~\cite{Inouye18}.
By density destruction, the authors in~\cite{Inouye18} mean a transform of the input PDF into a unit-covariance Gaussian or into a $d$-cube aligned with the axes. The considered Gaussianization~\cite{Laparra11,Johnson19} belongs to this family by definition.}
\blue{\emph{Total correlation} describes the redundancy within a vector, i.e. the information shared by the univariate variables~\cite{Watanabe60,Studeny98}. Note that strong relations between variables indicates a rich structure in the data.}

\blue{Density destruction together with differentiability is useful to estimate the total correlation within a vector $T(\vect{x}^{(0)})$.
Imagine that the considered RBIG transforms the PDF of the input $\vect{x}^{(0)}$ into a Gaussian through the application of $L$ layers ($L$ individual transforms).  As the redundancy of the Gaussianized signal, $g_{\vect{x}}(\vect{x}^{(0)}) = \vect{x}^{(L)}$, is zero, the redundancy of the original signal, $T(\vect{x}^{(0)})$, should correspond to the cumulative sum of the individual variations, $\Delta T^{(l)}$ with $l=1, \ldots L$, that take place along the $L$ layers of RBIG, while converting the original variable $\vect{x}$, into the Gaussianized variable $g_{\vect{x}}(\vect{x})$.}
Interestingly, the individual variation in each RBIG layer only depends on (easy to compute) univariate negentropies, therefore, after the $L$ layers of RBIG, the total correlation is \cite{Laparra11}:
\begin{equation}
    T(\vect{x}) = \sum_{l = 1}^L\Delta T(\vect{x}^{(l-1)},\vect{x}^{(l)}) = \sum_{l=1}^{L} J_{m}(\vect{x}^{(l)})
    \label{total_with_rbig}
\end{equation}
where the marginal negentropy of a $d$-dimensional random vector is given by a set of $d$ univariate divergences $J_m(\vect{v}) = \sum_{i=1}^d D_{\mathrm{KL}}(p(v_i)|\mathcal{N}(0,1))$. Therefore, using RBIG, the challenging problem of estimating one $d$-dimensional joint PDF to compute $T(\vect{x})$
reduces to solve $d\times L$ univariate problems.
Moreover, \blue{as opposed to the computation of variations of total correlation depending on the model transform, as for instance Eq.~\ref{deltaT} discussed
below in section \ref{discussion}, RBIG estimation is model-free and does not involve any averaging over the whole dataset.}

\blue{In the density destructor framework, where $T$ is easy to compute using RBIG, the transmitted information from the input LMS image, $\vect{r}$, to any of the considered layers downstream, $\vect{x}$, namely $I(\vect{r},\vect{x})$, could be obtained from $T$ in two ways:}
\blue{
\begin{eqnarray}
       I(\vect{r},\vect{x}) &=& T([\vect{r},\vect{x}]) - T(\vect{r}) - T(\vect{x})  \label{eq:mirbig2} \\
       I(\vect{r},\vect{x}) &=& T([g_{\vect{r}}(\vect{r}),g_{\vect{x}}(\vect{x})]) \label{eq:mirbig}
\end{eqnarray}
}
\blue{where the relation in Eq.~\ref{eq:mirbig2} is straightforward from the definitions of $I$ and $T$ in terms of entropies, and the sum would imply three Gaussianization transforms: one for the input, one for the responses, and an additional one for the concatenated input-response vectors $[\vect{r}, \vect{x}]$.}
\blue{Eq.~\ref{eq:mirbig} also implies three Gaussianization transforms:
the variables $g_{\vect{r}}$ and $g_{\vect{x}}$ come from the Gaussianization transforms of the images and the neural responses respectively, and then we make a single computation of total correlation for the concatenated variable $[g_{\vect{r}}(\vect{r}),g_{\vect{x}}(\vect{x})]$ through an extra Gaussianization.
It is important to note that, as opposed to Eq.~\ref{eq:mirbig2}, the strategy represented by Eq.~\ref{eq:mirbig} only implies \emph{one} computation of $T$, not three.}

Eq.~\ref{eq:mirbig} is possible because $I$ does not change under invertible transformations applied separately to each dataset~\cite{Kraskov04}.
Therefore, $I(\vect{r},\vect{x}) = I(g_{\vect{r}}(\vect{r}),g_{\vect{x}}(\vect{x}))$. Since we removed $T$ within each individual dataset by applying individual density destructors, the only redundant information that remains in the concatenated vectors is the one shared by the original datasets, therefore $I(g_{\vect{r}}(\vect{r}),g_{\vect{x}}(\vect{x})) = T([g_{\vect{r}}(\vect{r}),g_{\vect{x}}(\vect{x})])$, and hence Eq.~\ref{eq:mirbig}.
See \cite{Johnson19} for more elaborate proof.

\blue{These two strategies, Eqs.~\ref{eq:mirbig2} and~\ref{eq:mirbig}, involve different number of computations of $T$. Therefore their accuracy will depend on the nature of the data.
Appendix~\ref{appPerformance} compares the estimations of transmitted information given by Eqs.~\ref{eq:mirbig2} and~\ref{eq:mirbig} (and other estimators~\cite{Kozachenko87,ITE,Ivan13}) in scenarios where analytical solutions are available.
The result of such analysis is that for image-like heavy tailed variables the sum in Eq.~\ref{eq:mirbig2} leads to more error. Therefore, results in the experimental section are computed using Eq.~\ref{eq:mirbig}.}


\section{Experiments}
\label{experiments}

\blue{In this work we propose to use RBIG to study the information transmission in a standard psychophysically-tuned spatio-chromatic vision model.
We conduct experiments to quantify (1)~how much visual information is captured by the different image representations along the layers considered in Section~\ref{sect_plausible}, and (2)~how much information is transmitted for images with different visual features (e.g. different \emph{luminance}, \emph{achromatic contrast}, or \emph{chromatic contrast} as considered in Section~\ref{images}).}

\blue{Here we describe the magnitudes measured, then we describe the assumptions on the noise made throughout the experiments, and finally we describe \emph{global} and \emph{local} experiments (i.e. made for \emph{all} visual scenes, or for scenes with \emph{specific features}, respectively).}

\subsection{Measurements of transmitted information and redundancy reduction}
\blue{In the experiments, we measure the mutual information between the LMS input and different layers of the considered networks, $I(\vect{r}^{(1)},\vect{x})$, where $\vect{x}$ stands for one of the considered image representations (or layers in Eq.~\ref{cascade}). Therefore, $I(\vect{r}^{(1)},\vect{x})$ is the amount of information transmitted up to layer $\vect{x}$, or available at layer $\vect{x}$.
We also check how the input redundancy at the retinal representation, measured in terms of \emph{total correlation}, gets reduced along the different image representations, i.e. we compute $\Delta T(\vect{r}^{(1)},\vect{x}) = T(\vect{r}^{(1)}) - T(\vect{x})$ for the different layers, $\vect{x}$.}

\blue{The RBIG estimator of $I$ uses Eq.~\ref{eq:mirbig}. The RBIG estimator of $\Delta T$ uses Eq.~\ref{total_with_rbig} to estimate $T(\vect{r}^{(1)})$ and $T(\vect{x})$, and then it subtracts these results.}

\blue{Regarding the ability to interpret the visual world, measuring the information available at certain image representation, 
is more related to the visual function than measuring the (more technical) amount of redundancy reduced at that layer.
Redundancy reduction is sometimes taken as a convenient surrogate of transmitted information, but, as discussed below in Section~\ref{discussion}, $\Delta T$ and $I$ are not always aligned.
However, the analysis of $\Delta T$ is technically interesting because
alternative estimators of $\Delta T$  based on the analytical expression of the vision model can be used to check the accuracy of the RBIG estimators.}
\blue{As the alternative estimator of $\Delta T$ discussed in Section~\ref{discussion}, Eq.~\ref{deltaT}, only depends on the analytical expression of the model and on univariate entropies (for which very reliable estimators exist~\cite{ITE}), it will be referred to as \emph{theoretical estimation}. The RBIG estimations of $\Delta T$ will be compared to this  \emph{theoretical estimation} given by Eq.~\ref{deltaT}.
And checking $\Delta T$ is important because the proposed estimator of $I$ relies on measures of $T$, as seen in Eqs.~\ref{eq:mirbig2} and~\ref{eq:mirbig}.}

\subsection{Assumptions on the noise for illustration purposes}
\label{noise_ass}
\blue{Transmitted information between two layers of a network depends on the noise in the response.
The noise in psychophysical systems is related to the discrimination ability of the observers~\cite{Ahumada87,Burgess88},
but the nature of the noise is still under debate~\cite{Messe06,Neri10}.
Noise may have different sources and its amount may depend on attention~\cite{Goris14,MorenoBote14,Kanitscheider15}.}

\blue{The specific debate on the noise is out of the scope of this work.
However, the advantage of the RBIG estimate of transmitted information is that it doesn't rely on a specific analytical PDF of the noise, but only on the availability of noisy samples. Therefore, it can handle responses corrupted with arbitrary noise sources.}

\blue{For illustration purposes, the experiments to estimate $I$ consider a crude noise model using the following assumptions:}
\begin{itemize}
    \item \blue{\textbf{Assumption 1: single-step transforms.} In the path from the input up to the $i$-th layer we assume that noise is added to the signal \emph{only} at the $i$-th~layer. This choice is equivalent to assuming that the mapping into the $i$-th representation is a \emph{single-step transform} in which all the intermediate stages are noise-free.}
    \item \blue{\textbf{Assumption 2: fixed signal-to-noise ratio (SNR).} The additive noise is assumed to be Gaussian with diagonal covariance. Moreover, each marginal standard deviation is assumed to be proportional to the mean amplitude of the signal for that coefficient at the $i$-th layer. Specifically, we set the noise standard deviation at 5\% of the amplitude of the signal in every case.}
\end{itemize}
\vspace{0.3cm}
\noindent
\blue{\emph{Assumption 1} should be understood correctly to avoid misinterpretations of the results.
For instance, considering $\vect{r}^{(1)} \rightarrow \vect{x}^{(2)}$ and  $\vect{r}^{(1)} \rightarrow \vect{x}^{(3)}$ as \emph{single-step transforms}
means that noise is added
in $\vect{x}^{(2)}$ in the first case, and in $\vect{x}^{(3)}$ -but not in $\vect{x}^{(2)}$- in the second case.
This is not a problem to compare the representations $\vect{x}^{(2)}$~and~$\vect{x}^{(3)}$. Particularly if the mechanisms that perform these \emph{single-step transforms} have the same quality in terms of signal/noise ratio (set in \emph{Assumption 2}).
The design question under these assumptions is: if you were able to build a mechanism to lead you either to representation $\vect{x}^{(2)}$ or $\vect{x}^{(3)}$, but the quality of the mechanism could not be better than certain signal/noise ratio, which image representation would you prefer?.}

\blue{On the contrary, assuming that $\vect{r}^{(1)} \rightarrow \vect{x}^{(3)} \equiv \vect{r}^{(1)} \rightarrow \vect{x}^{(2)} \rightarrow \vect{x}^{(3)}$, where noise is injected \emph{both} in $\vect{x}^{(2)}$ \emph{and} in $\vect{x}^{(3)}$ is a quite different alternative scenario.
Under \emph{Assumption 1} the information captured by $\vect{x}^{(3)}$ may be bigger than the information captured by $\vect{x}^{(2)}$, but this is not be possible in the alternative scenario where noise is injected at every layer. In this latter case the \emph{data processing inequality}~\cite{Cover06} (chapt.2)
states that the information lost (due to the noise) at the second layer cannot be recovered afterwards.}

\blue{\emph{Assumption 1} (or the \emph{single-step transform} assumption) and the alternative \emph{multiple-step transform} are just different ways to consider where the noise comes from.
Given that no conclusive prescription for the amount of noise is available for \emph{all} the considered psychophysical stages~\cite{Ahumada87,Burgess88,Messe06,Neri10,Goris14,MorenoBote14,Kanitscheider15},
arbitrary assumptions would also be necessary in the \emph{multiple-step transform} case.}

\blue{In summary, for illustration purposes, \emph{Assumption~1} is as valid as the alternative \emph{multiple-stage transform} scenario, and it allows to compare different image representations with a clear design restriction (fixed signal/noise ratio). Nevertheless, it has to be clear that results obtained from \emph{Assumption 1} have not to be interpreted in terms of the data processing inequality. This is because in all the experiments we compare multiple \emph{single-step transforms}, and not cascades of noisy blocks in which information loss propagates though the network. Similarly, the specific nature of the noise and the specific noise level set by \emph{Assumption~2} are just convenient choices for illustration purposes.}

\blue{To conclude, lets remember again that RBIG estimation does not depend on the noise model. Therefore the procedure described below would not change using different assumptions (i.e. more sophisticated uncertainties in the sensors or noisy responses from actual measurements).}

\subsection{Global and local experiments}
\blue{Information measures are defined as integrals over the whole space of inputs $\vect{r}^{(1)}$ and the corresponding responses $\vect{x}$. Experiments considering
stimuli all across the input space will be referred to as \emph{global}.
Nevertheless, it is also interesting to consider how different regions of the image space are encoded. For instance, stimuli with specific visual features as those introduced in Section~\ref{images}, e.g. different \emph{luminance}, \emph{achromatic contrast}, or \emph{chromatic contrast}.
It is possible that a sensory system is specialized on stimuli with certain
features. This question could be quantified by computing the amount of transmitted information about samples belonging to specific regions of the image space.
Experiments considering specific regions of the image space will be referred to as \emph{local}, i.e. \emph{local} in the space of image features.}

\blue{In the \emph{global} experiments below, 10 realizations of each estimation are done using $0.5 \cdot 10^6$ randomly chosen stimuli from the image dataset that consists of $19\cdot 10^6$ samples. Of course, the corresponding responses are also considered in each case.}

\blue{In the \emph{local} experiments below, the estimations at each location of the
stimulus space are based on the images belonging to the corresponding luminance/contrast bin of the histogram shown in Fig.~\ref{jointPDF}.
In the \emph{local} experiments, 10 realizations of each estimation are done for the data in each bin.
In principle, each realization randomly selects 80\% of the samples available in the bin. However, note that the population in the low-luminance/low-contrast bins is very large, and considering that many samples slows down the estimation. Therefore, no more than $5\cdot10^5$ of these randomly chosen samples were considered in each case. Bins with less than 500 samples (the high-luminance / high-contrast corners of the domain) were discarded in the estimation because results may not be reliable.
In the results below, in those low-populated bins we plot a constant value from the
boundary of bins with population bigger than 500, but this assignation is arbitrary and these regions are not be considered in the discussion.}

\blue{In all the experiments, images and the corresponding responses are 27-dimensional as stated in Section~\ref{images}. The specific measurements shown is the following:}
\begin{itemize}
    \item \blue{\textbf{Global experiments:}}
           \begin{itemize}
                  \item \blue{\textbf{Global 1:} Redundancy reduction at different layers of different models.}
                  \item \blue{\textbf{Global 2:} Information available at different layers of different models.}
             \end{itemize}
    \item \blue{\textbf{Local experiments:}}
           \begin{itemize}
                  \item \blue{\textbf{Local 1:} Redundancy reduction in V1 for different kinds of stimuli.}
                  \item \blue{\textbf{Local 2:} Redundancy reduction at different layers of the model and at different locations of the image space.}
                  \item \blue{\textbf{Local 3:} Information available at different layers for different stimuli.}
                  \item \blue{\textbf{Local 4:} Information available at V1 and the PDF of natural images.}
           \end{itemize}
\end{itemize}

\section{Results}
\label{results}

\subsection{\blue{Global experiment 1: Redundancy reduction at different layers of different models.}}

\blue{In this experiment we compare the
RBIG estimates of $\Delta T$ with the so-called \emph{theoretical estimate} given by Eq.~\ref{deltaT}. In this work, redundancy reduction experiments are just instruments to illustrate the reliability of the RBIG estimations of $I$ which depend on estimating $T$. Therefore, in these instrumental experiments on $\Delta T$ only deterministic responses were considered to apply the estimate given by Eq.~\ref{deltaT}.}
The considered layers are those in Eq.~\ref{cascade}, including the linear and the nonlinear ones.
Redundancy reduction was computed in the baseline model (the one leading to the best correlation with human opinion in Fig.~\ref{biol_plaus1} and Table~\ref{biol_plaus2}), and also with the more rigid and more flexible versions of the model presented in Section~\ref{sect_plausible}.
\blue{As stated above, RBIG estimations subtract $T(\vect{x})$ from $T(\vect{r}^{(1)})$, while the theoretical estimates accumulate the reductions $\Delta T$ that happen at every intermediate layer.}
Results are displayed in Fig.~\ref{global1}.

\afterpage{
\begin{figure}[t!]
   \centering
   \vspace{-0.3cm}
   \begin{tabular}{cc}
    \includegraphics[width=0.45\textwidth]{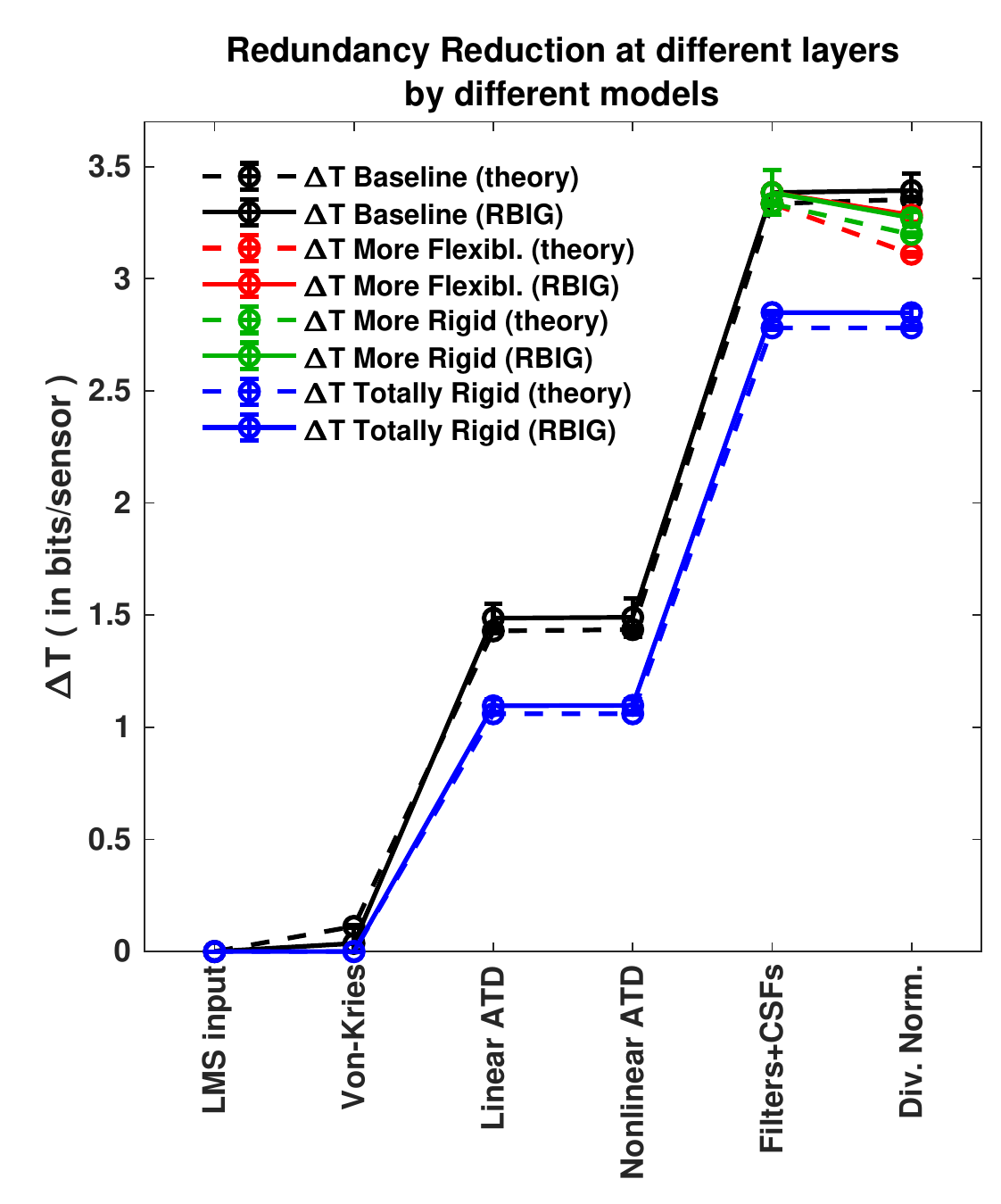} & \includegraphics[width=0.46\textwidth]{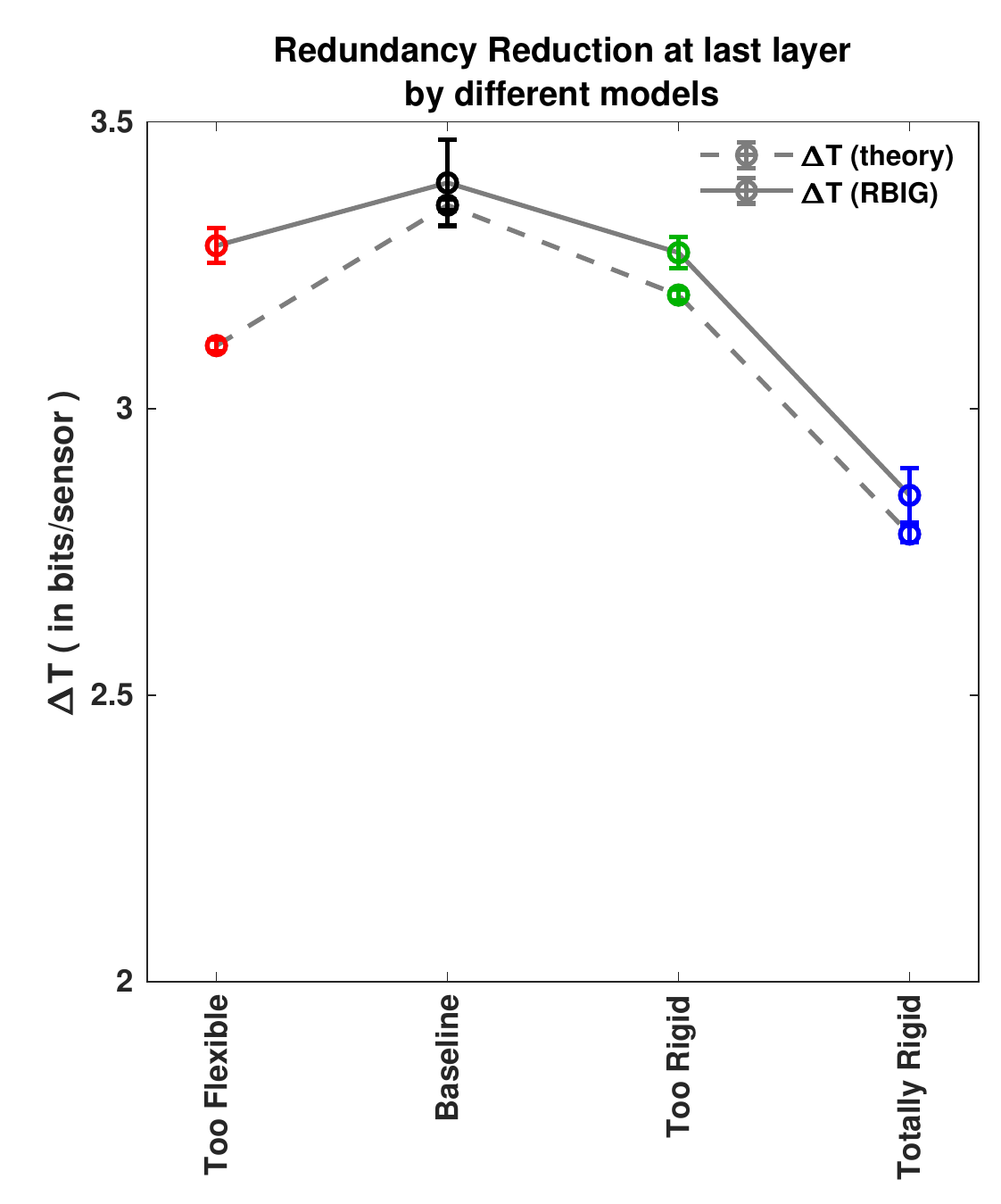} \\
    \end{tabular}
     \vspace{-0.8cm}
    \caption{\emph{\textbf{Left:}} Redundancy reduced at different layers of the network (labels in the horizontal axis) considering samples over the whole image manifold (\emph{global} experiment). Redundancy is measured in total correlation (in bits/sensor). Lines in different color refer to different models, and different line style (solid/dashed) refers to the empirical RBIG estimate given by Eq.~\ref{total_with_rbig} and the \emph{theoretical estimate} given by Eq.~\ref{deltaT}. Each result accumulates the reductions due to previous
    layers. \emph{\textbf{Right:}}~Redundancy reduced at the V1 layer (last layer) using different variations of the baseline model: more flexible and more rigid divisive normalization, and totally rigid (totally linear) model.
    In both plots (left and right) error bars stand for the standard deviation over the 10 realizations of the estimation. The \emph{theoretical estimation} also has error bars because the univariate entropies in Eq.~\ref{deltaT} were empirically  estimated.
    }
    \label{global1}
\end{figure}
}

\blue{The technical conclusion is that in general the RBIG estimate of $\Delta T$ is consistent with the \emph{theoretical estimate}.} Note that the only significant exception is for the case of the model with more flexible divisive normalization. Nevertheless, even in this case both estimates lead to the same qualitative conclusion: this too flexible model leads to less redundancy reduction than the baseline model.
This general agreement suggests that total correlation estimates (the core under RBIG information transmission estimates, as seen in Eqs.~\ref{eq:mirbig2} and~\ref{eq:mirbig}) can be trusted.

From the functional point of view it is interesting to note that redundancy reduction in the nonlinear stages of the baseline model is almost negligible. For instance see the plateau in $\vect{r}^{(2)} \rightarrow \vect{x}^{(2)}$ (Weber-like nonlinearities) or in $\vect{r}^{(3)} \rightarrow \vect{x}^{(3)}$ (Divisive Normalization of texture sensors).
However, it seems that this kind of nonlinear processing actually prepares the data so that subsequent linear stages can do a better job in removing redundancy. Note that the identical linear parts $\vect{x}^{(1)} \rightarrow \vect{r}^{(2)}$ (opponent transform) and $\vect{x}^{(2)} \rightarrow \vect{r}^{(3)}$ (Filters+CSFs) attain bigger reductions after the application of the Von-kries nonlinearity and the Weber-like nonlinearity. This is more clear by looking at the cumulative effect at the last layer: \blue{the baseline model (in black) is substantially better than a totally linear model (in blue)}.

In any case, as stated above, captured  information (analyzed in next experiment) is a more direct measure of visual function than redundancy reduction.

\subsection{\blue{Global experiment 2: Information available at different layers of different models.}}

\afterpage{
\begin{figure}[b!]
   \centering
   \vspace{-0.3cm}
   \begin{tabular}{cc}
\includegraphics[width=0.45\textwidth]{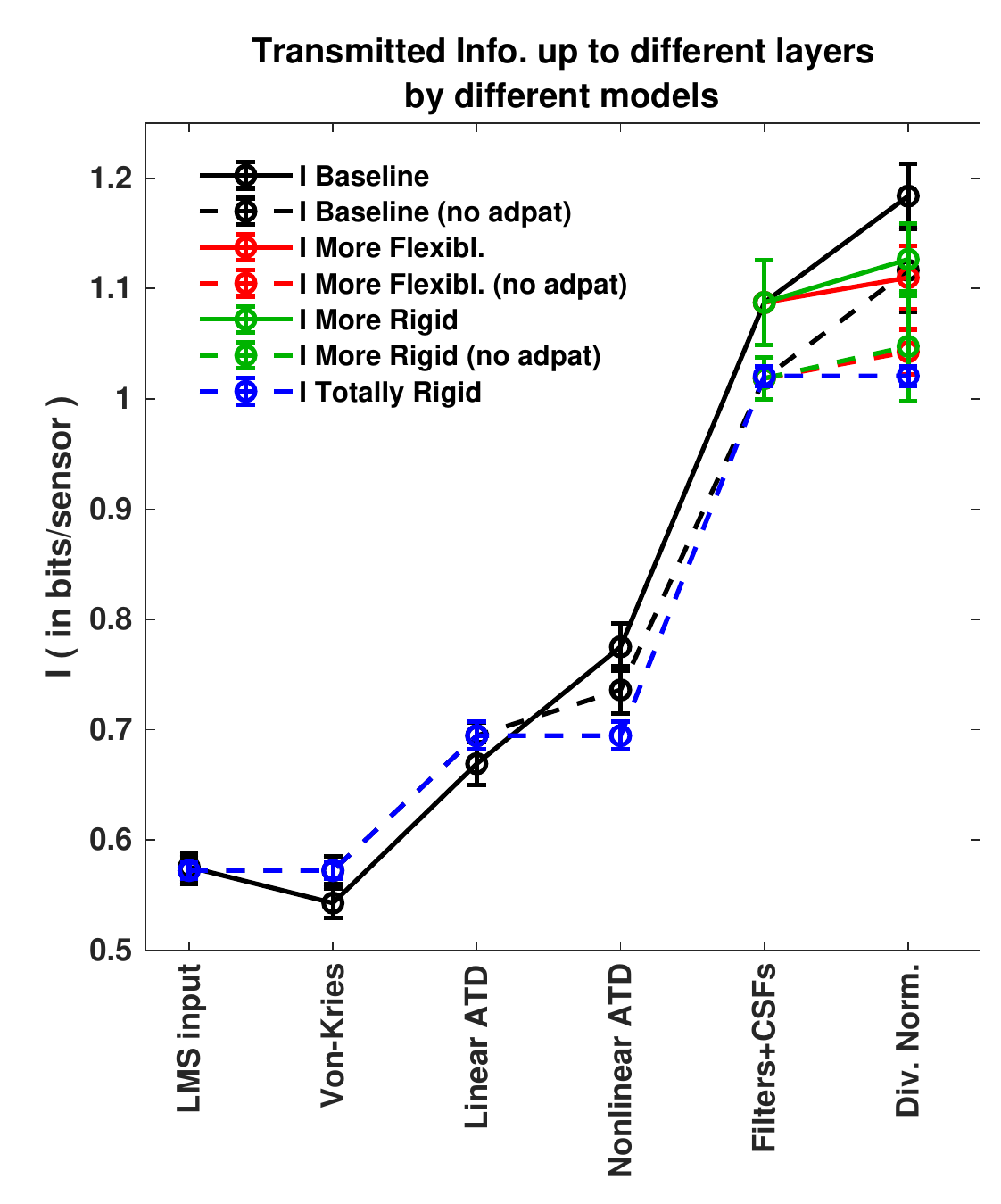} & \includegraphics[width=0.46\textwidth]{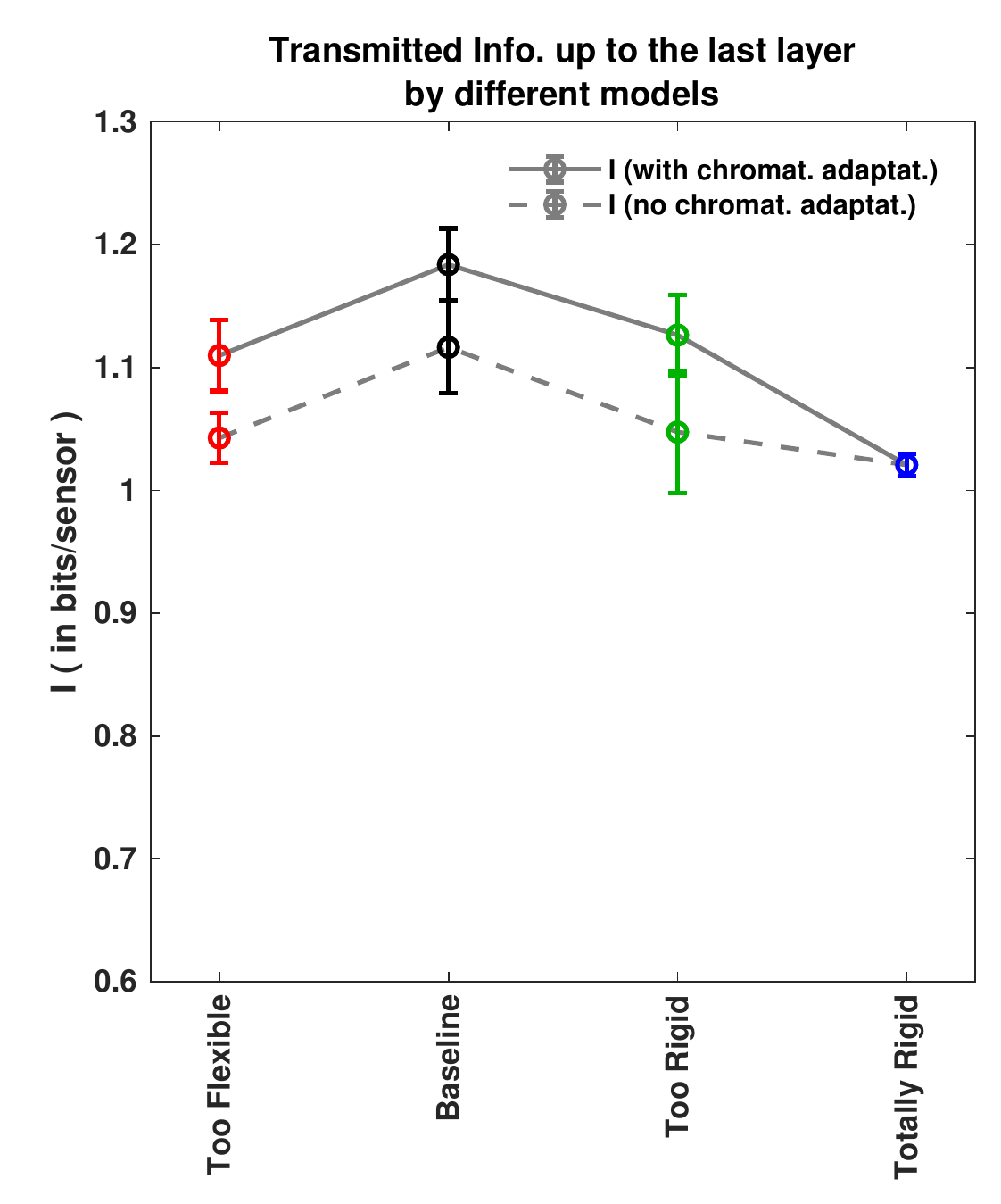} \\
    \end{tabular}
     \vspace{-0.8cm}
    \caption{Information available from different layers and different models.
    Estimations assume \emph{single-step transforms} and the same signal-to-noise ratio at every marginal sensor.  \emph{\textbf{Left:}} Mutual information between noise-free LMS and the noisy response at the considered layer (labels at the horizontal axis) for different variations of the model (different color/style). \emph{\textbf{Right:}} Information transmitted up to the cortical layer by different variations of the model: baseline model (in black), more flexible and more rigid divisive normalization (in red and green respectively), and totally rigid linear model (in blue). The corresponding versions of these models that neglect chromatic adaptation are represented in the same color but connected by a dashed line.
 }
    \label{global2}
\end{figure}
}

\blue{This experiment compares the different image representations along the network in terms of the amount of information they share with an ideal input (noise-free LMS image).
This global experiment for $I$ includes the baseline model and six interesting variations:} the more flexible and less flexible versions due to different divisive normalization semisaturation, the corresponding versions neglecting chromatic adaptation, and a totally linear version, which of course has no adaptation at all.

Measures at different layers include $I(\vect{r}^{(1)},\vect{r}^{(1)})$, $I(\vect{r}^{(1)},\vect{x}^{(1)})$, $\cdots$, $I(\vect{r}^{(1)},\vect{x}^{(3)})$.
As stated above in Section~\ref{noise_ass} we assume that in the different representations noise is added after a \emph{single-step transform},
and the relative amount of noise in all the representations is the same (the deviation of the noise in each dimension is 5\% of the dynamic range of that individual response). For instance, in the first case, referred to as $I(\vect{r}^{(1)},\vect{r}^{(1)})$, we compute the shared information between noise-free LMS images and LMS images corrupted with Gaussian noise with the selected SNR in each pixel and color channel.

\vspace{0.3cm}
Results in Fig.~\ref{global2}-left show that the cortical representation (adaptive contrast in local frequency in ATD) is much better than the original representation in the spatial domain and in the LMS color space.
The cortical representation doubles the amount of information captured by retinal sensors (assuming equivalent SNR).
It may seem that the Von-Kries adaptation is not worth it (if it was the last signal representation): see how it shares less information with the input than the LMS representation for the same noise level.
However, when combined with other processing blocks the resulting representations capture more information form the input than if it was not present.
\blue{These results indicate the overall progressive improvement of the stimulus representation along the pathway.}

\blue{Remember that, according to the noise assumptions discussed in Section~\ref{noise_ass}, each point along one of the curves in Fig.~\ref{global2}-left actually mean the amount of information available for a system that would perform the considered processing from the input in a single-step. That is, with all the noise added at that specific layer.
Therefore, increasing $I$ does not violate the \emph{data processing inequality}. It just points out that
these inner representations are intrinsically better because they would capture more information with sensors of the same SNR quality.}

\blue{These results also show that the baseline model (the one that gets the best correlation with human opinion in Table~\ref{biol_plaus2}) is also the one that captures more information about the input. Note that the different modifications of the model, i.e. doing it more rigid (by cancelling chromatic adaptation, by doing divisive normalization more rigid, or by considering a totally linear version) or more flexible (by increasing the adaptivity of divisive normalization) lead to poorer representations in terms of transmitted information.
This conclusion is more clear by looking at different versions of the last representation (after the divisive normalization) at Fig.~\ref{global2}-right}.


The nonlinear operations both in the saturation of ATD chromatic channels and in contrast adaptation lead to significant improvements in the amount of captured information.
Note that about 28\% of the improvement in available information at the final representation comes from the nonlinear operations,
while 72\% of the improvement comes from the linear operations. These increments in $I$
in Fig.~\ref{global2}-left are interesting taking into account that these nonlinear operations make almost negligible contributions to redundancy reduction (see $\Delta T \approx 0$ for these layers in Fig.~\ref{global1}-left).

It is also interesting to note that spatial transforms (texture filters and contrast adaptation) definitely have a bigger contribution to the amount of captured information than chromatic transforms (Von-Kries adaptation, opponent channels and Weber-like nonlinearities).
Note that Fig.~\ref{global2}-left shows that 67\% of the improvement in available information at the final representation comes from the spatial operations, as opposed to the 33\% that comes from the chromatic operations.
This is remarkable given the fact that tiny patches of 0.05 degrees were considered in our computations.


\subsection{\blue{Local experiment 1: Redundancy reduction at V1 for different kinds of stimuli.}}

In all local experiments (which involve computations over the different bins of the image space) only the baseline model was considered.

In this first local experiment RBIG estimations of $\Delta T(\vect{r}^{(1)},\vect{x}^{(3)})$ at different locations of the image space are compared to the corresponding \emph{theoretical estimates}.
\blue{This experiment is interesting to check that the global accuracy of the information estimates illustrated in Fig.~\ref{global1}, actually hold for every location across the stimulus space. As stated above, experiments involving comparison with the theoretical estimate, Eq.~\ref{deltaT}, require the use of deterministic responses.}


\afterpage{
\begin{landscape}
\vspace{-10cm}
\begin{figure}[!h]
\vspace{-0.3cm}
   \centering
    \includegraphics[width=1.75\textwidth,height=4.7cm]{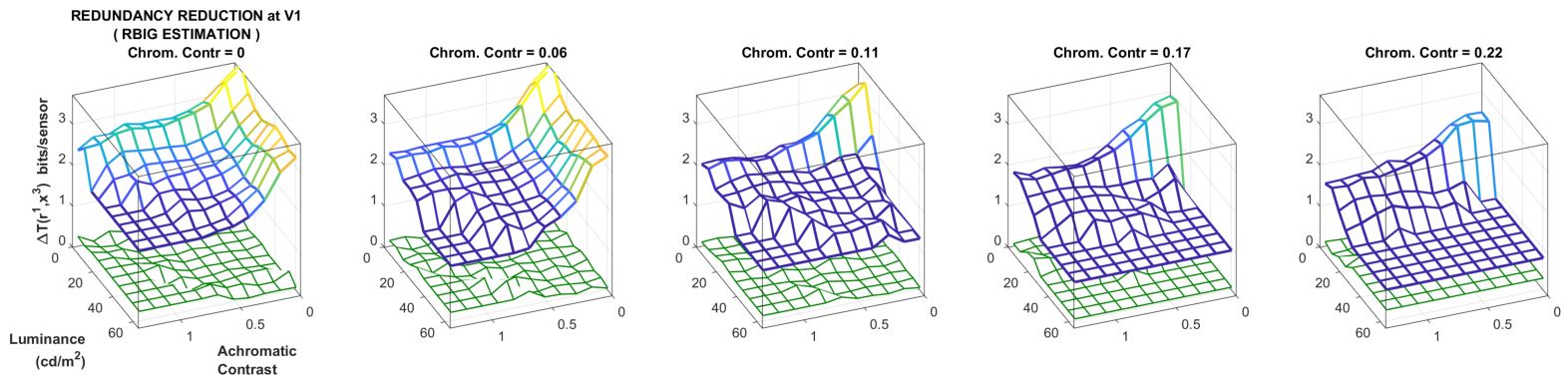}\\[0.15cm]
    \includegraphics[width=1.75\textwidth,height=4.7cm]{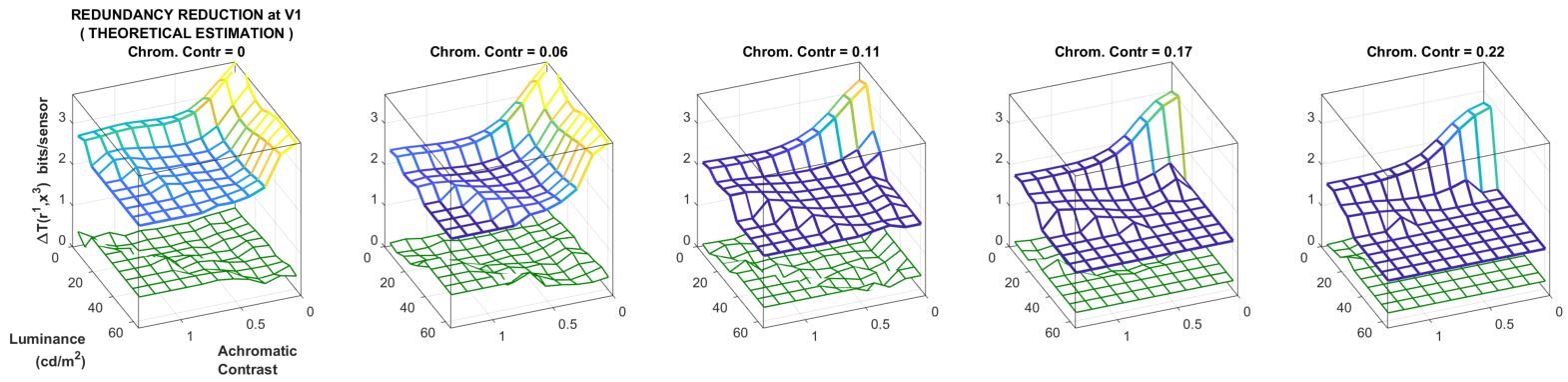}\\[0.15cm]
    \includegraphics[width=1.75\textwidth,height=4.7cm]{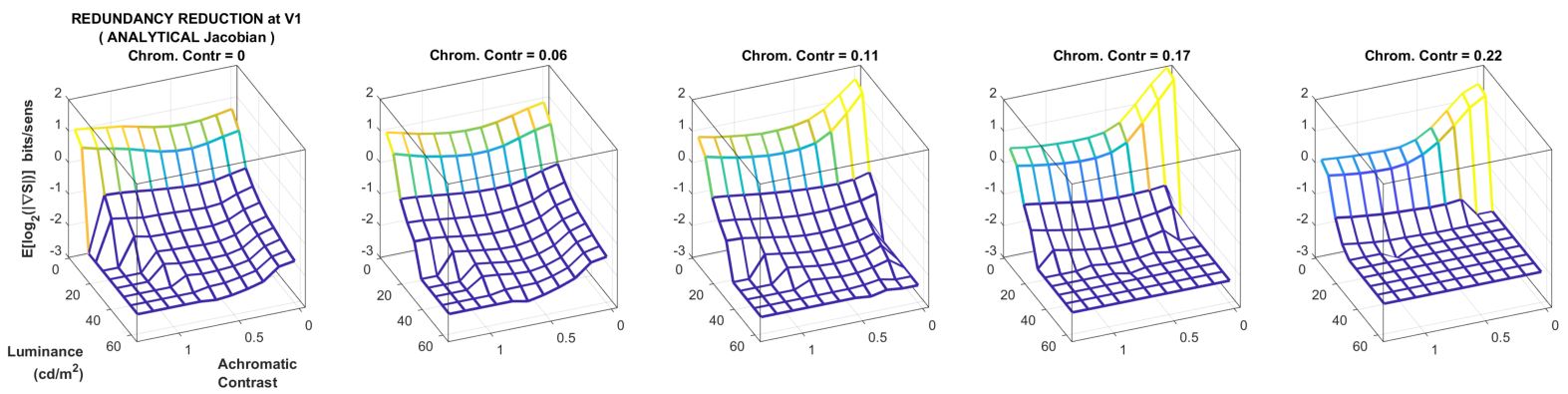}\\[-0.3cm]
    \caption{Agreement between estimations of redundancy reduction ($\Delta T$, \emph{in bits}) between the LMS input and the spatio-chromatic representation in V1, estimated via RBIG (\emph{top row}) and via the theoretical approach of Eq.~\ref{deltaT} (\emph{middle row}). The \emph{bottom row} shows the component of the theoretical $\Delta T$ that comes from the analytical Jacobian.
    Results are shown for every region of the achromatic contrast / luminance space for different chromatic contrast (in different columns).
    The surfaces in the blue-yellow colormap are the average of 10 estimations in each case.
    The green surfaces at the plots of the middle row represent the absolute difference between the theoretical and the RBIG estimates.
    The green surfaces at the plots of the top row represent the combination of the uncertainties of the estimates $(\sigma_{\mathrm{RBIG}}^2 + \sigma_{\mathrm{theor}}^2)^{1/2}$}
    \label{agreement}
\end{figure}
\end{landscape}
}

The results in Fig.~\ref{agreement} shows that the empirical RBIG estimate (top row) closely follows the theoretical estimate (middle row) all over the stimulus space.
Note also that the difference between the estimates is small (green surfaces in the plots of the theoretical result), and
this difference is similar to the uncertainties both estimates combined (green surfaces in the plots of the RBIG result).
The average values of the relative difference and relative standard deviation are 0.12 and 0.09 respectively.
Similar agreement is obtained for all the previous layers.
Agreement of both estimates not only points out the accuracy of the proposed estimator but also the correctness of the analytical Jacobian involved in the theoretical estimate since the Jacobian of the response at the last layer includes all the previous layers.

This confirms that information estimates based on RBIG computation of total correlation can be trusted for every location of the stimulus space (not only globally as pointed out by the agreement of the estimates in Fig.~\ref{global1}).

Finally, the term of Eq.~\ref{deltaT} that depends on the analytical Jacobian is shown in the bottom row.
The analytical Jacobian not always represents the trends of $\Delta T$.
This illustrates that analytical description of the model is not enough to get complete insight on the information-theoretic performance of the model.
In the specific example in Fig.~\ref{agreement} the analytical term roughly determines the behavior for high chromatic contrast,
but it is not the case for low chromatic contrasts.
This limitation of the knowledge that can be extracted from the analytical expression of the model emphasizes the need of empirical estimators such as RBIG. Section~\ref{discussion} elaborates on this point.


\subsection{\blue{Local experiment 2: Redundancy reduction at different layers and different locations of the image space.}}

\blue{This experiment extends the redundancy reduction values at the different layers presented in Fig.~\ref{global1} to specific regions of the image space. Note that specific operations may reduce more redundancy for some stimuli than for others.}

Fig.~\ref{T_reduction_per_bin} shows $\Delta T$ along the network in the achromatic contrast / luminance plane
for two chromatic contrasts (the minimum and the maximum in our images).
We can identify general trends which also apply to the other values of chromatic contrast not shown in the figure:
(1)~$\Delta T$ is always positive, i.e. redundancy is effectively reduced by the system.
(2)~$\Delta T$ always increases along the pathway, which suggests that inner representations are better
than earlier representations in terms of the information captured, and finally, (3)~the increments along the way mainly occur at the linear stages, i.e. the transform to opponent color representation, and
the transform from the spatial domain to the local-frequency domain.
This relevance of linear stages in redundancy reduction is consistent to what was found in Fig.~\ref{global1}.
Note that these two linear stages are the ones that rotate the representation similarly to PCA (see Appendix \ref{visual_manifolds}).

\afterpage{
\begin{landscape}
\begin{figure}[!h]
   \centering
    \includegraphics[width=1.75\textwidth]{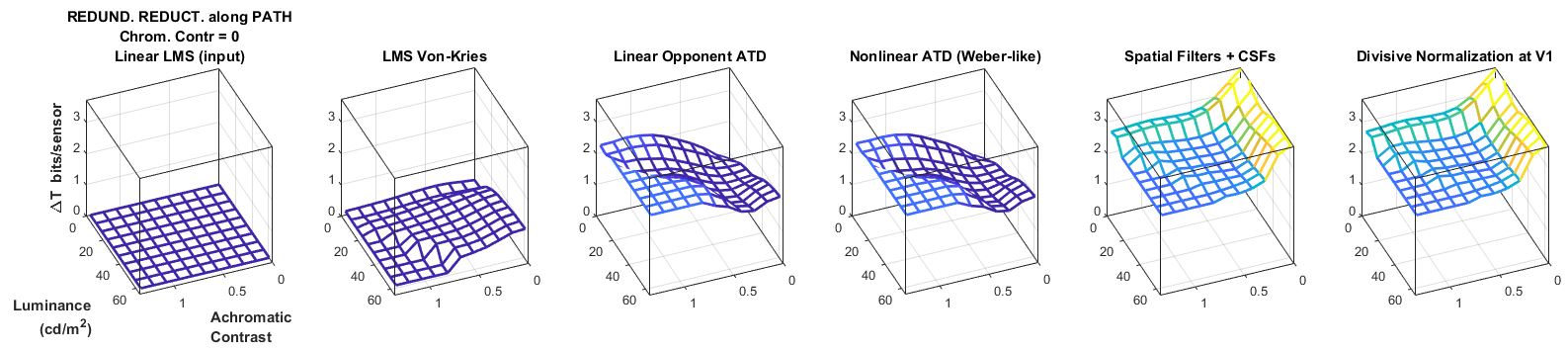}\\[0.4cm]
    \includegraphics[width=1.75\textwidth]{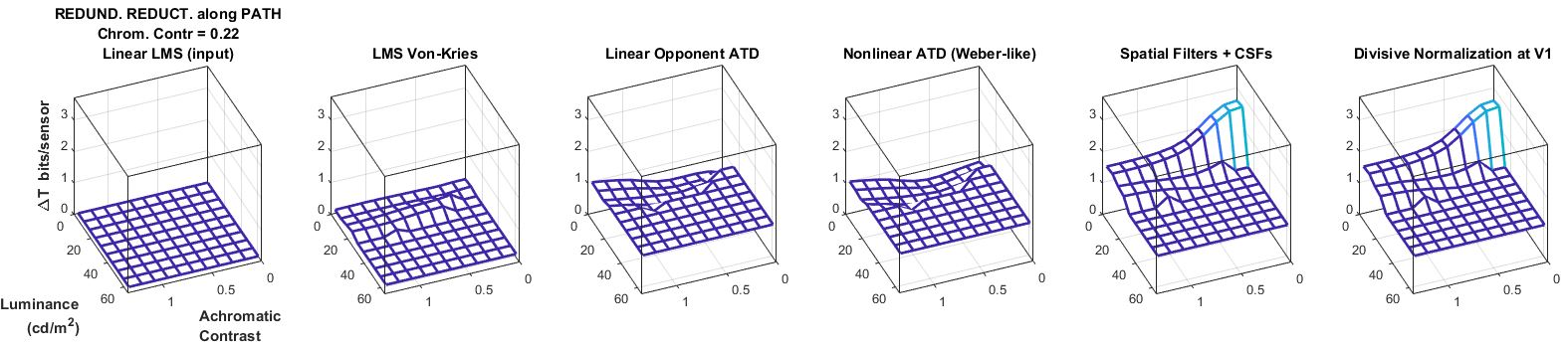}\\
    \caption{Redundancy reduction along the layers of the visual pathway. Results are shown over the achromatic contrast and luminance space for two fixed chromatic contrasts: the minimum (zero, on the top) and the maximum in out set (on the bottom).}
    \label{T_reduction_per_bin}
\end{figure}
\end{landscape}
}


\subsection{\blue{Local experiment 3: Information available at different layers and different stimuli.}}


\afterpage{
\begin{landscape}
\begin{figure}[!h]
   \centering
    \includegraphics[width=1.75\textwidth]{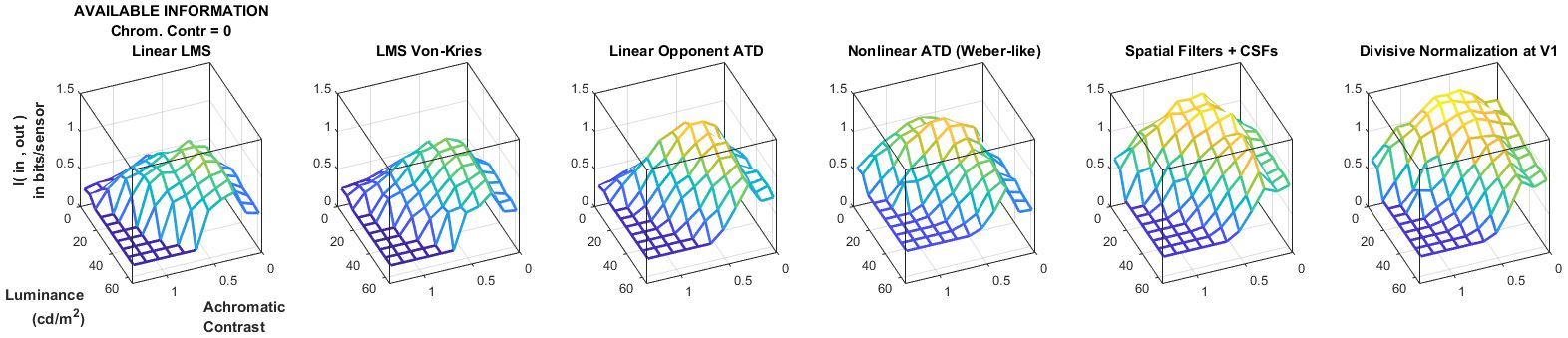}\\[0.4cm]
    \includegraphics[width=1.75\textwidth]{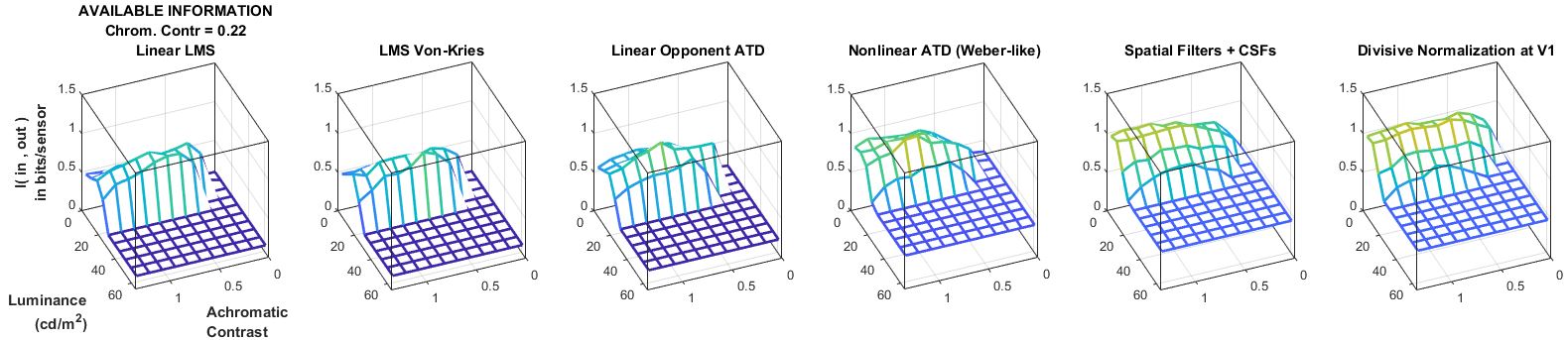}\\
    \caption{Information about the scene available from different layers of the visual pathway. Results are shown over the achromatic contrast and luminance space for two fixed chromatic contrasts: the minimum (zero, on the top) and the maximum in our set (on the bottom).
    In each case, we assume sensors at the different representations have equivalent noise level (5\% of the signal deviation).
    This result implies that using sensors of equivalent quality, the cortical representation is more appropriate because it captures more information from the input.}
    \label{I_reduction_per_bin}
\end{figure}
\end{landscape}
}


\blue{The object of this experiment (transmitted information per unit of volume, or \emph{per bin}, of the stimulus space) highlights the kind of stimuli better represented by the different layers.}
Fig.~\ref{I_reduction_per_bin} shows the information about the scenes (in bits/sensor) available from the different neural layers assuming the same signal to noise ratio of the sensors.

Only two chromatic contrast are shown in Fig.~\ref{I_reduction_per_bin}, but the following trends also hold of the
other chromatic contrasts omitted in the figure: assuming sensors with the same SNR,
(1)~the cortical representation is substantially better than the retinal representation since it captures more information about the scene,
(2)~different intermediate representations are progressively better from retina to cortex, and
(3)~improvements of the representation come both from the linear and the also the nonlinear stages.

Note that the total available information at certain layer (the \emph{global} result in Fig.~\ref{global2}) is not the integral of the corresponding surface of \emph{local} values in Fig.~\ref{I_reduction_per_bin}. However, progressive increase of the value for a region of the image space along the network can also be interpreted as an improvement of the representation. Similarly, the increments in $I$ do not violate the data processing inequality given the \emph{single-stage transform} assumption we are doing here.

It is important to note the differences of this $I$ result, Fig.~\ref{I_reduction_per_bin}, with regard to
the $\Delta T$ result, Fig.~\ref{T_reduction_per_bin}.
First, while no substantial gain is obtained through the nonlinear stages in terms of redundancy reduction, the improvements in transmitted information due to the nonlinearities are significant. This is consistent with what was found in Figs.~\ref{global1} and~\ref{global2}.
Second, the distribution of $I$ and $\Delta T$ over the stimulus space is substantially different, take for instance the bottom right plots in Fis.~\ref{T_reduction_per_bin} and~\ref{I_reduction_per_bin} respectively.
The distribution of transmitted information seems more consistent with the statistics of the natural scenes in Fig.~\ref{jointPDF}. That is the object of the comparison in the next section.

\subsection{\blue{Local experiment 4: Transmitted information and the PDF of natural images}}


Fig.~\ref{I_per_bin_compared to_pdf} explicitly compares the transmitted information up to the cortical layer, $\vect{x}^{(3)}$, for stimuli at all the locations of the considered image space (top) with the distribution of natural scenes over that space (bottom).

\blue{This comparison addresses these questions: \emph{Does the system transmit the same amount of information for different images?, and, if not,
is this uneven transmission similar to the PDF of natural scenes?}}

This result shows that the considered psychophysically-tuned network (no statistical knowledge was used in this crude biological model)
transmits more information in the more frequent regions of the image space: note that the peak of transmitted information shifts to higher
achromatic contrasts for bigger chromatic contrasts (as the PDF of natural images), and the amount of transmitted information decreases for high chromatic contrasts (as the PDF).

\blue{This result indicates a remarkable match between the properties of the cortical representation and the relative frequency of the stimuli in the image space. Note that the amount of available information for different images is not a by product of the relative probability of the images. For instance, if the luminance or contrast nonlinearities were expansive (as opposed to compressive as they are) the noise would have more impact in the low-luminance / low-contrast end. This would reduce the amount of available information for those signals using the same database to compute the results.}

\afterpage{
\begin{landscape}
\begin{figure}[!h]
   \centering
    \includegraphics[width=1.75\textwidth]{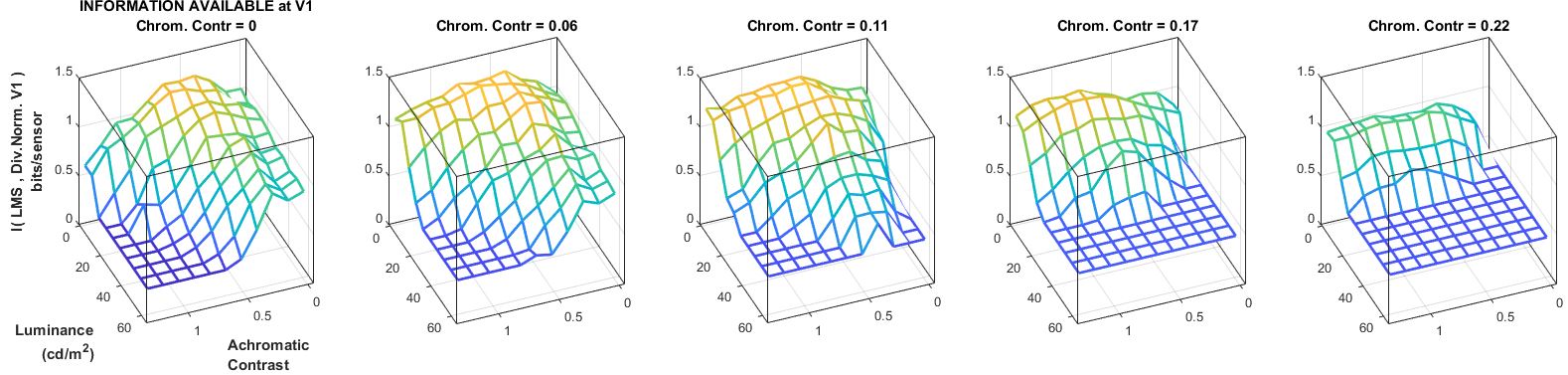}\\[0.4cm]
    \includegraphics[width=1.75\textwidth]{pdfs_joint2.JPG}\\
    \caption{Information available at the cortical representation after Divisive Normalization at different regions of the image space (top)
    compared to the PDF of natural images (bottom).}
    \label{I_per_bin_compared to_pdf}
\end{figure}
\end{landscape}
}

\section{Discussion and final remarks}
\label{discussion}

\paragraph{\textbf{Efficiency of the network and hierarchy of transforms.}}
The proposed method to estimate transmitted information allows a quantitative analysis of the different perceptual computations along the considered retina-cortex pathway.

The measurements presented above imply that the considered cortical representation
captures substantially more information about the scene (factor $\times 2$) than the
retinal representation with sensors of the same signal/noise quality.

Not all the perceptual computations contribute to the improvement of the signal representation in the same way.
Transforms acting on spatial content of the signal lead to bigger improvements in information transmission
than the purely chromatic transforms carried out in the first layers (67\% versus 33\% respectively).
The biggest contribution is due to the analysis of ATD images through local-oriented filters and Divisive Normalization.
The linear operations are responsible for most of the improvement in information transmission (about 72\%),
but the nonlinear/adaptive behavior of the system is responsible of a substantial 28\%.

\blue{Interestingly, if model adaptivity is modified so that it has lower correlation with human opinion,
the amount of transmitted information reduces too.
Moreover, local analysis on specific stimuli shows that the cortical representation captures relatively more information
in the regions of the image space where natural images are more frequent.
These two facts indicate that this biologically plausible model (not explicitly optimized for transmission) is efficient in terms of this information measure and it is well matched to the stimuli it faces.}

\blue{This is in line with the approaches to the Efficient Coding Hypothesis that go from \emph{biology-to-statistics}~\cite{Malo10,Gomez19}, with the advantage of using a more appropriate performance measure ($I(\vect{r},\vect{x})$ as opposed to $\Delta T(\vect{r},\vect{x})$). More generally, the proposed method can be used to estimate transmitted information in improved multi-goal cost functions where constraints of different nature (energy, wiring, reconstruction error, adaptation, etc.~\cite{Rehn+JCN+2007,Perge09,Laughlin13,Jolivet15,Laughlin15,Olshausen96,MacLeod03b,Laparra12,Laparra15,Gutmann14,Sporns12}) could be taken into account. These improved multi-goal cost functions could lead to interesting results in the conventional approach from \emph{statistics-to-biology}.}

\paragraph{\blue{\textbf{Transmitted information and analytical Jacobian.}}}
Here we discuss up to which point the trends of the transmitted information over the image space
can be predicted from the analytical response.
We will see that while the analytical Jacobian may be enough to understand the behavior of the system at a single layer (as done in \cite{Martinez18,Gomez19}), the relative relevance of the different layers cannot be easily inferred from the corresponding derivatives.
\blue{Therefore, a model-free tool such as the one presented is extremely useful in this context.}


First we recall the relation between \emph{transmitted information} ($I(\vect{r},\vect{x})$, or information about the input captured by the inner image representation) and \emph{total correlation} within the response ($T(\vect{x})$, or information shared by the responses of the different sensors, $x_i$). While maximizing $I(\vect{r},\vect{x})$ is a sensible goal for a decoder devoted to solve visual problems from the response $\vect{x}$, reduction of the redundancy $T(\vect{x})$ is an alternative goal which is related to the maximization of transmitted information.
Reasoning with $T(\vect{x})$ may be useful in some situations because it is related to the analytic expression of the perception model.

From the definitions of mutual information in terms of \emph{entropy} and \emph{conditional entropy}~\cite{Cover06} (chapt.2),
$I(\vect{r},\vect{x}) = h(\vect{r}) + h(\vect{x}) - h(\vect{r},\vect{x}) = h(\vect{x}) - h(\vect{x}|\vect{r})$;
and from the concept of \emph{total correlation}~\cite{Watanabe60,Studeny98},
$T(\vect{x}) = \sum_i h(x_i) - h(\vect{x})$, it is easy to see that the following expression holds\footnote{In the context of the noisy sensory system defined by Eq.~\ref{setting}, the transmitted information is: $I(\vect{r},\vect{x}) = h(\vect{x}) - h(\vect{x}|\vect{r})$, so, as the uncertainty of the response given the input is due to the noise, it holds $I(\vect{r},\vect{x}) = h(\vect{x}) - h(\vect{n})$.
Then, as the joint entropy is the sum of marginal entropies minus the shared information by all the components, $h(\vect{x}) = \sum_i h(x_i) - T(\vect{x})$, and by plugging this in the previous equation, we get Eq.~\ref{capacity_redundancy2}.}:
\begin{equation}
     I(\vect{r},\vect{x}) = \sum_i h(x_i) - T(\vect{x}) - h(\vect{n}) \label{capacity_redundancy2}
\end{equation}

Assuming that the intrinsic noise of the biological sensors cannot be reduced (and hence $h(\vect{n})$ is fixed), Eq.~\ref{capacity_redundancy2} points out \emph{what} the system should do to increase the transmitted information:
(i)~it should reduce the redundancy (the total correlation, $T$) within the response array, and
(ii)~it should increase the sum of entropies of the responses of the individual sensors.
\blue{On the one hand, in the particular case of independent noise in every sensor, the transform should
make the responses independent to obtain $T(\vect{x})=0$.
On the other hand, the entropies $h(x_i)$ can not be arbitrarily increased via trivial response amplification
because of energy constraints.
Therefore, once redundancy has been decreased as much as possible (ideally set to zero),
as the energy (or variance) of each coefficient has to be limited, the first term can be maximized by marginal equalization.
Note that marginal operations do not modify the total correlation~\cite{Lyu09}.
Therefore, marginal equalization could be used to maximize entropy with
constrained variance (for instance obtaining uniform or Gaussian PDFs~\cite{Cover06},~chapt.~12)
without increasing $T$, thus maximizing $I$.}

\vspace{0.3cm}

Eq.~\ref{capacity_redundancy2} identifies univariate and multivariate strategies for
information maximization.
When trying to assess the performance of a sensory system, reduction of the multivariate total correlation, $T(\vect{x})$, seems the relevant term to look at because univariate entropy maximization can always be performed after joint PDF factorization through a set of (easy-to-do) univariate equalizations.

Then, the \emph{reduction in redundancy}, $\Delta T(\vect{r},\vect{x}) = T(\vect{r}) - T(\vect{x})$, is a measure of performance that could be aligned with $I(\vect{r},\vect{x})$. Interestingly \blue{for deterministic responses}, this performance measure, $\Delta T$, can be written in terms of univariate quantities and the response model \cite{Lyu09}:
\begin{eqnarray}
     \Delta T(\vect{r},\vect{x}) & = & \sum_i h(r_i) - \sum_i h(x_i) + E_{\vect{r}}\Big\{ \log_2 |\nabla_{\vect{r}}S| \Big\} \nonumber \\
                                 & = & \Delta h_m(\vect{r},\vect{x}) + E_{\vect{r}}\Big\{ \log_2 |\nabla_{\vect{r}}S| \Big\} \label{deltaT}
\end{eqnarray}
Eq.~\ref{deltaT} is good for our purposes for two reasons:
(1)~\emph{in case the marginal difference, $\Delta h_m$, is approximately constant over the space of interest},
the performance is totally driven by the Jacobian of the response, so it can be theoretically studied from the model, and (2)~even if $\Delta h_m$ is not constant, the expression is still useful to get reliable estimates of $\Delta T$ because the multivariate contribution may be get analytically from the Jacobian of the model and the rest reduces to a set of univariate entropy estimations (for which reliable estimates do exist~\cite{ITE}).
In the Results section~\ref{results}, estimates of $\Delta T$ using Eq.~\ref{deltaT} were referred to
as \emph{theoretical estimation} (as opposed to model-agnostic empirical estimates purely based on samples Gaussianization) because of this second reason.
\blue{Note that, as the Jacobian of the composition is the product of Jacobians, Eq.~\ref{deltaT} implies that $\Delta T(\vect{r}^{(1)},\vect{x}^{(j)}) = \Delta T(\vect{r}^{(1)},\vect{x}^{(i)}) + \Delta T(\vect{x}^{(i)},\vect{x}^{(j)})$}.

In previous works, Eq.~\ref{deltaT} has been used to describe the communication performance
either in linear systems~\cite{Bethge06} and in nonlinear transforms such as Divisive Normalization~\cite{Lyu09,Martinez18}
and Wilson-Cowan interaction~\cite{Gomez19}.
In those cases interesting insight could be gained from the analytical expressions of the corresponding Jacobian.
In the linear case the Jacobian is constant over the image space, therefore it is relevant when considering different
linear models, but it is irrelevant when comparing the transmitted information for different stimuli.
In the nonlinear cases, authors used Eq.~\ref{deltaT} to analyze the performance at a single layer, and
$\Delta h_m$ was explicitly shown to be roughly constant over the stimulus domain.
Therefore the analytical Jacobian certainly explained the behavior of the system.

\emph{However}, $\Delta h_m$ may not be constant in general, and hence, the trends obtained from the Jacobian
of the model can be counteracted by the variation of $\Delta h_m$.

In particular, the situation gets complicated if one wants to study the relative effect of different layers in the cascade. At a single layer (whose Jacobian accumulates the effect of all previous
layers~\cite{Martinez18}) the marginal difference of entropies \emph{with regard to the input} may be constant
over the domain, and hence negligible. However, reasoning only with the Jacobians in the case of comparisons between
multiple layers will only be valid if \emph{all} the marginal differences of entropy between every layer are
constant over the domain. This more strict condition is harder to fulfil in a specific network.
For instance, in the illustrative model considered in our experiments, this condition
not always holds (see the differences between the Jacobian and $\Delta T$ in Fig.~\ref{agreement}).


Therefore, since the intuition from the analytical response is conclusive only in restricted situations, there is a need for empirical methods to estimate the
transmitted information directly from sets of stimuli and the responses they elicit.


\paragraph{\textbf{Relation to previous work.}}

The analysis done here has a number of relevant differences with
previous work that already analyzed the statistical performance of psychophysically-plausible transforms.

For instance, previous works were limited either because explored a limited range of models or because they used limited performance measures.
In~\cite{Bethge06} the authors addressed the interesting study of the gain that can be obtained from different
redundancy reduction transforms, basically using Eq.~\ref{deltaT}, but restricted the analysis
to linear cases to neglect the term that depends on the Jacobian.
In~\cite{Malo10} authors do consider a more general (nonlinear) model, but their analysis is limited because
they didn't use a multivariate measure for the redundancy, but a set of mutual information measures
between pairs of coefficients at the considered layer.

From the technical point of view our analysis is related to
the work of Foster et al. who are also concerned about the use of accurate
information-theoretic measures to study human vision \cite{FosterJOSA18}.
The similarity is that they also use non-parametric measures of mutual information that operate directly on natural samples.
The main difference is their focus on color vision, and specifically on characterizing
the performance of humans in different illumination conditions (e.g. determining the number of discriminable colors) \cite{FosterVisNeuro04,FosterJOSA09,FosterJoV10}.
This is related to the amount of color information in a scene that can be extracted from color measurements under other illumination \cite{Ivan13,FosterJOSA18}. These problems are related to entropy and mutual-information measures (which is the same problem that we address with RBIG), but they do not quantify the information flow through the visual pathway (mutual information between layers and redundancy within layers) that we address here to identify the most relevant computations.
As an example, in \cite{FosterJOSA09,FosterCIC08} the redundancy is considered only because of its impact on the available information for illumination compensation, not as a measure of information transmission in the visual pathway.

An example of the conceptual difference is that chromatic transforms actually are less important for information transmission than spatial transforms when considering spatio-chromatic aspects at the same time.
Moreover, note that the Von-Kries color adaptation transform is actually the only transform that leads to a representation that captures less information than the input representation (for sensors with the same SNR).
Color adaptation may be more related to manifold alignment to improve
color-based classification (the kind of goal studied by Foster et al.) than to improve information transmission (the goal we study here).
Despite these differences, the interesting improvements of Kozachenko-Leonenko entropy estimator~\cite{Kozachenko87} proposed in~\cite{Ivan13}
should be compared in the future with RBIG because these two alternative estimators may be applicable to other problems of visual neuroscience~\cite{Victor02}.

This work originated from the analytical results for total correlation developed for cascades of linear+nonlinear networks~\cite{Martinez18}, and from the analysis of redundancy reduction in Wilson-Cowan networks~\cite{Gomez19}.
In both cases the analysis was restricted to achromatic stimuli. In~\cite{Martinez18} the approach was totally analytical, while~\cite{Gomez19} included RBIG estimations for the first time.
However, the main difference is that those works didn't considered the transmitted information,
but the redundancy reduction surrogate. In this work we have shown that, in general, $\Delta T$ may not be a good descriptor for $I$.


\paragraph{\textbf{Consistency between different databases and models.}}
Despite~\cite{Gomez19} is purely achromatic and does not consider $I$ (which are crucial conceptual differences),
comparison with those results is interesting for different reasons:
(1)~results of the redundancy at the input retinal representation are comparable (beyond the achromatic/chromatic difference) and some interesting consequences can be extracted.
(2)~the small gain in redundancy reduction at the Weber saturation and at the Divisive Normalization saturation
(also obtained in~\cite{Gomez19}) has been better explained here.

First, it is interesting to note that this work and~\cite{Gomez19} use different databases: the colorimetrically calibrated IPL database~\cite{Laparra12,Gutmann14}, and the radiometrically calibrated database by Foster-Nascimento-Amano~\cite{Foster15,Foster16} respectively.
Interestingly (for the users of the databases), the redundancy measures at the retinal input are comparable, which means that the statistics of both databases is similar.
Specifically, for achromatic patches subtending 0.06 deg in the Foster et al. database the total correlation is about 3.8 bits/sensor \cite{Gomez19}.
Here, for color patches subtending a smaller angle (0.05 deg in the IPL database) the total correlation is 4.1 bits/sensor. This redundancy is a little bit bigger because it includes color, which is redundant, but not that big because the size is smaller and hence less spatial structure is present, which should increase redundancy too.
Moreover, this suggest that consideration of color on top of spatial information increases the redundacy by a small amount, which is consistent with the fact that spatial operations are more relevant in removing redundancy and transmitting information.

Second, the models considered in both works have similar structure but they are not exactly the same: the one here doesn't have a specific layer for contrast computation and preserves the dimension in the local frequency transform.
However in both cases small gain in $\Delta T$ is obtained at the Weber-like saturation and at the cortical Divisive Normalization.
This consistency of behavior is a safety check for the models, but more interestingly, the analysis of $I$ proposed here and the point made about the advantage of using $I$ as descriptor of performance explains the benefits of these saturations even though they do not contribute to the reduction of redundancy.

\paragraph{\textbf{Consequences in image quality metrics and image coding}}

The \emph{Visual Information Fidelity} (VIF)~\cite{Sheikh05,Sheikh06} is an original approach to
characterize the distortion introduced in an image which is based in comparing the information about
the scene that \emph{a human} could extract from the distorted image wrt the information that he/she could
extract from the original image.

The results presented here can be incorporated in that attractive framework in different ways.
On the one hand, one may improve the perceptual model including nonlinearities and more sophisticated
noise schemes with no restriction because the non-parametric RBIG estimation is insensitive to the complexity
of the model.
On the other hand, estimations of mutual information in the original VIF scheme made crude approximations
on the PDF of the signals to apply analytical estimations, which may be too biased. Better measures of $I$ not subject to approximated models could certainly improve the results.

Following previous tradition of improvements of JPEG/MPEG compression based on Divisive Normalization~\cite{Malo01a,Malo06a},
current state-of-the-art in image coding also uses this kind of perceptually inspired linear+nonlinear architectures~\cite{ICLR17}.
The difference is that current architectures are optimized through the powerful automatic differentiation tools refined for deep-learning~\cite{Goodfellow16}.
In this case, the encoding and decoding transforms are optimized to minimize simultaneously the \emph{bitrate} and
the \emph{perceptual distortion}. Nowadays, these two magnitudes have different nature. However, with the considerations
done here, VIF distortion, which is expressed in information-theoretic units, could have more meaningful values, and the
rate in the image coder could be bounded or modulated by the amount of information transmitted by the perceptual system,
thus leading to a better optimization goal.

%
%
%

\appendix

\section{Manifold of natural images through the visual pathway}
\label{visual_manifolds}

Here we present scatter plots of the spatio-chromatic samples at the different layers of the considered model and the marginal PDFs throughout the network.
These transforms illustrate in practice the comments made in
Section~\ref{discussion} on Eq.~\ref{capacity_redundancy2}:
in order to increase the transmitted information, redundancy between the responses
at each layer throughout the pathway should be reduced, and the marginal PDFs should be equalized.

Moreover, visualization of the marginal PDFs and the corresponding joint PDFs is interesting
in case one wants to propose models for the marginals, as in~\cite{Simoncelli98,Malo10}, to make analytical estimations of the marginal entropies in Eqs.~\ref{capacity_redundancy2} or~\ref{deltaT}.

Fig.~\ref{transform_A} shows the geometrical effect of the point-wise chromatic transforms that occur at the first layers of the network.
We display selected projections of the $27$-dimensional arrays in order to illustrate (1) the distribution of the chromatic information (response of color sensors, LMS or ATD, at a fixed spatial location, in the scatter plots of the first row), and (2) the distribution of the spatio-chromatic information (response of color sensors, A, T and D, tuned to different  spatial locations, either close spatial neighbors -middle row-, or more distant neighbors -bottom row-).

\afterpage{
\begin{landscape}
\vspace{3cm}
\begin{figure}
   \centering
    \hspace{-2.21cm}
    \includegraphics[height=0.97\textheight,trim={0 3cm 0 3cm},clip]{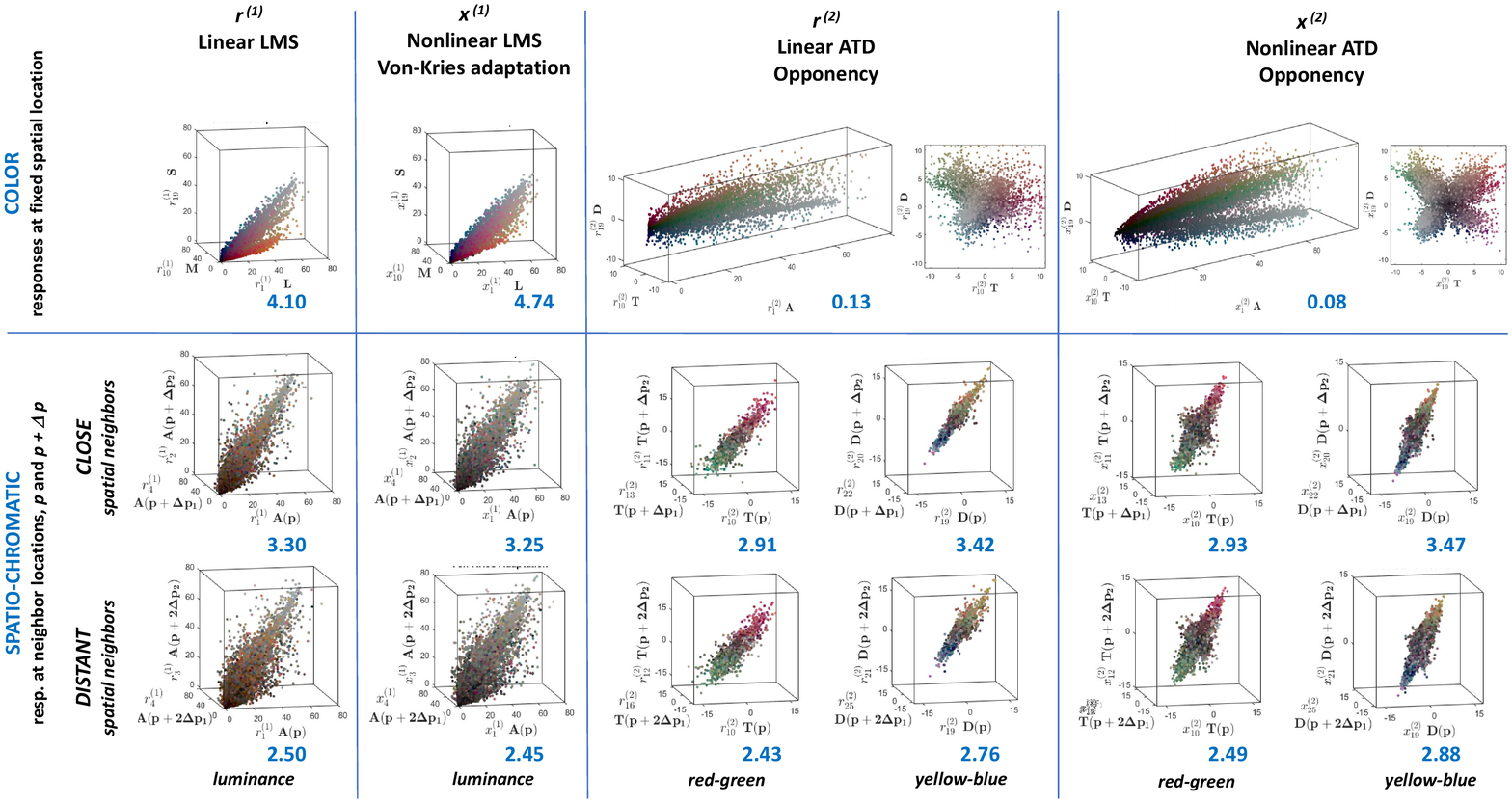}
    \caption{Image manifold transformed through the chromatic layers.
    Different rows display projections that represent different aspects of the signal, and different columns show how these aspects transform over the series of layers of the network.
    In the first row the samples are colored according to the actual color they represent
    (appropriate rendering was done with Colorlab \cite{Colorlab}).
    The scatter plots in the spatio-chromatic rows actually represent 4 dimensional data:
    the sample in a point in the 3d-axes represents the response of the corresponding chromatic sensor at the 3 corners of a square, and the color of the sample corresponds to the color of the 4th corner in the square.
    Correspondence of the color of the sample with the location in the space implies strong correlation between spatial neighbors.
    Numbers in blue display the 2nd order correlation measure proposed in \cite{Cardoso03}: $C = \frac{1}{2}log_2(\frac{\prod_i \Sigma_{ii}}{|\Sigma|})$, where $\Sigma$ is the covariance matrix and hence, $C$ is the difference between the entropy of the marginal distributions and the joint entropy assuming Gaussian approximations.}
    \label{transform_A}
\end{figure}
\end{landscape}
}

In the input representation, LMS or $\vect{r}^{(1)}$ vectors, the responses are strongly correlated, both between sensors of different spectral sensitivity and between sensors tuned to different spatial locations (see the alignment of the scatter plots and the numbers \emph{in blue} representing 2nd order correlation).
This represents the spectral~\cite{Maloney99,jimenez2014role} and spatial~\cite{Simoncelli01} smoothness of natural scenes.
As no chromatic adaptation has been applied yet, the scenes under reddish illumination lead to a distinctly separated blob in the color manifold (top plot of the first column). Note also the lower length of this cluster due to the lower luminance of the CIE A illumination in the experimental setting. In the spatio-chromatic rows (still first column) note that alignment of the scatter plot corresponding to sensors that are closer in space is bigger, consistently with a bigger correlation measure. Not surprisingly, spatial smoothness decays with distance.

Divisive normalization by the \emph{white} (Von-Kries adaptation, 2nd column) aligns the distinct clusters corresponding to the CIE D65 and CIE A in the input representation (see scatter plot and increased correlation measure), but it does not introduce qualitative changes in the distributions with spatial information.

Transform to opponent channels (3rd column) substantially reduces the correlation between color sensors at a fixed location. It is remarkable how Jameson and Hurvich opponent transform rotates the manifold as a sort of PCA even though it is not based on any statistical consideration. This is consistent with efficient coding interpretations of this stage~\cite{Buchsbaum83}. Nevertheless, responses of chromatic sensors T and D are still strongly correlated
to their spatial neighbors. In fact the spatial correlation of the chromatic sensors is similar to the achromatic counterpart (in previous columns). In the case of chromatic sensors spatial correlation also decays with distance.

Nonlinear saturation of the response at each chromatic sensor (4th column) reduces the correlation even further. Note that here we just picked a crude exponential saturation with fixed exponent following the Weber-like curves of the brightness and opponent mechanisms \cite{Krauskopf92,Romero93}.
However, this psychophysically inspired choice reduced the correlation again, consistently with efficient coding interpretations of this nonlinearity \cite{MacLeod03b,Laparra12}.
However, as in previous layers, this chromatic transform also has small effect on the spatial interaction between the sensors (mid and bottom rows of the 4th column).
Finally, it is important to mention the effect of this
saturation in the joint PDF: note the \emph{four-leaf-clover} shape of the color manifold projected onto the nonlinear T-D plane (top-right scatter plot). This \emph{four-leaf-clover} shape is a characteristic consequence of the saturation since it also appears in the spatial transforms based on Divisive Normalization, and has consequences on the bimodal nature of the marginal PDFs (see Figs.~\ref{transform_B}-\ref{marginals}),
that had been reported before~\cite{Malo10}.

Fig.~\ref{transform_B} illustrates the effect of the spatial transforms: texture sensors with local oriented receptive fields tuned to different frequencies weighted by achromatic and chromatic CSFs (linear responses $\vect{r}^{(3)}$), and Divisive Normalization (nonlinear responses  $\vect{x}^{(3)}$).

In each case, linear and nonlinear, on the left and right panels respectively, we display samples in $3d$~spaces corresponding to sensors tuned to zero frequency and to
two low-frequency components of vertical and horizontal orientation. We also represent samples in $2d$ projections for a better assessment of the shape of the joint PDF.
This is done for the three kinds of chromatic sensors: achromatic, red-green and yellow-blue (on the top, med, and bottom rows respectively).

\afterpage{
\begin{landscape}
\vspace{3cm}
\begin{figure}[!h]
   \centering
    \includegraphics[height=1.1\textheight,trim={3cm 2.5cm 3cm 2.5cm},clip]{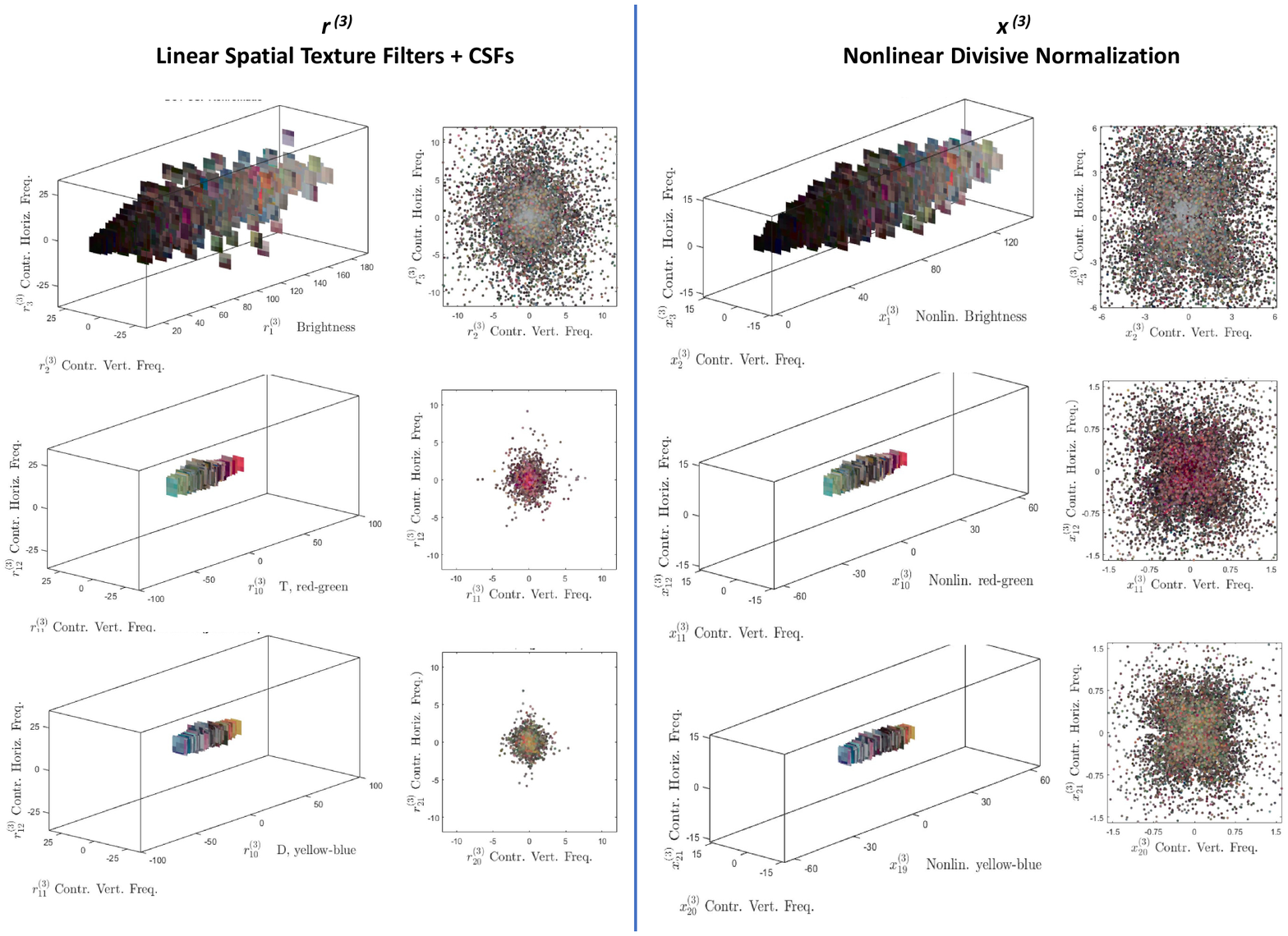}
    \caption{Transform of the image manifold through the spatial layers.
    In the $3d$ scatter plots the stimuli (image patches) are displayed at the locations given by the responses they elicit in the three considered sensors.
    This illustrates the texture-chromatic meaning of each region of the spaces.
    On the contrary, $2d$ projections only display dots of the average color of the corresponding patches so
    that the \emph{elliptically symmetric} shape or the \emph{four-leaf-clover} shape of the densities can be seen.
    }
    \label{transform_B}
\end{figure}
\end{landscape}
}

In this case, the 2nd-order correlation measure, $C$, given in the previous figure (numbers in blue) has not been included because here differences are more subtle: $C$ is basically negligible in all cases and differences are in the level of the estimation error. Actually, differences between these representations have to be described by more appropriate higher-order (multivariate) estimates proposed in this work: the total correlation and the transmitted information.

The scatter plots in the first column illustrate how the texture sensors reorient the PDFs of the spatial neighbors of previous layers.
In the bottom rows of the previous figure the PDFs were systematically correlated regardless of the chromatic manipulations. The spatial transforms applied to the A, T and D parts of the array $\vect{x}^{(2)}$ virtually remove the 2nd order correlation.
This is consistent with the efficient coding interpretation of this linear filter bank \cite{Hancock92,Olshausen96,Ruderman98color,Doi03,Gutmann14}.

The top left scatter plot shows that stimuli with specific visual
features actually give the expected response because it is located in the corresponding region of the space: note for instance the horizontal/vertical patterns of high contrast and different polarity aligned along their corresponding axis, and located along the zero-frequency axis according to their brightness.

As always throughout this Appendix, equivalent plots represent the same sets of samples.
Therefore the scatter plots of the second and third rows (still left column) represent the very same set of images as the top-left scatter plot.
However, the location of these images in the chromatic texture axes
is markedly different because of two reasons:
(1) chromatic contrast is smaller than achromatic contrast in natural images (note that images with high chromatic contrast in Fig.~\ref{jointPDF} are less frequent), and on top of that, (2) the chromatic CSFs has lower bandwidth than the achromatic filter, so the high frequency color oscillations are strongly attenuated by the system. As a result, the images are clustered along the DC color axis.
This difference in amplitude of the oscillations can be seen in the extent of the samples over the $2d$~scatter plots on the left panel.
More interestingly, in these $2d$ plots we can see that the elliptically symmetric distribution that had been reported before for natural achromatic images~\cite{Lyu09}, also happens in the T and D channels.

\afterpage{
\begin{figure}[t]
   \begin{center}
    \includegraphics[width=1.05\textwidth,trim={2.2cm 2.6cm 0 2.8cm},clip]{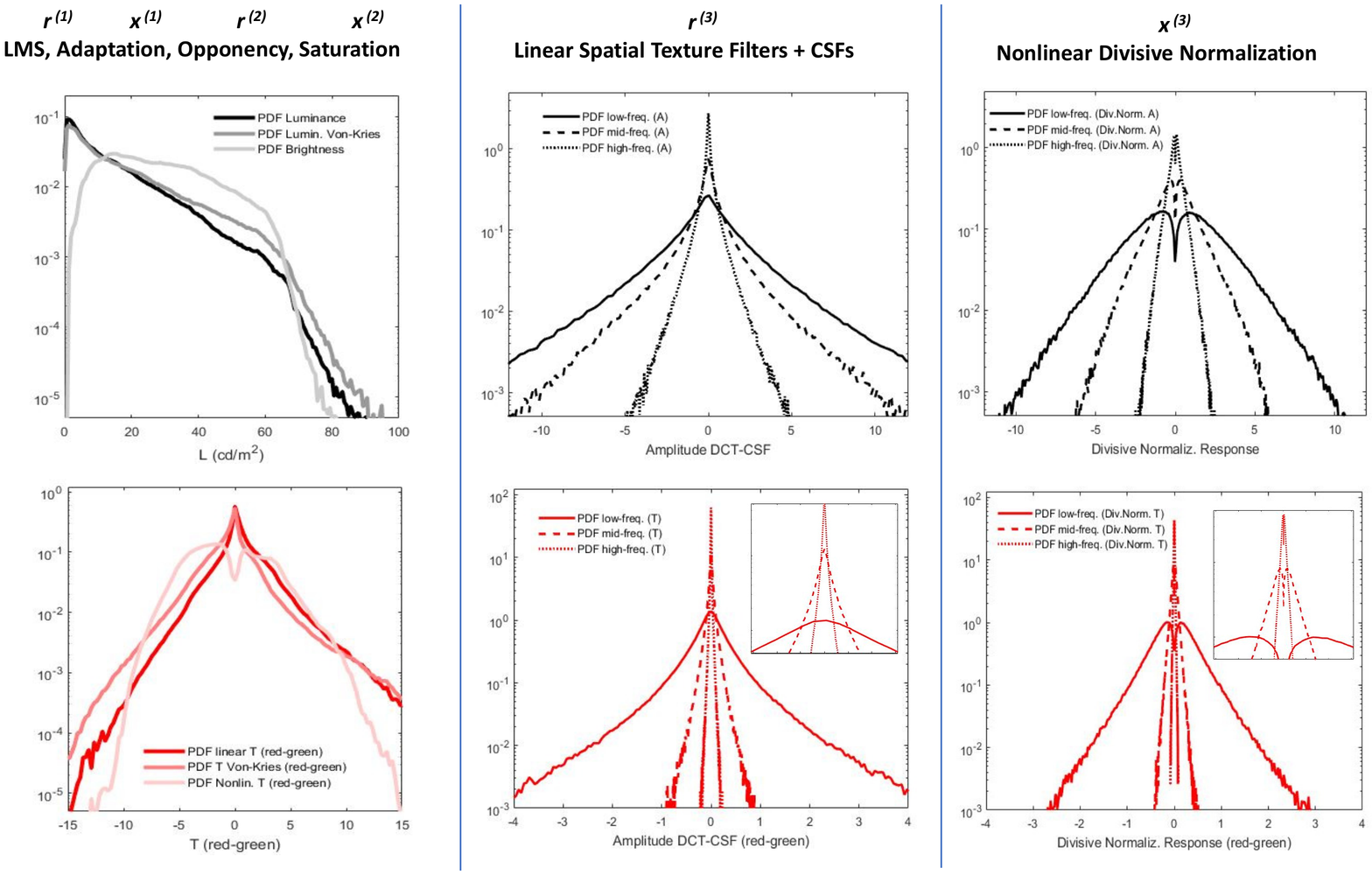}
    \end{center}
    \caption{Marginal PDFs of chromatic and spatial-texture sensors along the visual pathway.}
    \label{marginals}
\end{figure}
}

The right panel shows the result of a Divisive Normalization transform applied to the responses of the linear sensors.
Divisive Normalization implies saturation in each dimension but (as opposed to the crude dimension-wise saturation done in the chromatic channels)
here the saturation of the response of a sensor for certain stimulus depends on the activity of the sensors tuned to other
textures~\cite{Watson97,Malo06a}.
Saturation is apparent in the $3d$ scatter plots because the higher contrast patterns in the achromatic
plot have been pushed towards the DC axis and the low-brightness region has been expanded.
In the chromatic cases saturation is also apparent since the low-saturation region has been expanded.

The $2d$~scatter plots show how, due to Divisive Normalization, the elliptically symmetric distribution characteristic of local-frequency
representations
change to this \emph{four-leaf-clover} shape, as in the case of the color distribution
after the saturation (previous figure, top-right scatter plot).

Fig.~\ref{marginals} shows that saturation operations also lead to bimodal marginal PDFs for chromatic sensors together with classical results for marginal PDFs~\cite{Laughlin83,Mumford99,Simoncelli98,malo2000role,Simoncelli01}. This bimodal marginal PDFs had only been reported for achromatic texture sensors~\cite{Malo10,Martinez18}.

Regarding well known results, Fig.~\ref{marginals} (top left) shows how the first layers contribute to equalization of the PDF of the responses associated to brightness: Von-Kries adaptation expands the range of the responses to the stimuli under darker illumination, and the saturation contributes to reduce the peak at the dark region~\cite{Laughlin83,Mumford99}. In the chromatic case (bottom left), before Von-Kries adaptation the tail in the reddish region is heavier (half of the samples are under the reddish CIE A illumination). Then, adaptation reduces the bias in the PDF, and saturation leads to a bimodal PDF of reduced support in the red-green sensors (yellow-blue sensors behave similarly). The column at the center displays heavy tailed PDFs in the case of linear texture sensors, either achromatic (center top) or chromatic (center bottom) with decreasing variance depending on the frequency~\cite{Simoncelli98,malo2000role,Simoncelli01}.

On the other hand, the right column and bottom-left plot of Fig.~\ref{marginals} show how the joint distributions
with a mode in each quadrant (\emph{four-leaf-clover} shapes) in Figs.~\ref{transform_A} and \ref{transform_B}, lead to the
bimodal marginal PDFs.
These marginal PDFs are not specific of the considered dataset: they appear in the Van Hateren radiance calibrated dataset
and in the Foster-Nascimento-Amano dataset~\cite{Foster15,Foster16}, as has been reported in~\cite{Malo10,Gomez19} respectively.
This shape appears when applying Divisive Normalization saturations with the appropriate exponent:
in~\cite{Malo10} we proposed a functional form for this marginal related to the parameters of the Divisive Normalization,
and this kind of \emph{four-leaf-clover} joint PDFs and resulting bimodal marginal also appear when optimizing
Divisive Normalization for image coding \cite{ICLR17} (personal communication from the authors).

According to Eq.~\ref{capacity_redundancy2} analytical modelling of these marginal PDFs
can be helpful to characterize the performance. However, in this illustrative section we just want to show new evidences (now for chromatic sensors) about the fact that the family of possible PDFs after Divisive Normalization is not restricted to Gaussian, as sometimes assumed~\cite{BovikGaussianization}.


The suggestions made in this Appendix are actually quantified using the proposed Gaussianization tool in the Results section \ref{results}.

\section{\blue{Performance of RBIG estimators of $I$ for t-student and Gaussian sources}}
\label{appPerformance}

\blue{In this appendix we justify the use of a specific RBIG estimator for the transmitted information, Eq.~\ref{eq:mirbig}, for image-like sources,
$\vect{r}$, and sensory-like systems $\vect{x} = s(\vect{r}) + \vect{n}$.
As a convenient reference we also consider the transmitted information in the
trivial case $\vect{x} = \vect{r} + \vect{n}$ when both signal and noise are Gaussian.}

\blue{Natural images are known to be non-Gaussian since empirical image histograms are found to have heavier tails than Gaussian PDFs (see Appendix~\ref{visual_manifolds} and~\cite{malo2000role,Simoncelli98,Simoncelli01}).
This motivated the use of different sorts of models (such as Gaussian Scale Mixtures~\cite{Portilla01}, mixtures of heavy tailed distributions with controlled correlation \cite{Hyvarinen09}, $L_p$-symmetric densities~\cite{Sinz10}, Gaussian densities with signal dependent covariance~\cite{Malo10}, etc). In this appendix we assume a simpler image model based on \emph{t-student} densities defined in a local frequency domain.
We do this for convenience in sample generation and in the computation of an analytical ground truth for transmitted information.
Image models based on \emph{t-student} have been used in natural image coding~\cite{Oord14}, and they are a common illustrative example when discussing the
properties of natural images~\cite{Lyu09,Malo10,Gutmann14}.
Another advantage of \emph{t-student} models is that they can be turned into Gaussian PDFs by increasing the degrees-of-freedom parameter $\nu$ that controls the kurtosis of the density.}

\begin{figure}[t]
   \centering
    \hspace{0cm}
    \includegraphics[width=0.95\textwidth]{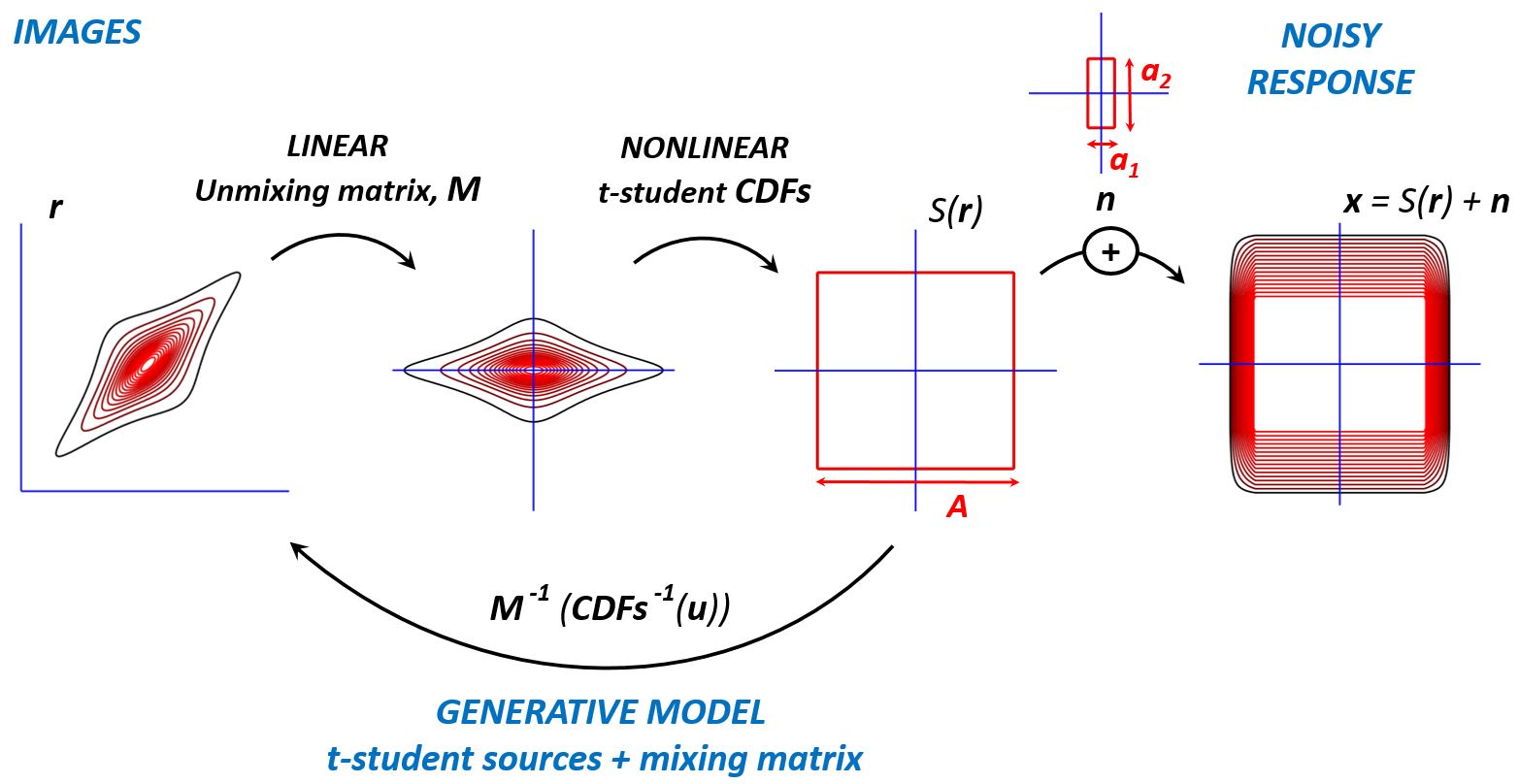}
    \vspace{-0.8cm}
    \caption{Idealized image model and sensory system with (1)~a simple analytical expression for transmitted information, $I(\vect{r},\vect{x}) =  \sum_{i=1}^d \left(\frac{A-a_i}{A} log_2(A) + \frac{a_i}{A} log_2(e^{1/2} A) - log_2(a_i) \right) $,
    and (2)~a simple procedure to generate image samples to check the information estimators.}
    \label{ideal_infomax}
\end{figure}

\blue{Fig.~\ref{ideal_infomax} describes both a simple linear+nonlinear noisy sensory system, and a procedure for sample generation. For a simple computation of the transmitted information, the linear and nonlinear stages of the sensory system match the elements of the image model.
First, the linear transform identifies the independent components of the images (because it uses the inverse of the selected mixing matrix), and then, it marginally equalizes the outputs (because it uses the marginal CDFs of the selected \emph{t-student} PDFs to lead to a joint uniform PDF, a $d$-cube of size $A$).
Finally the response in this uniform domain is subject to uniform noise, eventually with different amplitudes, $a_i$, for each coefficient.
The advantage of this idealized setting is that the transmitted information can be computed analytically and samples are easy to obtain with the proper choices for the transforms.}

\blue{In this setting, synthetic images are obtained in the following way.
First, samples drawn from the uniform density in the $d$-cube, are passed through the inverse univariate CDFs of the selected \emph{t-student} PDFs, to obtain independent univariate heavy tailed samples.
Then, these independent sources are mixed using a $d\times d$ \emph{mixing} matrix.
In order to have sensible image samples the mixing matrix for this illustration is the inverse of 3 sets of 2D-DCT filters of size $3\times 3$ ($d=27$ as in the experiments done with real images).
The t-student sources are defined in a color-opponent space (the same used in the model). Therefore, the resulting opponent images after the inverse DCTs are linearly mixed again according to the ATD-to-LMS matrix, to finally obtain $9\times 9$ image patches
in the LMS color space. Note that, given the known frequency and color meaning of the $i$-th column of the mixing matrix, if the amplitudes of the \emph{t-student} variables are multiplied by a frequency and color dependent factor, one can reproduce the classical $1/f$ spectrum of natural images~\cite{Ruderman94,Simoncelli98,Mumford99,malo2000role} and the relative energy of the opponent channels~\cite{Ruderman98color}.}

\begin{table}[b!]
\caption{Performance of $I$ estimators for \emph{t-student} and \emph{Gaussian} samples (Relative Error, in \%).}
\label{performance}
      \begin{tabular}{c}
            T-STUDENT source / Ideal Infomax System
      \end{tabular}
      \begin{tabular}{lc|ccccc}
        \hline\\[-0.1cm]
        Shape & \# Samples & Eq.~\ref{eq:mirbig} & Eq.~\ref{eq:mirbig2} & KL+offset~\cite{Ivan13}  & KL~\cite{Kozachenko87} & k-NN~\cite{ITE} \\ \hline\\[-0.2cm]
        & $1 \cdot 10^3$            & $\mathbf{24 \pm 7}$ & $40 \pm 30$ & $58 \pm 2$ & $278 \pm 2$ & $239 \pm 1$  \\
        $\nu = 3$& $4 \cdot 10^3$   & $\mathbf{5 \pm 2}$  & $16 \pm 7 $ & $56 \pm 1$ & $259 \pm 9$ & $225 \pm 1$  \\
        & $1 \cdot 10^4$            & $\mathbf{4 \pm 2}$  & $16 \pm 6 $ & $55 \pm 1$ & $246 \pm 1$ & $215 \pm 1$  \\
        & $4 \cdot 10^4$            & $\mathbf{12 \pm 2}$ & $18 \pm 4 $ & $53 \pm 1$ & $229 \pm 1$ & $203 \pm 1$  \\
        \hline\\[-0.1cm]
        & $1 \cdot 10^3$            & $\mathbf{ 17 \pm 8} $ & $25 \pm 15$ & $43 \pm 2 $    & $255 \pm 1$ & $222 \pm 1$  \\
        $\nu = 4$& $4 \cdot 10^3$   & $\mathbf{  3 \pm 3} $ & $16 \pm 8 $ & $43 \pm 1 $    & $238 \pm 1$ & $210 \pm 1$  \\
        & $1 \cdot 10^4$            & $\mathbf{  7 \pm 3} $ & $15 \pm 6 $ & $41.1 \pm 0.5$ & $224 \pm 1$ & $199 \pm 1$  \\
        & $4 \cdot 10^4$            & $\mathbf{ 17 \pm 2} $ & $21 \pm 7 $ & $39.4 \pm 0.2$ & $208 \pm 1$ & $188 \pm 1$  \\
        \hline\\[-0.1cm]
        & $1 \cdot 10^3$            & $\mathbf{15 \pm 7 }$ & $25 \pm 14$ & $20 \pm 1$     & $203 \pm 1$ & $185 \pm 1$  \\
        $\nu = 100$& $4 \cdot 10^3$ & $\mathbf{5 \pm 2 }$ & $18 \pm 4 $ & $21.4 \pm 0.7 $& $187 \pm 1$ & $173 \pm 1$  \\
        & $1 \cdot 10^4$            & $\mathbf{9 \pm 2 }$ & $21 \pm 4 $ & $21.1 \pm 0.1$ & $174 \pm 1$ & $164 \pm 1$  \\
        & $4 \cdot 10^4$            & $23 \pm 3$ & $29 \pm 4 $ & $\mathbf{20.4 \pm 0.1}$ & $160 \pm 1$ & $153 \pm 1$  \\
        \hline\\[0.0cm]
      \end{tabular}
      \begin{tabular}{c}
            GAUSSIAN source / Trivial System
      \end{tabular}
      \begin{tabular}{lc|ccccc}
        \hline\\[-0.1cm]
      \blanco{Shape}&\# Samples & Eq.~\ref{eq:mirbig} & Eq.~\ref{eq:mirbig2} & KL+offset~\cite{Ivan13}  & KL~\cite{Kozachenko87} & k-NN~\cite{ITE} \\ \hline\\[-0.2cm]
       \blanco{$\nu = \inf$} & $1 \cdot 10^3$   & $ 24 \pm 1 $ & $12 \pm 2 $ & $\mathbf{6.0 \pm 0.2}$ & $91.5 \pm 0.2$ & $95.6 \pm 0.1$  \\
        & $4 \cdot 10^3$                        & $ 16 \pm 3 $ & $ 9 \pm 3 $ & $\mathbf{6.4 \pm 0.4}$ & $90.4 \pm 0.4$ & $95.1 \pm 0.2$  \\
        & $1 \cdot 10^4$                        & $ 10 \pm 2 $ & $\mathbf{ 4 \pm 1 }$ & $6.4 \pm 0.2$ & $88.5 \pm 0.2$ & $93.8 \pm 0.2$  \\
        & $4 \cdot 10^4$                        & $  7 \pm 3 $ & $\mathbf{ 3 \pm 2 }$ & $6.1 \pm 0.2$ & $86.3 \pm 0.2$ & $92.3 \pm 0.2$  \\
      \hline
      \end{tabular}
\end{table}

\blue{The analytical transmitted information is also easy to compute using Eq.~\ref{capacity_redundancy2}.
First, the joint entropy of the noise is $h(\vect{n}) = \sum_{i_1}^d \,\, log_2(a_i)$ because it is drawn from a \mbox{$d$-rectangle} of sizes $a_i$.
Second, we see that $T(\vect{x})=0$ because the components of the noisy response are independent since the PDFs of $s(\vect{r})$ and $\vect{n}$ are uniform $d$-rectangles aligned with the axes.
Finally, the computation of $\sum_{i=1}^d h(x_i)$ is also easy because noise additivity implies that the marginal PDFs are the convolution of the PDFs of the deterministic response and the noise.
That means the convolution of two unit-volume rectangles of support $A$, for the response, and $a_i$, for the noise.
This leads to marginal PDFs with truncated-pyramid shape. Therefore, the entropy integral can be broken into the known entropy integrals
of a \emph{rectangle} and a \emph{triangle}, and hence:
$h(x_i) = \frac{A-a_i}{A} log_2(A) + \frac{a_i}{A} log_2(e^{1/2} A)$, for the case where $a_i<A$.
These amplitudes have to be interchanged if $a_i>A$.}

\blue{An analytical solution is also possible in the trivial case (Gaussian source, identity transform and Gaussian noise independent of the response).
In that case, using that $I(\vect{r},\vect{x}) = h(\vect{x}) - h(\vect{n})$, and the known entropy of the Gaussian, we
have: $I(\vect{r},\vect{x}) = \frac{1}{2} log_2\left(  |2 \pi e (\Sigma_r + \Sigma_n)| \right) - \frac{1}{2} log_2 \left( |2 \pi e \Sigma_n| \right)$}.

\blue{Table~\ref{performance} shows the relative error with regard to the analytical value (in percentage of the actual information) in the two cases: \emph{t-student} and \emph{Gaussian}. We compare the RBIG-based expressions, Eqs.~\ref{eq:mirbig} and~\ref{eq:mirbig2},
with different implementations of classical mutual information estimators based on k-NN estimators of entropy~\cite{ITE,Kozachenko87}, and a recent improvement of the Kozachenko-Leonenko estimator~\cite{Ivan13}.
Different values of the kurtosis (from more image-like $\nu = 3, 4$ to more Gaussian $\nu=100$) are explored, as well as different number of samples in the estimation.}

\blue{The results show that while the method proposed by Marin-Franch et al.~\cite{Ivan13} shows better
behavior for Gaussian sources, the estimator proposed here, Eq.~\ref{eq:mirbig}, gives better results with heavy tailed samples
and the kind of systems considered in this work.}

\begin{backmatter}

\section*{Declarations}

\paragraph{Availability of data and materials:}

Color-calibrated image database is available at http://isp.uv.es/data\_calibrated.html

Code for the visual model is available at http://isp.uv.es/code/visioncolor/vistamodels.html

Code for the RBIG estimator is available at http://isp.uv.es/rbig.html

\vspace{-0.4cm}
\paragraph{Acknowledgements:}
I thank the attendees to my lectures on \emph{Information theory for Visual Neuroscience} and
particularly to \emph{Vir} for encouraging me to write this with the promise of a travel.

\vspace{-0.4cm}
\paragraph{Funding:}
This work was partially funded by the Spanish Government through the MINECO
grant DPI2017-89867-C2-2-R and by the Generalitat Valenciana through the grant GrisoliaP/2019/035.

\vspace{-0.4cm}
\paragraph{Authors' contributions:}
The corresponding (and only) author completed all the work.

\vspace{-0.4cm}
\paragraph{Consent for publication:} The author supports the publication.

\vspace{-0.4cm}
\paragraph{Competing interests:}
The author declares that he has no competing interests.

\vspace{-0.4cm}
\paragraph{Ethics approval and consent to participate:}
Not applicable.


\bibliographystyle{bmc-mathphys} 


\newcommand{\BMCxmlcomment}[1]{}

\BMCxmlcomment{

<refgrp>

<bibl id="B1">
  <title><p>The limiting capacity of a neuronal link</p></title>
  <aug>
    <au><snm>MacKay</snm><fnm>D.</fnm></au>
    <au><snm>McCulloch</snm><fnm>W.</fnm></au>
  </aug>
  <source>Bull.Math.Biophys.</source>
  <pubdate>1952</pubdate>
  <volume>14</volume>
  <fpage>127</fpage>
  <lpage>135</lpage>
</bibl>

<bibl id="B2">
  <title><p>Sensory mechanisms, the reduction of redundancy, and
  intelligence</p></title>
  <aug>
    <au><snm>Barlow</snm><fnm>H.B.</fnm></au>
  </aug>
  <source>Proc. of the Nat. Phys. Lab. Symposium on the Mechanization of
  Thought Process</source>
  <pubdate>1959</pubdate>
  <issue>10</issue>
  <fpage>535</fpage>
  <lpage>539</lpage>
</bibl>

<bibl id="B3">
  <title><p>Redundancy reduction revisited</p></title>
  <aug>
    <au><snm>Barlow</snm><fnm>H.</fnm></au>
  </aug>
  <source>Network: Comp. Neur. Syst.</source>
  <pubdate>2001</pubdate>
  <volume>12</volume>
  <issue>3</issue>
  <fpage>241</fpage>
  <lpage>253</lpage>
</bibl>

<bibl id="B4">
  <title><p>Information theory in neuroscience</p></title>
  <aug>
    <au><snm>Dimitrov</snm><fnm>AG.</fnm></au>
    <au><snm>Lazar</snm><fnm>AA.</fnm></au>
    <au><snm>Victor</snm><fnm>JD.</fnm></au>
  </aug>
  <source>J. Comput. Neurosci.</source>
  <pubdate>2011</pubdate>
  <volume>30</volume>
  <issue>1</issue>
  <fpage>1</fpage>
  <lpage>-5</lpage>
</bibl>

<bibl id="B5">
  <title><p>The free-energy principle: a rough guide to the brain?</p></title>
  <aug>
    <au><snm>Friston</snm><fnm>K.</fnm></au>
  </aug>
  <source>Trends Cognit. Sci.</source>
  <publisher>Elsevier</publisher>
  <pubdate>2009</pubdate>
  <volume>13</volume>
  <issue>7</issue>
  <fpage>293</fpage>
  <lpage>-301</lpage>
</bibl>

<bibl id="B6">
  <title><p>Information Processing in Living Systems</p></title>
  <aug>
    <au><snm>Tkacik</snm><fnm>G.</fnm></au>
    <au><snm>Bialek</snm><fnm>W.</fnm></au>
  </aug>
  <source>Ann.Rev.Condens.Mat.Phys.</source>
  <pubdate>2016</pubdate>
  <volume>7</volume>
  <fpage>89</fpage>
  <lpage>-117</lpage>
</bibl>

<bibl id="B7">
  <title><p>Entropy and Information in Neural Spike Trains</p></title>
  <aug>
    <au><snm>Strong</snm><fnm>S.P.</fnm></au>
    <au><snm>Koberle</snm><fnm>R.</fnm></au>
    <au><snm>Steveninck</snm><fnm>R.R.</fnm></au>
    <au><snm>Bialek</snm><fnm>W.</fnm></au>
  </aug>
  <source>Phys. Rev. Lett.</source>
  <pubdate>1998</pubdate>
  <volume>80</volume>
  <fpage>197–</fpage>
  <lpage>200</lpage>
</bibl>

<bibl id="B8">
  <title><p>A network that uses few active neurones to code visual input
  predicts the diverse shapes of cortical receptive fields.</p></title>
  <aug>
    <au><snm>Rehn</snm><fnm>M.</fnm></au>
    <au><snm>Sommer</snm><fnm>FT.</fnm></au>
  </aug>
  <source>J Comp. Neurosci.</source>
  <pubdate>2007</pubdate>
  <volume>22</volume>
  <fpage>135</fpage>
  <lpage>146</lpage>
</bibl>

<bibl id="B9">
  <title><p>How the Optic Nerve Allocates Space, Energy Capacity, and
  Information</p></title>
  <aug>
    <au><snm>Perge</snm><fnm>JA.</fnm></au>
    <au><snm>Koch</snm><fnm>K.</fnm></au>
    <au><snm>Miller</snm><fnm>R.</fnm></au>
    <au><snm>Sterling</snm><fnm>P.</fnm></au>
    <au><snm>Balasubramanian</snm><fnm>V.</fnm></au>
  </aug>
  <source>J. Neurosci.</source>
  <pubdate>2009</pubdate>
  <volume>29</volume>
  <issue>24</issue>
  <fpage>7917–</fpage>
  <lpage>7928</lpage>
</bibl>

<bibl id="B10">
  <title><p>The effect of cell size and channel density on neuronal information
  encoding and energy efficiency</p></title>
  <aug>
    <au><snm>Sengupta</snm><fnm>B.</fnm></au>
    <au><snm>Faisal</snm><fnm>AA.</fnm></au>
    <au><snm>Laughlin</snm><fnm>SB.</fnm></au>
    <au><snm>Niven</snm><fnm>JE.</fnm></au>
  </aug>
  <source>J. Cereb. Blood Flow Metabol.</source>
  <pubdate>2013</pubdate>
  <volume>33</volume>
  <fpage>1465–</fpage>
  <lpage>1473</lpage>
</bibl>

<bibl id="B11">
  <title><p>Energy-Efficient Information Transfer by Visual Pathway
  Synapses</p></title>
  <aug>
    <au><snm>Harris</snm><fnm>JJ.</fnm></au>
    <au><snm>Jolivet</snm><fnm>R.</fnm></au>
    <au><snm>Engl</snm><fnm>E.</fnm></au>
    <au><snm>Attwell</snm><fnm>D.</fnm></au>
  </aug>
  <source>Current Biology</source>
  <pubdate>2015</pubdate>
  <volume>25</volume>
  <issue>24</issue>
  <fpage>3151–</fpage>
  <lpage>3160</lpage>
</bibl>

<bibl id="B12">
  <title><p>Principles of neural design</p></title>
  <aug>
    <au><snm>Sterling</snm><fnm>P.</fnm></au>
    <au><snm>Laughlin</snm><fnm>S.</fnm></au>
  </aug>
  <publisher>London, England: The MIT Press</publisher>
  <pubdate>2015</pubdate>
</bibl>

<bibl id="B13">
  <title><p>Design of a Neuronal Array</p></title>
  <aug>
    <au><snm>Borghuis</snm><fnm>BG.</fnm></au>
    <au><snm>Ratliff</snm><fnm>CP.</fnm></au>
    <au><snm>Smith</snm><fnm>RG.</fnm></au>
    <au><snm>Sterling</snm><fnm>P.</fnm></au>
    <au><snm>Balasubramanian</snm><fnm>V.</fnm></au>
  </aug>
  <source>J. Neurosci.</source>
  <pubdate>2008</pubdate>
  <volume>28</volume>
  <issue>12</issue>
  <fpage>3178</fpage>
  <lpage>–3189</lpage>
</bibl>

<bibl id="B14">
  <title><p>How much the eye tells the brain</p></title>
  <aug>
    <au><snm>Koch</snm><fnm>K.</fnm></au>
    <au><snm>McLean</snm><fnm>J.</fnm></au>
    <au><snm>Segev</snm><fnm>R.</fnm></au>
    <au><snm>Freed</snm><fnm>MA.</fnm></au>
    <au><snm>Berry</snm><fnm>MJ.</fnm></au>
    <au><snm>Balasubramanian</snm><fnm>V.</fnm></au>
    <au><snm>Sterling</snm><fnm>P.</fnm></au>
  </aug>
  <source>Curr. Biol.</source>
  <pubdate>2006</pubdate>
  <volume>16</volume>
  <issue>14</issue>
  <fpage>1428</fpage>
  <lpage>-1434</lpage>
</bibl>

<bibl id="B15">
  <title><p>Normalization of cell responses in cat striate cortex</p></title>
  <aug>
    <au><snm>Heeger</snm><fnm>DJ</fnm></au>
  </aug>
  <source>Visual Neuroscience</source>
  <publisher>Cambridge University Press</publisher>
  <pubdate>1992</pubdate>
  <volume>9</volume>
  <issue>2</issue>
  <fpage>181–197</fpage>
</bibl>

<bibl id="B16">
  <title><p>Summation and Division by Neurons in Visual Cortex</p></title>
  <aug>
    <au><snm>Carandini</snm><fnm>M.</fnm></au>
    <au><snm>Heeger</snm><fnm>D.</fnm></au>
  </aug>
  <source>Science</source>
  <pubdate>1994</pubdate>
  <volume>264</volume>
  <issue>5163</issue>
  <fpage>1333</fpage>
  <lpage>-6</lpage>
</bibl>

<bibl id="B17">
  <title><p>Habituation reveals fundamental chromatic mechanisms in striate
  cortex of macaque</p></title>
  <aug>
    <au><snm>Tailby</snm><fnm>C.</fnm></au>
    <au><snm>Solomon</snm><fnm>SG.</fnm></au>
    <au><snm>Dhruv</snm><fnm>NT.</fnm></au>
    <au><snm>P.</snm><fnm>L</fnm></au>
  </aug>
  <source>J Neurosci.</source>
  <pubdate>2008</pubdate>
  <volume>28</volume>
  <issue>5</issue>
  <fpage>1131‐1139</fpage>
</bibl>

<bibl id="B18">
  <title><p>Normalization as a canonical neural computation</p></title>
  <aug>
    <au><snm>Carandini</snm><fnm>M.</fnm></au>
    <au><snm>Heeger</snm><fnm>D. J</fnm></au>
  </aug>
  <source>Nature Rev. Neurosci.</source>
  <publisher>Nature Publishing Group</publisher>
  <pubdate>2012</pubdate>
  <volume>13</volume>
  <issue>1</issue>
  <fpage>51</fpage>
  <lpage>-62</lpage>
</bibl>

<bibl id="B19">
  <title><p>The Relation Between Color Discrimination and Color Constancy: When
  Is Optimal Adaptation Task Dependent?</p></title>
  <aug>
    <au><snm>Abrams</snm><fnm>AB</fnm></au>
    <au><snm>Hillis</snm><fnm>JM</fnm></au>
    <au><snm>Brainard</snm><fnm>DH</fnm></au>
  </aug>
  <source>Neural Computation</source>
  <pubdate>2007</pubdate>
  <volume>19</volume>
  <issue>10</issue>
  <fpage>2610</fpage>
  <lpage>2637</lpage>
</bibl>

<bibl id="B20">
  <title><p>Color Appearance Models</p></title>
  <aug>
    <au><snm>Fairchild</snm><fnm>M.D.</fnm></au>
  </aug>
  <publisher>Sussex, UK: Wiley</publisher>
  <series><title><p>The Wiley-IS\&T Series in Imaging Science and
  Technology</p></title></series>
  <pubdate>2013</pubdate>
</bibl>

<bibl id="B21">
  <title><p>Perceptual-components architecture for digital video</p></title>
  <aug>
    <au><snm>Watson</snm><fnm>BA</fnm></au>
  </aug>
  <source>Journal of The Optical Society of America A-optics Image Science and
  Vision</source>
  <pubdate>1990</pubdate>
  <volume>7</volume>
  <issue>10</issue>
  <fpage>1943</fpage>
  <lpage>1954</lpage>
</bibl>

<bibl id="B22">
  <title><p>Model of visual contrast gain control and pattern
  masking</p></title>
  <aug>
    <au><snm>Watson</snm><fnm>A. B.</fnm></au>
    <au><snm>Solomon</snm><fnm>J. A.</fnm></au>
  </aug>
  <source>JOSA A</source>
  <publisher>Optical Society of America</publisher>
  <pubdate>1997</pubdate>
  <volume>14</volume>
  <issue>9</issue>
  <fpage>2379</fpage>
  <lpage>-2391</lpage>
</bibl>

<bibl id="B23">
  <title><p>A Model of Neuronal Reponses in Visual Area {MT}</p></title>
  <aug>
    <au><snm>Simoncelli</snm><fnm>E.P.</fnm></au>
    <au><snm>Heeger</snm><fnm>D.</fnm></au>
  </aug>
  <source>Vis. Res.</source>
  <pubdate>1998</pubdate>
  <volume>38</volume>
  <issue>5</issue>
  <fpage>743</fpage>
  <lpage>761</lpage>
</bibl>

<bibl id="B24">
  <title><p>Trichromacy, opponent colours coding and optimum colour information
  transmission in the retina</p></title>
  <aug>
    <au><snm>Buchsbaum</snm><fnm>G.</fnm></au>
    <au><snm>Gottschalk</snm><fnm>A.</fnm></au>
  </aug>
  <source>Proc. Roy. Soc. Lond. B Biol. Sci.</source>
  <pubdate>1983</pubdate>
  <volume>220</volume>
  <issue>1218</issue>
  <fpage>89</fpage>
  <lpage>113</lpage>
</bibl>

<bibl id="B25">
  <title><p>Matching coding to scenes to enhance efficiency</p></title>
  <aug>
    <au><snm>Laughlin</snm><fnm>S. B.</fnm></au>
  </aug>
  <source>In Braddick, O.J. \& Sleigh, A.C. (Eds) Physical and Biological
  Processing of Images</source>
  <publisher>Berlin, Germany: Springer</publisher>
  <pubdate>1983</pubdate>
  <fpage>42</fpage>
  <lpage>-52</lpage>
</bibl>

<bibl id="B26">
  <title><p>The pleistochrome: optimal opponent codes for natural
  colors</p></title>
  <aug>
    <au><snm>MacLeod</snm><fnm>D.</fnm></au>
    <au><snm>Twer</snm><fnm>T.</fnm></au>
  </aug>
  <source>Color Perception: From Light to Object</source>
  <publisher>Oxford, UK: Oxford Univ. Press</publisher>
  <editor>Heyer, D. and Mausfeld, R.</editor>
  <pubdate>2003</pubdate>
</bibl>

<bibl id="B27">
  <title><p>Nonlinearities and adaptation of color vision from sequential
  principal curves analysis</p></title>
  <aug>
    <au><snm>Laparra</snm><fnm>V.</fnm></au>
    <au><snm>Jim{\'e}nez</snm><fnm>S.</fnm></au>
    <au><snm>Camps Valls</snm><fnm>G.</fnm></au>
    <au><snm>Malo</snm><fnm>J.</fnm></au>
  </aug>
  <source>Neural Computation</source>
  <publisher>MIT Press 55 Hayward Street, Cambridge, MA 02142-1315 email:
  journals-info@ mit. edu</publisher>
  <pubdate>2012</pubdate>
  <volume>24</volume>
  <issue>10</issue>
  <fpage>2751</fpage>
  <lpage>-2788</lpage>
</bibl>

<bibl id="B28">
  <title><p>The Principal Components of Natural Images</p></title>
  <aug>
    <au><snm>Hancock</snm><fnm>P.</fnm></au>
    <au><snm>Baddeley</snm><fnm>R.</fnm></au>
    <au><snm>Smith</snm><fnm>L.</fnm></au>
  </aug>
  <source>Network</source>
  <pubdate>1991</pubdate>
  <volume>3</volume>
  <fpage>61</fpage>
  <lpage>70</lpage>
</bibl>

<bibl id="B29">
  <title><p>Emergence of simple-cell receptive field properties by learning a
  sparse code for natural images</p></title>
  <aug>
    <au><snm>Olshausen</snm><fnm>B.</fnm></au>
    <au><snm>Field</snm><fnm>D.</fnm></au>
  </aug>
  <source>Nature</source>
  <pubdate>1996</pubdate>
  <volume>281</volume>
  <fpage>607</fpage>
  <lpage>609</lpage>
</bibl>

<bibl id="B30">
  <title><p>Statistics of Cone Responses to Natural Images: Implications for
  Visual Coding</p></title>
  <aug>
    <au><snm>Ruderman</snm><fnm>DL</fnm></au>
    <au><snm>Cronin</snm><fnm>TW</fnm></au>
    <au><snm>Chiao</snm><fnm>CC</fnm></au>
  </aug>
  <source>Journal of the Optical Society of America A</source>
  <pubdate>1998</pubdate>
  <volume>15</volume>
  <fpage>2036</fpage>
  <lpage>-2045</lpage>
</bibl>

<bibl id="B31">
  <title><p>Spatiochromatic Receptive Field Properties Derived from
  Information-Theoretic Analyses of Cone Responses to Natural
  Scenes</p></title>
  <aug>
    <au><snm>Doi</snm><fnm>E.</fnm></au>
    <au><snm>Inui</snm><fnm>T.</fnm></au>
    <au><snm>Lee</snm><fnm>TW.</fnm></au>
    <au><snm>Wachtler</snm><fnm>T.</fnm></au>
    <au><snm>Sejnowski</snm><fnm>TJ.</fnm></au>
  </aug>
  <source>Neural Comput.</source>
  <pubdate>2003</pubdate>
  <volume>15</volume>
  <issue>2</issue>
  <fpage>397–</fpage>
  <lpage>417</lpage>
</bibl>

<bibl id="B32">
  <title><p>Spatio-chromatic adaptation via higher-order canonical correlation
  analysis of natural images</p></title>
  <aug>
    <au><snm>Gutmann</snm><fnm>M. U.</fnm></au>
    <au><snm>Laparra</snm><fnm>V.</fnm></au>
    <au><snm>Hyv{\"a}rinen</snm><fnm>A.</fnm></au>
    <au><snm>Malo</snm><fnm>J.</fnm></au>
  </aug>
  <source>{PloS} {ONE}</source>
  <publisher>Public Library of Science</publisher>
  <pubdate>2014</pubdate>
  <volume>9</volume>
  <issue>2</issue>
  <fpage>e86481</fpage>
</bibl>

<bibl id="B33">
  <title><p>{N}atural image statistics: A probabilistic approach to early
  computational vision</p></title>
  <aug>
    <au><snm>Hyv{\"a}rinen</snm><fnm>A.</fnm></au>
    <au><snm>Hurri</snm><fnm>J.</fnm></au>
    <au><snm>Hoyer</snm><fnm>P.O.</fnm></au>
  </aug>
  <publisher>Heidelberg, Germany: Springer-Verlag</publisher>
</bibl>

<bibl id="B34">
  <title><p>Natural signal statistics and sensory gain control</p></title>
  <aug>
    <au><snm>Schwartz</snm><fnm>O.</fnm></au>
    <au><snm>Simoncelli</snm><fnm>E.P.</fnm></au>
  </aug>
  <source>Nature Neurosci.</source>
  <pubdate>2001</pubdate>
  <volume>4</volume>
  <issue>8</issue>
  <fpage>819</fpage>
  <lpage>-825</lpage>
</bibl>

<bibl id="B35">
  <title><p>V1 non-linear properties emerge from local-to-global non-linear
  {ICA}</p></title>
  <aug>
    <au><snm>Malo</snm><fnm>J.</fnm></au>
    <au><snm>Guti{\'e}rrez</snm><fnm>J.</fnm></au>
  </aug>
  <source>Network: Computation in Neural Systems</source>
  <publisher>Taylor \& Francis</publisher>
  <pubdate>2006</pubdate>
  <volume>17</volume>
  <issue>1</issue>
  <fpage>85</fpage>
  <lpage>-102</lpage>
</bibl>

<bibl id="B36">
  <title><p>Visual aftereffects and sensory nonlinearities from a single
  statistical framework</p></title>
  <aug>
    <au><snm>Laparra</snm><fnm>V.</fnm></au>
    <au><snm>Malo</snm><fnm>J.</fnm></au>
  </aug>
  <source>Frontiers in Human Neuroscience</source>
  <pubdate>2015</pubdate>
  <volume>9</volume>
  <fpage>557</fpage>
  <url>http://journal.frontiersin.org/article/10.3389/fnhum.2015.00557</url>
</bibl>

<bibl id="B37">
  <title><p>Iterative gaussianization: from {ICA} to random
  rotations</p></title>
  <aug>
    <au><snm>Laparra</snm><fnm>V.</fnm></au>
    <au><snm>Camps Valls</snm><fnm>G.</fnm></au>
    <au><snm>Malo</snm><fnm>J.</fnm></au>
  </aug>
  <source>IEEE Trans. Neural Networks</source>
  <publisher>IEEE</publisher>
  <pubdate>2011</pubdate>
  <volume>22</volume>
  <issue>4</issue>
  <fpage>537</fpage>
  <lpage>-549</lpage>
</bibl>

<bibl id="B38">
  <title><p>Information Theory in Density Destructors</p></title>
  <aug>
    <au><snm>Johnson</snm><fnm>J.E.</fnm></au>
    <au><snm>Laparra</snm><fnm>V.</fnm></au>
    <au><snm>Santos</snm><fnm>R.</fnm></au>
    <au><snm>Camps</snm><fnm>G.</fnm></au>
    <au><snm>Malo</snm><fnm>J.</fnm></au>
  </aug>
  <source>7th Int. Conf. Mach. Learn., {ICML} 2019, Workshop on Invertible
  Normalization Flows</source>
  <pubdate>2019</pubdate>
</bibl>

<bibl id="B39">
  <title><p>Derivatives and inverse of cascaded linear+nonlinear neural
  models</p></title>
  <aug>
    <au><snm>Martinez Garcia</snm><fnm>M.</fnm></au>
    <au><snm>Cyriac</snm><fnm>P.</fnm></au>
    <au><snm>Batard</snm><fnm>T.</fnm></au>
    <au><snm>Bertalm{\'\i}o</snm><fnm>M.</fnm></au>
    <au><snm>Malo</snm><fnm>J.</fnm></au>
  </aug>
  <source>PLOS ONE</source>
  <publisher>Public Library of Science</publisher>
  <pubdate>2018</pubdate>
  <volume>13</volume>
  <issue>10</issue>
  <fpage>1</fpage>
  <lpage>49</lpage>
  <url>https://doi.org/10.1371/journal.pone.0201326</url>
</bibl>

<bibl id="B40">
  <title><p>In Paraise of Artifice Reloaded: Caution with Natural Image
  Databases in Modeling Vision</p></title>
  <aug>
    <au><snm>Martinez</snm><fnm>M.</fnm></au>
    <au><snm>Bertalm\'io</snm><fnm>M.</fnm></au>
    <au><snm>Malo</snm><fnm>J.</fnm></au>
  </aug>
  <source>Frontiers in Neuroscience</source>
  <pubdate>2019</pubdate>
  <volume>13</volume>
  <issue>8</issue>
  <url>http://frontiersin.org/articles/10.3389/fnins.2019.00008</url>
</bibl>

<bibl id="B41">
  <title><p>Nonlinear extraction of independent components of natural images
  using radial gaussianization</p></title>
  <aug>
    <au><snm>Lyu</snm><fnm>S.</fnm></au>
    <au><snm>Simoncelli</snm><fnm>E. P</fnm></au>
  </aug>
  <source>Neural Comp.</source>
  <publisher>MIT Press</publisher>
  <pubdate>2009</pubdate>
  <volume>21</volume>
  <issue>6</issue>
  <fpage>1485</fpage>
  <lpage>1519</lpage>
</bibl>

<bibl id="B42">
  <title><p>Visual Information Flow in Wilson-Cowan Networks</p></title>
  <aug>
    <au><snm>Gomez Villa</snm><fnm>A.</fnm></au>
    <au><snm>Bertalmio</snm><fnm>M.</fnm></au>
    <au><snm>Malo</snm><fnm>J.</fnm></au>
  </aug>
  <source>\url{https://doi.org/10.1152/jn.00487.2019}</source>
  <pubdate>2020</pubdate>
</bibl>

<bibl id="B43">
  <title><p>Psychophysically tuned divisive normalization approximately
  factorizes the {PDF} of natural images</p></title>
  <aug>
    <au><snm>Malo</snm><fnm>J.</fnm></au>
    <au><snm>Laparra</snm><fnm>V.</fnm></au>
  </aug>
  <source>Neural computation</source>
  <pubdate>2010</pubdate>
  <volume>22</volume>
  <issue>12</issue>
  <fpage>3179</fpage>
  <lpage>-3206</lpage>
</bibl>

<bibl id="B44">
  <title><p>The economy of brain network organization</p></title>
  <aug>
    <au><snm>Bullmore</snm><fnm>E.</fnm></au>
    <au><snm>Sporns</snm><fnm>O.</fnm></au>
  </aug>
  <source>Nat.Rev.Neurosci.</source>
  <pubdate>2012</pubdate>
  <volume>13</volume>
  <fpage>336</fpage>
  <lpage>349</lpage>
</bibl>

<bibl id="B45">
  <title><p>The spectral sensitivities of the middle- and
  long-wavelength-sensitive cones derived from measurements in observers of
  known genotype</p></title>
  <aug>
    <au><snm>Stockman</snm><fnm>A.</fnm></au>
    <au><snm>Sharpe</snm><fnm>L.T.</fnm></au>
  </aug>
  <source>Vision Research</source>
  <pubdate>2000</pubdate>
  <volume>40</volume>
  <issue>13</issue>
  <fpage>1711</fpage>
  <lpage>1737</lpage>
</bibl>

<bibl id="B46">
  <title><p>An opponent-process theory of color vision</p></title>
  <aug>
    <au><snm>Hurvich</snm><fnm>L.M.</fnm></au>
    <au><snm>Jameson</snm><fnm>D.</fnm></au>
  </aug>
  <source>Psychol. Rev.</source>
  <pubdate>1957</pubdate>
  <volume>64</volume>
  <issue>6</issue>
  <fpage>384</fpage>
  <lpage>-404</lpage>
</bibl>

<bibl id="B47">
  <title><p>Colour representation spaces at different physiological levels: a
  comparative analysis</p></title>
  <aug>
    <au><snm>Capilla</snm><fnm>P</fnm></au>
    <au><snm>Malo</snm><fnm>J.</fnm></au>
    <au><snm>Luque</snm><fnm>MJ</fnm></au>
    <au><snm>Artigas</snm><fnm>J.M.</fnm></au>
  </aug>
  <source>Journal of optics</source>
  <publisher>IOP Publishing</publisher>
  <pubdate>1998</pubdate>
  <volume>29</volume>
  <issue>5</issue>
  <fpage>324</fpage>
</bibl>

<bibl id="B48">
  <title><p>OSA Handbook of Optics (3rd. Ed.)</p></title>
  <aug>
    <au><snm>Stockman</snm><fnm>A.</fnm></au>
    <au><snm>Brainard</snm><fnm>D.H.</fnm></au>
  </aug>
  <publisher>NY: McGraw-Hill</publisher>
  <editor>Bass, M.</editor>
  <section><title><p>Color vision mechanisms</p></title></section>
  <pubdate>2010</pubdate>
  <fpage>147</fpage>
  <lpage>-152</lpage>
</bibl>

<bibl id="B49">
  <title><p>Color discrimination and adaptation</p></title>
  <aug>
    <au><snm>Krauskopf</snm><fnm>J.</fnm></au>
    <au><snm>Gegenfurtner</snm><fnm>K.</fnm></au>
  </aug>
  <source>Vision Res.</source>
  <pubdate>1992</pubdate>
  <volume>32</volume>
  <issue>11</issue>
  <fpage>2165</fpage>
  <lpage>2175</lpage>
</bibl>

<bibl id="B50">
  <title><p>Evaluation of color-discrimination ellipsoids in two-color
  spaces</p></title>
  <aug>
    <au><snm>Romero</snm><fnm>J.</fnm></au>
    <au><snm>Garc\'{i}a</snm><fnm>J.A.</fnm></au>
    <au><snm>Barco</snm><fnm>L.</fnm></au>
    <au><snm>Hita</snm><fnm>E.</fnm></au>
  </aug>
  <source>J. Opt. Soc. Am. A</source>
  <publisher>OSA</publisher>
  <pubdate>1993</pubdate>
  <volume>10</volume>
  <issue>5</issue>
  <fpage>827</fpage>
  <lpage>-837</lpage>
  <url>http://josaa.osa.org/abstract.cfm?URI=josaa-10-5-827</url>
</bibl>

<bibl id="B51">
  <title><p>Colorimetry - {P}art 4: {CIE} 1976 {L}*a*b* colour
  space</p></title>
  <aug>
    <au><cnm>CIE Commission</cnm></au>
  </aug>
  <pubdate>1976</pubdate>
  <issue>ISO/CIE 11664-4:2019</issue>
</bibl>

<bibl id="B52">
  <title><p>The structure of the {CIE} 1997 colour appearance model
  ({CIECAM}97s)</p></title>
  <aug>
    <au><snm>Luo</snm><fnm>MR.</fnm></au>
    <au><snm>Hunt</snm><fnm>RWG.</fnm></au>
  </aug>
  <source>Color Res. Appl.</source>
  <pubdate>1998</pubdate>
  <volume>22</volume>
  <fpage>138</fpage>
  <lpage>-146</lpage>
</bibl>

<bibl id="B53">
  <title><p>Spatial structure and symmetry of simple-cell receptive fields in
  macaque primary visual cortex</p></title>
  <aug>
    <au><snm>Ringach</snm><fnm>D.L.</fnm></au>
  </aug>
  <source>J. Neurophysiol.</source>
  <pubdate>2002</pubdate>
  <volume>88</volume>
  <issue>1</issue>
  <fpage>455‐463</fpage>
</bibl>

<bibl id="B54">
  <title><p>Color in the cortex: single- and double-opponent cells</p></title>
  <aug>
    <au><snm>Shapley</snm><fnm>R.</fnm></au>
    <au><snm>Hawken</snm><fnm>M.</fnm></au>
  </aug>
  <source>Vis.Res.</source>
  <pubdate>2011</pubdate>
  <volume>51</volume>
  <issue>7</issue>
  <fpage>701</fpage>
  <lpage>-717</lpage>
</bibl>

<bibl id="B55">
  <title><p>Application of {F}ourier Analysis to the Visibility of
  Gratings</p></title>
  <aug>
    <au><snm>Campbell</snm><fnm>F.W.</fnm></au>
    <au><snm>Robson</snm><fnm>J.G.</fnm></au>
  </aug>
  <source>Journal of Physiology</source>
  <pubdate>1968</pubdate>
  <volume>197</volume>
  <fpage>551</fpage>
  <lpage>-566</lpage>
</bibl>

<bibl id="B56">
  <title><p>The {CSF} of Human Colour Vision to Red-Green and Yellow-Blue
  Chromatic Gratings</p></title>
  <aug>
    <au><snm>Mullen</snm><fnm>K. T.</fnm></au>
  </aug>
  <source>J. Physiol.</source>
  <pubdate>1985</pubdate>
  <volume>359</volume>
  <fpage>381</fpage>
  <lpage>-400</lpage>
</bibl>

<bibl id="B57">
  <title><p>Video quality measures based on the standard spatial
  observer</p></title>
  <aug>
    <au><snm>Watson</snm><fnm>A. B</fnm></au>
    <au><snm>Malo</snm><fnm>J.</fnm></au>
  </aug>
  <source>Image Processing. 2002. Proceedings. 2002 International Conference
  on</source>
  <pubdate>2002</pubdate>
  <volume>3</volume>
  <fpage>III</fpage>
  <lpage>-41</lpage>
</bibl>

<bibl id="B58">
  <title><p>Characterization of the human visual system threshold performance
  by a weighting function in the Gabor domain</p></title>
  <aug>
    <au><snm>Malo</snm><fnm>J.</fnm></au>
    <au><snm>Pons</snm><fnm>AM.</fnm></au>
    <au><snm>Felipe</snm><fnm>A.</fnm></au>
    <au><snm>Artigas</snm><fnm>JM.</fnm></au>
  </aug>
  <source>J. Mod. Opt.</source>
  <pubdate>1997</pubdate>
  <volume>44</volume>
  <issue>1</issue>
  <fpage>127</fpage>
  <lpage>-148</lpage>
</bibl>

<bibl id="B59">
  <title><p>Vis. Sci. and Eng.: Models and Appl.</p></title>
  <aug>
    <au><snm>Martinez Uriegas</snm><fnm>E.</fnm></au>
  </aug>
  <publisher>NY: Marcel Dekker Inc.</publisher>
  <editor>Kelly, D.H.</editor>
  <section><title><p>Chromatic-achromatic multiplexing in human color
  vision</p></title></section>
  <pubdate>1994</pubdate>
  <fpage>117</fpage>
  <lpage>187</lpage>
</bibl>

<bibl id="B60">
  <title><p>Spatiotemporal receptive field organization in the {LGN} of cats
  and kittens</p></title>
  <aug>
    <au><snm>Cai</snm><fnm>D.</fnm></au>
    <au><snm>DeAngelis</snm><fnm>G.C.</fnm></au>
    <au><snm>Freeman</snm><fnm>R.D.</fnm></au>
  </aug>
  <source>J. Neurophysiol.</source>
  <pubdate>1997</pubdate>
  <volume>78</volume>
  <issue>2</issue>
  <fpage>1045‐1061</fpage>
</bibl>

<bibl id="B61">
  <title><p>Nonlinear image representation for efficient perceptual
  coding</p></title>
  <aug>
    <au><snm>Malo</snm><fnm>J.</fnm></au>
    <au><snm>Epifanio</snm><fnm>I.</fnm></au>
    <au><snm>Navarro</snm><fnm>R</fnm></au>
    <au><snm>Simoncelli</snm><fnm>EP</fnm></au>
  </aug>
  <source>IEEE Transactions on Image Processing</source>
  <publisher>IEEE</publisher>
  <pubdate>2006</pubdate>
  <volume>15</volume>
  <issue>1</issue>
  <fpage>68</fpage>
  <lpage>-80</lpage>
</bibl>

<bibl id="B62">
  <title><p>A mathematical theory of the functional dynamics of cortical and
  thalamic nervous tissue</p></title>
  <aug>
    <au><snm>Wilson</snm><fnm>HR</fnm></au>
    <au><snm>Cowan</snm><fnm>JD</fnm></au>
  </aug>
  <source>Kybernetik</source>
  <publisher>Springer</publisher>
  <pubdate>1973</pubdate>
  <volume>13</volume>
  <issue>2</issue>
  <fpage>55</fpage>
  <lpage>-80</lpage>
</bibl>

<bibl id="B63">
  <title><p>Divisive Normalization from {W}ilson-{C}owan Dynamics</p></title>
  <aug>
    <au><snm>Malo</snm><fnm>J</fnm></au>
    <au><snm>Esteve Taboada</snm><fnm>JJ</fnm></au>
    <au><snm>Bertalm{\'\i}o</snm><fnm>M</fnm></au>
  </aug>
  <source>ArXiv: Quant. Biol. https://arxiv.org/abs/1906.08246</source>
  <pubdate>2019</pubdate>
</bibl>

<bibl id="B64">
  <title><p>Color Image Database for Evaluation of Image Quality
  Metrics</p></title>
  <aug>
    <au><snm>Ponomarenko</snm><fnm>N.</fnm></au>
    <au><snm>Carli</snm><fnm>M.</fnm></au>
    <au><snm>Lukin</snm><fnm>V.</fnm></au>
    <au><snm>Egiazarian</snm><fnm>K.</fnm></au>
    <au><snm>Astola</snm><fnm>J.</fnm></au>
    <au><snm>Battisti</snm><fnm>F.</fnm></au>
  </aug>
  <source>Proc. Int. Workshop on Multimedia Signal Processing</source>
  <pubdate>2008</pubdate>
  <fpage>403</fpage>
  <lpage>408</lpage>
</bibl>

<bibl id="B65">
  <title><p>COLORLAB: The Matlab tootbox for colorimetry and color
  vision</p></title>
  <aug>
    <au><snm>Malo</snm><fnm>J.</fnm></au>
    <au><snm>Luque</snm><fnm>M.J.</fnm></au>
  </aug>
  <source>Internet site:
  http://isp.uv.es/code/visioncolor/colorlab.html</source>
  <pubdate>2002</pubdate>
</bibl>

<bibl id="B66">
  <title><p>VISTALAB: the Matlab toolbox for Spatio-Temporal Vision</p></title>
  <aug>
    <au><snm>Malo</snm><fnm>J.</fnm></au>
    <au><snm>Gutierrez</snm><fnm>J.</fnm></au>
  </aug>
  <source>Internet site:
  http://isp.uv.es/code/visioncolor/vistalab.html</source>
  <pubdate>1997</pubdate>
</bibl>

<bibl id="B67">
  <title><p>Image quality assessment: from error visibility to structural
  similarity</p></title>
  <aug>
    <au><snm>Wang</snm><fnm>Z.</fnm></au>
    <au><snm>Bovik</snm><fnm>A. C.</fnm></au>
    <au><snm>Sheikh</snm><fnm>H. R.</fnm></au>
    <au><snm>Simoncelli</snm><fnm>E. P.</fnm></au>
  </aug>
  <source>IEEE Trans. Im. Proc.</source>
  <publisher>IEEE</publisher>
  <pubdate>2004</pubdate>
  <volume>13</volume>
  <issue>4</issue>
  <fpage>600</fpage>
  <lpage>-612</lpage>
</bibl>

<bibl id="B68">
  <title><p>Divisive normalization image quality metric revisited</p></title>
  <aug>
    <au><snm>Laparra</snm><fnm>V.</fnm></au>
    <au><snm>Mu{\~n}oz Mar{\'\i}</snm><fnm>J.</fnm></au>
    <au><snm>Malo</snm><fnm>J.</fnm></au>
  </aug>
  <source>JOSA A</source>
  <publisher>Optical Society of America</publisher>
  <pubdate>2010</pubdate>
  <volume>27</volume>
  <issue>4</issue>
  <fpage>852</fpage>
  <lpage>-864</lpage>
</bibl>

<bibl id="B69">
  <title><p>Image denoising with kernels based on natural image
  relations</p></title>
  <aug>
    <au><snm>Laparra</snm><fnm>V.</fnm></au>
    <au><snm>Guti{\'e}rrez</snm><fnm>J.</fnm></au>
    <au><snm>Camps Valls</snm><fnm>G.</fnm></au>
    <au><snm>Malo</snm><fnm>J.</fnm></au>
  </aug>
  <source>The Journal of Machine Learning Research</source>
  <publisher>JMLR. org</publisher>
  <pubdate>2010</pubdate>
  <volume>11</volume>
  <fpage>873</fpage>
  <lpage>-903</lpage>
</bibl>

<bibl id="B70">
  <title><p>PerceptNet: A Human Visual System Inspired Neural Net for
  Estimating Perceptual Distance</p></title>
  <aug>
    <au><snm>Hepburn</snm><fnm>A.</fnm></au>
    <au><snm>Laparra</snm><fnm>V.</fnm></au>
    <au><snm>Malo</snm><fnm>J.</fnm></au>
    <au><snm>McConville</snm><fnm>R.</fnm></au>
    <au><snm>Santos</snm><fnm>R.</fnm></au>
  </aug>
  <source>Proc. IEEE ICIP (2020)
  \texttt{https://arxiv.org/abs/1910.12548}</source>
</bibl>

<bibl id="B71">
  <title><p>Time-lapse ratios of cone excitations in natural scenes</p></title>
  <aug>
    <au><snm>Foster</snm><fnm>DH</fnm></au>
    <au><snm>Amano</snm><fnm>K</fnm></au>
    <au><snm>Nascimento</snm><fnm>SM</fnm></au>
  </aug>
  <source>Vision research</source>
  <publisher>Elsevier</publisher>
  <pubdate>2016</pubdate>
  <volume>120</volume>
  <fpage>45</fpage>
  <lpage>-60</lpage>
</bibl>

<bibl id="B72">
  <title><p>Spatial distributions of local illumination color in natural
  scenes</p></title>
  <aug>
    <au><snm>Nascimento</snm><fnm>SM</fnm></au>
    <au><snm>Amano</snm><fnm>K</fnm></au>
    <au><snm>Foster</snm><fnm>DH</fnm></au>
  </aug>
  <source>Vision Research</source>
  <publisher>Elsevier</publisher>
  <pubdate>2016</pubdate>
  <volume>120</volume>
  <fpage>39</fpage>
  <lpage>-44</lpage>
</bibl>

<bibl id="B73">
  <title><p>Color Constancy Algorithms: Psychophysical Evaluation on a New
  Dataset</p></title>
  <aug>
    <au><snm>Vazquez Corral</snm><fnm>J.</fnm></au>
    <au><snm>P{\'a}rraga</snm><fnm>C.</fnm></au>
    <au><snm>Baldrich</snm><fnm>R.</fnm></au>
    <au><snm>Vanrell</snm><fnm>M.</fnm></au>
  </aug>
  <source>Journal of Imaging Science and Technology</source>
  <pubdate>2009</pubdate>
  <volume>53</volume>
  <issue>3</issue>
  <fpage>31105</fpage>
  <lpage>1-31105-9</lpage>
</bibl>

<bibl id="B74">
  <title><p>Elements of Information Theory, 2nd Edition</p></title>
  <aug>
    <au><snm>Cover</snm><fnm>T. M.</fnm></au>
    <au><snm>Thomas</snm><fnm>J. A.</fnm></au>
  </aug>
  <source>Hardcover</source>
  <publisher>Hoboken, NJ, USA: Wiley-Interscience</publisher>
  <edition>2</edition>
  <pubdate>2006</pubdate>
</bibl>

<bibl id="B75">
  <title><p>{Sample estimate of the entropy of a random vector}</p></title>
  <aug>
    <au><snm>Kozachenko</snm><fnm>L.F.</fnm></au>
    <au><snm>Leonenko</snm><fnm>N.N.</fnm></au>
  </aug>
  <source>Probl. Inf. Trans.</source>
  <pubdate>1987</pubdate>
  <volume>23</volume>
  <fpage>95</fpage>
  <lpage>101</lpage>
</bibl>

<bibl id="B76">
  <title><p>Estimating Information from Image Colors: An Application to Digital
  Cameras and Natural Scenes</p></title>
  <aug>
    <au><snm>Marin Franch</snm><fnm>I</fnm></au>
    <au><snm>Foster</snm><fnm>DH</fnm></au>
  </aug>
  <source>IEEE Trans. Pattern Anal. Mach. Intell.</source>
  <publisher>Washington, DC, USA: IEEE Computer Society</publisher>
  <pubdate>2013</pubdate>
  <volume>35</volume>
  <issue>1</issue>
  <fpage>78</fpage>
  <lpage>-91</lpage>
</bibl>

<bibl id="B77">
  <title><p>The Verriest Lecture: Color vision in an uncertain
  world</p></title>
  <aug>
    <au><snm>Foster</snm><fnm>DH</fnm></au>
  </aug>
  <source>JOSA A</source>
  <pubdate>2018</pubdate>
  <volume>35</volume>
  <issue>4</issue>
  <fpage>192</fpage>
  <lpage>-201</lpage>
</bibl>

<bibl id="B78">
  <title><p>Proc. ICML Workshop on Invertible Neural Nets and Normalizing
  Flows</p></title>
  <aug>
    <au><snm>Huang</snm><fnm>C.W.</fnm></au>
    <au><snm>Kruger</snm><fnm>D.</fnm></au>
  </aug>
  <source>Int. Conf. Mach. Learn.
  \texttt{https://invertibleworkshop.github.io/INNF\_2019/accepted\_papers/}</source>
  <pubdate>2019</pubdate>
</bibl>

<bibl id="B79">
  <title><p>Proc. ICML Workshop on Invertible Neural Nets and Normalizing
  Flows</p></title>
  <aug>
    <au><snm>Huang</snm><fnm>C.W.</fnm></au>
    <au><snm>Kruger</snm><fnm>D.</fnm></au>
  </aug>
  <source>Int. Conf. Mach. Learn.
  \texttt{https://invertibleworkshop.github.io}</source>
  <pubdate>2020</pubdate>
</bibl>

<bibl id="B80">
  <title><p>Deep Density Destructors</p></title>
  <aug>
    <au><snm>Inouye</snm><fnm>D</fnm></au>
    <au><snm>Ravikumar</snm><fnm>P</fnm></au>
  </aug>
  <source>35th ICML</source>
  <series><title><p>Proc. Mach. Learn. Res.</p></title></series>
  <pubdate>2018</pubdate>
  <volume>80</volume>
  <fpage>2167</fpage>
  <lpage>-2175</lpage>
</bibl>

<bibl id="B81">
  <title><p>Information theoretical analysis of multivariate
  correlation</p></title>
  <aug>
    <au><snm>Watanabe</snm><fnm>S</fnm></au>
  </aug>
  <source>IBM Journal of research and development</source>
  <publisher>IBM</publisher>
  <pubdate>1960</pubdate>
  <volume>4</volume>
  <issue>1</issue>
  <fpage>66</fpage>
  <lpage>-82</lpage>
</bibl>

<bibl id="B82">
  <title><p>The Multi-information function as a tool for measuring stochastic
  dependence</p></title>
  <aug>
    <au><snm>Studeny</snm><fnm>M.</fnm></au>
    <au><snm>Vejnarova</snm><fnm>J.</fnm></au>
  </aug>
  <source>Learning in graphical models</source>
  <publisher>Kluwer</publisher>
  <editor>M. I. Jordan</editor>
  <pubdate>1998</pubdate>
  <fpage>261</fpage>
  <lpage>298</lpage>
</bibl>

<bibl id="B83">
  <title><p>Estimating mutual information</p></title>
  <aug>
    <au><snm>Kraskov</snm><fnm>A</fnm></au>
    <au><snm>St\"ogbauer</snm><fnm>H</fnm></au>
    <au><snm>Grassberger</snm><fnm>P</fnm></au>
  </aug>
  <source>Phys. Rev. E</source>
  <publisher>American Physical Society</publisher>
  <pubdate>2004</pubdate>
  <volume>69</volume>
  <fpage>066138</fpage>
</bibl>

<bibl id="B84">
  <title><p>Information theoretical estimators toolbox</p></title>
  <aug>
    <au><snm>Szabó</snm><fnm>Z.</fnm></au>
  </aug>
  <source>J. Mach. Learn. Res.</source>
  <pubdate>2014</pubdate>
  <volume>15</volume>
  <fpage>283–287</fpage>
</bibl>

<bibl id="B85">
  <title><p>Puting the visual system noise back in the picture</p></title>
  <aug>
    <au><snm>Ahumada</snm><fnm>A</fnm></au>
  </aug>
  <source>J. Opt. Soc. Am. A</source>
  <pubdate>1987</pubdate>
  <volume>4</volume>
  <issue>12</issue>
  <fpage>2372</fpage>
  <lpage>-2378</lpage>
</bibl>

<bibl id="B86">
  <title><p>Visual signal detection. IV. Observer inconsistency</p></title>
  <aug>
    <au><snm>Burgess</snm><fnm>A. E.</fnm></au>
    <au><snm>Colborne</snm><fnm>B.</fnm></au>
  </aug>
  <source>J. Opt. Soc. Am. A</source>
  <publisher>OSA</publisher>
  <pubdate>1988</pubdate>
  <volume>5</volume>
  <issue>4</issue>
  <fpage>617</fpage>
  <lpage>-627</lpage>
</bibl>

<bibl id="B87">
  <title><p>Fixed or variable noise in contrast discrimination? The jury's
  still out...</p></title>
  <aug>
    <au><snm>Georgeson</snm><fnm>{Mark A.}</fnm></au>
    <au><snm>Meese</snm><fnm>{Timothy S.}</fnm></au>
  </aug>
  <source>Vision Research</source>
  <publisher>Elsevier</publisher>
  <pubdate>2006</pubdate>
  <volume>46</volume>
  <issue>25</issue>
  <fpage>4294</fpage>
  <lpage>-4303</lpage>
</bibl>

<bibl id="B88">
  <title><p>How inherently noisy is human sensory processing?</p></title>
  <aug>
    <au><snm>Neri</snm><fnm>P.</fnm></au>
  </aug>
  <source>Psychon. Bull. Rev.</source>
  <pubdate>2010</pubdate>
  <volume>17</volume>
  <fpage>802–</fpage>
  <lpage>808</lpage>
</bibl>

<bibl id="B89">
  <title><p>Partitioning neuronal variability</p></title>
  <aug>
    <au><snm>Goris</snm><fnm>LTR.</fnm></au>
    <au><snm>Movshon</snm><fnm>JA.</fnm></au>
    <au><snm>Simoncelli</snm><fnm>EP.</fnm></au>
  </aug>
  <source>Nat. Neurosci.</source>
  <pubdate>2014</pubdate>
  <volume>17</volume>
  <issue>6</issue>
  <fpage>858–</fpage>
  <lpage>865</lpage>
</bibl>

<bibl id="B90">
  <title><p>Information-limiting correlations</p></title>
  <aug>
    <au><snm>Moreno Bote</snm><fnm>R.</fnm></au>
    <au><snm>Beck</snm><fnm>J.</fnm></au>
    <au><snm>Kanitscheider</snm><fnm>I.</fnm></au>
    <au><snm>Pitkow</snm><fnm>X.</fnm></au>
    <au><snm>Latham</snm><fnm>P.</fnm></au>
    <au><snm>Pouget</snm><fnm>A.</fnm></au>
  </aug>
  <source>Nat. Neurosci.</source>
  <pubdate>2014</pubdate>
  <volume>17</volume>
  <issue>10</issue>
  <fpage>1410–</fpage>
  <lpage>1417</lpage>
</bibl>

<bibl id="B91">
  <title><p>Origin of information-limiting noise correlations</p></title>
  <aug>
    <au><snm>Kanitscheider</snm><fnm>I.</fnm></au>
    <au><snm>Coen Cagli</snm><fnm>R.</fnm></au>
    <au><snm>Pouget</snm><fnm>A.</fnm></au>
  </aug>
  <source>Proc. Nat. Acad. Sci.</source>
  <pubdate>2015</pubdate>
  <volume>112</volume>
  <issue>50</issue>
  <fpage>E6973</fpage>
  <lpage>-E6982</lpage>
</bibl>

<bibl id="B92">
  <title><p>Factorial coding of natural images: how effective are linear models
  in removing higher-order dependencies?</p></title>
  <aug>
    <au><snm>Bethge</snm><fnm>M</fnm></au>
  </aug>
  <source>JOSA A</source>
  <publisher>Optical Society of America</publisher>
  <pubdate>2006</pubdate>
  <volume>23</volume>
  <issue>6</issue>
  <fpage>1253</fpage>
  <lpage>-1268</lpage>
</bibl>

<bibl id="B93">
  <title><p>Information limits on neural identification of colored surfaces in
  natural scenes</p></title>
  <aug>
    <au><snm>Foster</snm><fnm>D.H.</fnm></au>
    <au><snm>Nascimento</snm><fnm>S.M.C.</fnm></au>
    <au><snm>Amano</snm><fnm>K.</fnm></au>
  </aug>
  <source>Visual Neuroscience</source>
  <publisher>Cambridge University Press</publisher>
  <pubdate>2004</pubdate>
  <volume>21</volume>
  <issue>3</issue>
  <fpage>331</fpage>
  <lpage>-336</lpage>
</bibl>

<bibl id="B94">
  <title><p>Approaching ideal observer efficiency in using color to retrieve
  information from natural scenes</p></title>
  <aug>
    <au><snm>Foster</snm><fnm>DH</fnm></au>
    <au><snm>Mar\'{i}n Franch</snm><fnm>I</fnm></au>
    <au><snm>Amano</snm><fnm>K</fnm></au>
    <au><snm>Nascimento</snm><fnm>SMC</fnm></au>
  </aug>
  <source>J. Opt. Soc. Am. A</source>
  <pubdate>2009</pubdate>
  <volume>26</volume>
  <issue>11</issue>
  <fpage>14</fpage>
  <lpage>-24</lpage>
</bibl>

<bibl id="B95">
  <title><p>{Number of perceptually distinct surface colors in natural
  scenes}</p></title>
  <aug>
    <au><snm>Marin Franch</snm><fnm>I.</fnm></au>
    <au><snm>Foster</snm><fnm>D.H.</fnm></au>
  </aug>
  <source>Journal of Vision</source>
  <pubdate>2010</pubdate>
  <volume>10</volume>
  <issue>9</issue>
  <fpage>9</fpage>
  <lpage>9</lpage>
</bibl>

<bibl id="B96">
  <title><p>Coding efficiency of CIE color spaces</p></title>
  <aug>
    <au><snm>Foster</snm><fnm>D.H.</fnm></au>
    <au><snm>Marin Franch</snm><fnm>I.</fnm></au>
    <au><snm>Nascimento</snm><fnm>S.M.C.</fnm></au>
  </aug>
  <source>Proc. 16th Color Imag. Conf.</source>
  <pubdate>2008</pubdate>
  <fpage>285</fpage>
  <lpage>-288</lpage>
</bibl>

<bibl id="B97">
  <title><p>Binless strategies for estimation of information from neural
  data</p></title>
  <aug>
    <au><snm>Victor</snm><fnm>JD.</fnm></au>
  </aug>
  <source>Phys. Rev. E</source>
  <pubdate>2002</pubdate>
  <volume>66</volume>
  <issue>5</issue>
  <fpage>051903</fpage>
</bibl>

<bibl id="B98">
  <title><p>An information fidelity criterion for image quality assessment
  using natural scene statistics</p></title>
  <aug>
    <au><snm>{Sheikh}</snm><fnm>H. R.</fnm></au>
    <au><snm>{Bovik}</snm><fnm>A. C.</fnm></au>
    <au><snm>{de Veciana}</snm><fnm>G.</fnm></au>
  </aug>
  <source>IEEE Trans. Im. Proc.</source>
  <pubdate>2005</pubdate>
  <volume>14</volume>
  <issue>12</issue>
  <fpage>2117</fpage>
  <lpage>2128</lpage>
</bibl>

<bibl id="B99">
  <title><p>Image information and visual quality</p></title>
  <aug>
    <au><snm>{Sheikh}</snm><fnm>H. R.</fnm></au>
    <au><snm>{Bovik}</snm><fnm>A. C.</fnm></au>
  </aug>
  <source>IEEE Trans. Im. Proc.</source>
  <pubdate>2006</pubdate>
  <volume>15</volume>
  <issue>2</issue>
  <fpage>430</fpage>
  <lpage>444</lpage>
</bibl>

<bibl id="B100">
  <title><p>Perceptual feedback in multigrid motion estimation using an
  improved DCT quantization</p></title>
  <aug>
    <au><snm>Malo</snm><fnm>J.</fnm></au>
    <au><snm>Guti{\'e}rrez</snm><fnm>J.</fnm></au>
    <au><snm>Epifanio</snm><fnm>I.</fnm></au>
    <au><snm>Ferri</snm><fnm>F. J</fnm></au>
    <au><snm>Artigas</snm><fnm>JM</fnm></au>
  </aug>
  <source>IEEE Trans. Im. Proc.</source>
  <publisher>IEEE</publisher>
  <pubdate>2001</pubdate>
  <volume>10</volume>
  <issue>10</issue>
  <fpage>1411</fpage>
  <lpage>-1427</lpage>
</bibl>

<bibl id="B101">
  <title><p>End-to-end Optimized Image Compression</p></title>
  <aug>
    <au><snm>Ball{\'{e}}</snm><fnm>J</fnm></au>
    <au><snm>Laparra</snm><fnm>V</fnm></au>
    <au><snm>Simoncelli</snm><fnm>EP</fnm></au>
  </aug>
  <source>5th Int. Conf. Learn. Repres., {ICLR} 2017</source>
  <pubdate>2017</pubdate>
</bibl>

<bibl id="B102">
  <title><p>Deep Learning</p></title>
  <aug>
    <au><snm>Goodfellow</snm><fnm>I.</fnm></au>
    <au><snm>Bengio</snm><fnm>Y.</fnm></au>
    <au><snm>Courville</snm><fnm>A.</fnm></au>
  </aug>
  <publisher>Cambridge, MA: MIT Press</publisher>
  <pubdate>2016</pubdate>
  <note>\url{http://www.deeplearningbook.org}</note>
</bibl>

<bibl id="B103">
  <title><p>Statistical models for images: Compression, restoration and
  synthesis</p></title>
  <aug>
    <au><snm>Simoncelli</snm><fnm>E.</fnm></au>
  </aug>
  <source>{IEEE} Asilomar Conf. Sign. Syst. Comp.</source>
  <publisher>Asilomar, CA, USA</publisher>
  <editor>M.P. Farques</editor>
  <pubdate>1998</pubdate>
  <volume>1</volume>
  <fpage>673</fpage>
  <lpage>-678</lpage>
</bibl>

<bibl id="B104">
  <title><p>Physics-based approaches to modeling surface color
  perception</p></title>
  <aug>
    <au><snm>Maloney</snm><fnm>L</fnm></au>
  </aug>
  <source>Color vision</source>
  <publisher>Cambridge, UK: Cambridge University Press</publisher>
  <editor>K.R. Gegenfurtner and L.T. Sharpe</editor>
  <pubdate>1999</pubdate>
  <fpage>387</fpage>
  <lpage>-422</lpage>
</bibl>

<bibl id="B105">
  <title><p>The Role of Spatial Information in Disentangling the
  Irradiance--Reflectance--Transmittance Ambiguity</p></title>
  <aug>
    <au><snm>Jimenez</snm><fnm>S</fnm></au>
    <au><snm>Malo</snm><fnm>J.</fnm></au>
  </aug>
  <source>Geoscience and Remote Sensing, IEEE Transactions on</source>
  <publisher>IEEE</publisher>
  <pubdate>2014</pubdate>
  <volume>52</volume>
  <issue>8</issue>
  <fpage>4881</fpage>
  <lpage>-4894</lpage>
</bibl>

<bibl id="B106">
  <title><p>Natural Image Statistics and Neural Representation</p></title>
  <aug>
    <au><snm>Simoncelli</snm><fnm>EP</fnm></au>
    <au><snm>Olshausen</snm><fnm>BA</fnm></au>
  </aug>
  <source>Annual Review of Neuroscience</source>
  <pubdate>2001</pubdate>
  <volume>24</volume>
  <issue>1</issue>
  <fpage>1193</fpage>
  <lpage>1216</lpage>
</bibl>

<bibl id="B107">
  <title><p>{ColorLab: A Matlab Toolbox for Color Science and Calibrated Color
  Image Processing}</p></title>
  <aug>
    <au><snm>Malo</snm><fnm>J.</fnm></au>
    <au><snm>Luque</snm><fnm>M.J.</fnm></au>
  </aug>
  <source>\url{http://isp.uv.es/code/visioncolor/colorlab.html}</source>
  <pubdate>2002</pubdate>
</bibl>

<bibl id="B108">
  <title><p>Dependence, correlation and Gaussianity in independent component
  analysis</p></title>
  <aug>
    <au><snm>Cardoso</snm><fnm>J.</fnm></au>
  </aug>
  <source>J. Mach. Learn. Res.</source>
  <publisher>JMLR.org</publisher>
  <pubdate>2003</pubdate>
  <volume>4</volume>
  <fpage>1177</fpage>
  <lpage>-1203</lpage>
</bibl>

<bibl id="B109">
  <title><p>Statistics of natural images and models</p></title>
  <aug>
    <au><snm>Huang</snm><fnm>J.</fnm></au>
    <au><snm>Mumford</snm><fnm>D.</fnm></au>
  </aug>
  <source>IEEE CVPR</source>
  <pubdate>1999</pubdate>
  <volume>1</volume>
  <fpage>541</fpage>
  <lpage>-547</lpage>
</bibl>

<bibl id="B110">
  <title><p>The role of perceptual contrast non-linearities in image transform
  quantization</p></title>
  <aug>
    <au><snm>Malo</snm><fnm>J.</fnm></au>
    <au><snm>Ferri</snm><fnm>F</fnm></au>
    <au><snm>Albert</snm><fnm>J</fnm></au>
    <au><snm>Soret</snm><fnm>J</fnm></au>
    <au><snm>Artigas</snm><fnm>J.M.</fnm></au>
  </aug>
  <source>Image and Vision Computing</source>
  <publisher>Elsevier</publisher>
  <pubdate>2000</pubdate>
  <volume>18</volume>
  <issue>3</issue>
  <fpage>233</fpage>
  <lpage>-246</lpage>
</bibl>

<bibl id="B111">
  <title><p>Blind Image Quality Assessment: From Natural Scene Statistics to
  Perceptual Quality</p></title>
  <aug>
    <au><snm>{Moorthy}</snm><fnm>A. K.</fnm></au>
    <au><snm>{Bovik}</snm><fnm>A. C.</fnm></au>
  </aug>
  <source>IEEE Transactions on Image Processing</source>
  <pubdate>2011</pubdate>
  <volume>20</volume>
  <issue>12</issue>
  <fpage>3350</fpage>
  <lpage>3364</lpage>
</bibl>

<bibl id="B112">
  <title><p>Image denoising using scale mixtures of Gaussians in the wavelet
  domain</p></title>
  <aug>
    <au><snm>Portilla</snm><fnm>J.</fnm></au>
    <au><snm>Strela</snm><fnm>V.</fnm></au>
    <au><snm>Wainwright</snm><fnm>M.</fnm></au>
    <au><snm>Simoncelli</snm><fnm>E.</fnm></au>
  </aug>
  <source>IEEE Trans. Im. Proc.</source>
  <pubdate>2003</pubdate>
  <volume>12</volume>
  <issue>11</issue>
  <fpage>1338</fpage>
  <lpage>1351</lpage>
</bibl>

<bibl id="B113">
  <title><p>The Student-t Mixture as a Natural Image Patch Prior with
  Application to Image Compression</p></title>
  <aug>
    <au><snm>Sinz</snm><fnm>F.</fnm></au>
    <au><snm>Bethge</snm><fnm>M.</fnm></au>
  </aug>
  <source>J. Mach. Learn. Res.</source>
  <pubdate>2014</pubdate>
  <volume>15</volume>
  <fpage>2061</fpage>
  <lpage>2086</lpage>
</bibl>

<bibl id="B114">
  <title><p>The Student-t Mixture as a Natural Image Patch Prior with
  Application to Image Compression</p></title>
  <aug>
    <au><snm>Oord</snm><fnm>A.</fnm></au>
    <au><snm>Schrauwen</snm><fnm>B.</fnm></au>
  </aug>
  <source>J. Mach. Learn. Res.</source>
  <pubdate>2014</pubdate>
  <volume>15</volume>
  <fpage>2061</fpage>
  <lpage>2086</lpage>
</bibl>

<bibl id="B115">
  <title><p>The statistics of natural images</p></title>
  <aug>
    <au><snm>Ruderman</snm><fnm>DL.</fnm></au>
  </aug>
  <source>Network: Comp. Neur. Syst.</source>
  <pubdate>1994</pubdate>
  <volume>5</volume>
  <issue>4</issue>
  <fpage>517</fpage>
  <lpage>-548</lpage>
</bibl>

</refgrp>
} 

\end{backmatter}
\end{document}